\documentclass[preprint2,numberedappendix]{emulateapj-rtx4}
\usepackage{graphicx}
\usepackage{times}
\usepackage{url}
\usepackage{bm}
\usepackage{natbib}
\graphicspath{{./fig/}{./png/}}
\usepackage{color}

\newcommand{\EQ}{\begin{equation}}
\newcommand{\EE}{\end{equation}}
\newcommand{\EQA}{\begin{eqnarray}}
\newcommand{\EEA}{\end{eqnarray}}
\newcommand{\brac}[1]{\langle #1 \rangle}
\newcommand{\pd}{\partial}

\newcommand{\ve}[1]{\boldsymbol{#1}}
\newcommand{\mean}[1]{\overline{#1}}

\newcommand{\mOmega}{\overline{\Omega}}

\newcommand{\etatz}{\eta_{\rm t0}}

\newcommand{\urms}{u_{\rm rms}}

\newcommand{\Beq}{B_{\rm eq}}

\newcommand{\kef}{k_{\rm f}}

\newcommand{\chit}{\chi_{\rm SGS}}
\newcommand{\chitm}{\overline{\chi}_{\rm SGS}}

\newcommand{\Pm}{{\rm Pm}}
\newcommand{\Rm}{{\rm Rm}}
\newcommand{\Rey}{{\rm Re}}

\newcommand{\Pra}{{\rm Pr}}
\newcommand{\Ta}{{\rm Ta}}

\newcommand{\Rat}{{\rm Ra}_{\rm t}}

\newcommand{\Co}{{\rm Co}}

\def\onethird{{\textstyle{1\over3}}}
\def\onehalf{{\textstyle{1\over2}}}

\begin{document}

\title{Effects of enhanced stratification on equatorward dynamo wave propagation}

\author{Petri J.\ K\"apyl\"a$^{1,2}$, Maarit J.\ Mantere$^{3,1}$, Elizabeth Cole$^{1}$, J\"orn Warnecke$^{2,4}$ and Axel Brandenburg$^{2,4}$}
\affil{$^1$Physics Department, Gustaf H\"allstr\"omin katu 2a, PO Box 64,
FI-00014 University of Helsinki, Finland\\
$^2$NORDITA, KTH Royal Institute of Technology and Stockholm University,
Roslagstullsbacken 23, SE-10691 Stockholm, Sweden\\
$^3$Aalto University, Department of Information and Computer Science, 
PO Box 15400, FI-00076 Aalto, Finland \\
$^4$Department of Astronomy, AlbaNova University Center,
Stockholm University, SE-10691 Stockholm, Sweden}
\email{petri.kapyla@helsinki.fi
($ $Revision: 1.334 $ $)
}

\begin{abstract}
  We present results from simulations of rotating magnetized
  turbulent convection in spherical wedge geometry representing parts
  of the latitudinal and longitudinal extents of a star.
  Here we consider a set of runs for which the density stratification is
  varied, keeping the
  Reynolds and Coriolis numbers at similar values. In the case of weak
  stratification, we find quasi-steady dynamo solutions for moderate rotation
  and oscillatory ones with poleward migration of activity belts
  for more rapid rotation. For stronger
  stratification, the growth rate tends to become smaller.
  Furthermore, a transition from quasi-steady to oscillatory dynamos
  is found as the Coriolis number is increased,
  but now there is an equatorward migrating branch near the equator.
  The breakpoint where this happens corresponds to a rotation rate
  that is about 3--7 times the solar value.
  The phase relation of the magnetic field is such that the toroidal
  field lags behind the radial field by about $\pi/2$, which can be
  explained by an oscillatory $\alpha^2$ dynamo caused by the sign
  change of the $\alpha$-effect about the equator.
  We test the domain size dependence of our results for a rapidly
  rotating run with equatorward migration by varying the longitudinal
  extent of our wedge. The energy of the axisymmetric mean magnetic
  field decreases as the domain size increases and we find that an
  $m=1$ mode is excited for a full $2\pi$ azimuthal extent, reminiscent of the
  field configurations deduced from observations of rapidly rotating
  late-type stars.
\end{abstract}


\section{Introduction}
The large-scale magnetic field of the Sun, manifested by the 11 year
sunspot cycle, is generally believed to be generated
within or just below the turbulent
convection zone \citep[e.g.,][and references therein]{O03}. The latter
concept is based on the idea that strong shear in the tachocline
near the bottom of the convection zone amplifies the toroidal magnetic field
which then becomes buoyantly unstable and erupts to the surface
\citep[e.g.,][]{Pa55a}. This process has been adopted in many
mean-field models of the solar cycle in the form of a non-local
$\alpha$-effect \citep[e.g.,][]{KO11}, which is based on early
ideas of \cite{Bab61} and \cite{Lei69} that the source term for
poloidal field can be explained through the tilt of active regions.
Such models assume a reduced
turbulent diffusivity within the convection
zone and a single cell anti-clockwise meridional circulation which
acts as a conveyor belt for the magnetic field. These so-called flux
transport models \citep[e.g.,][]{DC99} are now widely used to study the
solar cycle and to predict its future course \citep{DG06,CCJ07}.

The flux transport paradigm is, however, facing several theoretical
challenges: $10^5$\,gauss magnetic fields are expected to reside in the
tachocline \citep{DSC93}, but such fields are difficult to explain with
dynamo theory \citep{GK11} and may have become unstable at much lower
field strengths \citep{ASR05}.
Furthermore, flux transport dynamos require a rather low value
of the turbulent diffusivity within the convection zone
\citep[several $10^{11}\,{\rm cm}^2\,{\rm s}^{-1}$; see][]{BERB02},
which is much less than the standard estimate of several
$10^{12}\,{\rm cm}^2\,{\rm s}^{-1}$ based on mixing length theory,
which, in turn, is also verified numerically \citep[e.g.,][]{KKB09a}.
Several other issues have already been addressed within this
paradigm, for example, the parity
of the dynamo \citep{BERB02,CNC04,DdTGAW04} and the possibility of a
multicellular structure of the meridional circulation \citep{JB07},
which may be more complicated than that required in the flux transport
models \citep{Ha11,MFRT12,ZBKDH13}.
These difficulties have led to
a revival of the distributed dynamo \citep[e.g.,][]{Br05,Pi13} in which
magnetic fields are generated throughout the convection zone due to
turbulent effects \citep[e.g.,][]{KR80,KKT06,PS09}.

Early studies of self-consistent three-dimensional magnetohydrodynamic
(MHD) simulations of convection in spherical coordinates produced
oscillatory large-scale dynamos \citep{Gi83,Gl85}, but the dynamo wave
was found to propagate toward the poles rather than the equator---as in the
Sun.
These models are referred to as direct numerical simulations (DNS), i.e.,
all operators of viscous and diffusive terms are just the original ones,
but with vastly increased viscosity and diffusivity coefficients.
More recent anelastic large-eddy simulations (LES) with rotation
rates somewhat higher than that of the Sun have produced non-oscillatory
\citep{BBBMT10} and oscillatory \citep{BMBBT11,NBBMT13} large-scale
magnetic fields, depending essentially on the rotation rate and
the vigor of the turbulence.
However, similar models with the solar rotation rate have
either failed to produce an appreciable large-scale component
\citep{BMT04} or, more recently, oscillatory solutions with almost no
latitudinal propagation of the activity belts \citep{GCS10,RCGBS11}.
These simulations covered a full spherical shell
and used realistic values for solar
luminosity and rotation rate, necessitating the use of anelastic
solvers and spherical harmonics \citep[e.g.,][]{BMT04} or implicit
methods \citep[e.g.][]{GCS10}.
Here we exploit an alternative approach by modeling
fully compressible convection
in wedge geometry \citep[see also][]{RC01} with a finite-difference
method. We omit the polar
regions and usually cover only a part of the longitudinal extent,
e.g., $90\degr$ instead of the full $360\degr$. At the cost of
omitting connecting flows across the poles and introducing artificial
boundaries there, the gain is that higher spatial resolution can be
achieved. Furthermore, retaining the sound waves can be beneficial
when considering possible helio- or asteroseismic applications.
Our model is a hybrid between DNS and LES in that we supplement
the thermal energy flux by an additional subgrid scale (SGS) term
to stabilize the scheme and to further reduce the radiative background flux.
Recent hydrodynamic \citep{KMB11,KMGBC11} and MHD
\citep{KKBMT10} studies have shown that this approach
produces results that are in accordance with fully spherical models.
Moreover, the first turbulent dynamo solution with solar-like migration
properties of the magnetic field was recently obtained using this type
of setup \citep{KMB12a}. 
Extended setups that include a coronal layer as a
more realistic upper radial boundary have been successful in producing
dynamo-driven coronal ejections \citep{WKMB12}.  As we show in a
companion paper \citep{WKMB13}, a solar-like differential rotation
pattern might be another consequence of including an outer coronal
layer.

Here we concentrate on exploring further the recent discovery of
equatorward migration in spherical wedge simulations \citep{KMB12a}. In
particular, we examine a set of runs for which the rotational influence on
the fluid, measured by the Coriolis number, which is also called the
inverse Rossby number, is kept approximately constant while the
density stratification of the simulations is gradually increased.

\section{The model}
\label{sec:model}

Our model is the same as that in \cite{KMB12a}.
We consider a wedge in spherical polar coordinates, where
$(r,\theta,\phi)$ denote radius, colatitude, and longitude. The
radial, latitudinal, and longitudinal extents of the wedge are $r_0
\leq r \leq R$, $\theta_0 \leq \theta \leq \pi-\theta_0$, and $0 \leq
\phi \leq \phi_0$, respectively, where $R$ is the radius of the
star and $r_0=0.7\,R$ denotes the position of the bottom of the convection zone.
Here we take $\theta_0=\pi/12$ and in most of our models
we use $\phi_0=\pi/2$, so we cover a quarter of the azimuthal extent
between $\pm75^\circ$ latitude.
We solve the compressible hydromagnetic equations\footnote{Note that
in Equation~(4) of \cite{KMB12a} the Ohmic heating term $\mu_0 \eta {\bm J}^2$
and a factor $\rho$ in the viscous dissipation term
$2\nu \bm{\mathsf{S}}^2$ were missing, but they were
actually included in the calculations.},
\begin{equation}
\frac{\pd \bm A}{\pd t} = {\bm u}\times{\bm B} - \mu_0\eta {\bm J},
\end{equation}
\begin{equation}
\frac{D \ln \rho}{Dt} = -\bm\nabla\cdot\bm{u},
\end{equation}
\begin{equation}
\frac{D\bm{u}}{Dt} = \bm{g} -2\bm\Omega_0\times\bm{u}+\frac{1}{\rho}
\left(\bm{J}\times\bm{B}-\bm\nabla p
+\bm\nabla \cdot 2\nu\rho\bm{\mathsf{S}}\right),
\end{equation}
\begin{equation}
T\frac{D s}{Dt} = \frac{1}{\rho}\left[-\bm\nabla \cdot
\left({\bm F^{\rm rad}}+ {\bm F^{\rm SGS}}\right) +
\mu_0 \eta {\bm J}^2\right] +2\nu \bm{\mathsf{S}}^2,
\label{equ:ss}
\end{equation}
where ${\bm A}$ is the magnetic vector potential, $\bm{u}$ is the
velocity, ${\bm B} =\bm\nabla\times{\bm A}$ is the magnetic field,
${\bm J} =\mu_0^{-1}\bm\nabla\times{\bm B}$ is the current density,
$\mu_0$ is the vacuum
permeability, $D/Dt = \pd/\pd t + \bm{u} \cdot \bm\nabla$ is the
advective time derivative, $\rho$ is the density, 
$\nu$ is the kinematic viscosity, $\eta$ is the magnetic diffusivity,
both assumed constant,
\begin{equation}
{\bm F^{\rm rad}}=-K\ve{\nabla} T\quad\mbox{and}\quad
{\bm F^{\rm SGS}} =-\chit \rho  T\ve{\nabla} s
\end{equation}
are radiative and SGS heat fluxes,
where $K$ is the radiative heat conductivity and $\chit$ is the turbulent heat
conductivity, which represents the unresolved convective transport of heat
and was referred to as $\chi_{\rm t}$ in \cite{KMB12a},
$s$ is the specific entropy, $T$ is the temperature, and $p$ is the
pressure. The fluid obeys the ideal gas law with $p=(\gamma-1)\rho e$,
where $\gamma=c_{\rm P}/c_{\rm V}=5/3$ is the ratio of specific heats
at constant pressure and volume, respectively, and $e=c_{\rm V} T$ is
the specific internal energy.
The rate of strain tensor
$\bm{\mathsf{S}}$ is given by
\begin{equation}
\mathsf{S}_{ij}=\onehalf(u_{i;j}+u_{j;i})
-\onethird \delta_{ij}\bm\nabla\cdot\bm{u},
\end{equation}
where the semicolons denote covariant differentiation \citep{MTBM09}.

The gravitational acceleration is given by
$\bm{g}=-GM\hat{\bm{r}}/r^2$,
where $G$ is the gravitational constant, $M$ is the mass of the star
(without the convection zone),
and $\hat{\bm{r}}$ is the unit vector in the radial direction.
Furthermore, the rotation vector $\bm\Omega_0$ is given by
$\bm\Omega_0=(\cos\theta,-\sin\theta,0)\Omega_0$.

\subsection{Initial and boundary conditions}
\label{sec:initcond}
The initial state is isentropic and the hydrostatic temperature
gradient is given by
\begin{equation}
\frac{\pd T}{\pd r}=-\frac{GM/r^2}{c_{\rm V}(\gamma-1)(n_{\rm ad}+1)},
\end{equation}
where $n_{\rm ad}=1.5$ is the polytropic index for an adiabatic stratification.
We fix the value of $\pd T/\pd r$ on the lower boundary.
The density profile follows from hydrostatic equilibrium.
The heat conduction profile is chosen so that
radiative diffusion is responsible for supplying the energy flux in
the system, with $K$ decreasing more than two orders of magnitude from
bottom to top \citep{KMB11}.
We do this by choosing a variable polytropic index $n=2.5\,(r/r_0)^{-15}-1$,
which equals 1.5 at the bottom of the convection zone and approaches $-1$
closer to the surface.
This means that $K=(n+1)K_0$ decreases toward the surface like $r^{-15}$
such that most of the flux is carried by convection \citep{BCNS05}.
Here, $K_0$ is a constant that will be defined below.

Our simulations are defined by the energy flux imposed at the bottom
boundary, $F_{\rm b}=-(K \pd T/\pd r)|_{r=r_0}$
as well as the values of $\Omega_0$, $\nu$, $\eta$, and
$\chitm=\chit(r_{\rm m}=0.85\, R)$.
Furthermore, the radial profile of $\chit$ is piecewise constant above
$r>0.75R$ with $\chit=\chitm$ at $0.75 < r <0.98$, and
$\chit=12.5\chitm$ above $r=0.98R$. Below $r=0.75R$, $\chit$ tends
smoothly to zero; see Fig.~1 of \cite{KMB11}.

The radial and latitudinal boundaries are assumed to be impenetrable and
stress free, i.e.,
\begin{eqnarray}
&&\!\!\!
u_r=0,\quad \frac{\pd u_\theta}{\pd r}=\frac{u_\theta}{r},\quad \frac{\pd
u_\phi}{\pd r}=\frac{u_\phi}{r} \quad (r=r_0, R),\\
&&\!\!\!
\frac{\pd u_r}{\pd \theta}=u_\theta=0,\quad \frac{\pd u_\phi}{\pd
\theta}=u_\phi \cot \theta \quad (\theta=\theta_0,\pi-\theta_0).
\quad
\end{eqnarray}
For the magnetic field we assume perfect conductors on the latitudinal
and lower radial boundaries, and radial field on the outer
radial boundary. In terms of the magnetic vector potential these
translate to
\begin{eqnarray}
&&\!\!\!
\frac{\pd A_r}{\pd r}= A_\theta=A_\phi =0 \,\quad
(r=r_0),\\
&&\!\!\!
A_r=0, \;\; \frac{\pd A_{\theta}}{\pd r}=-\frac{A_{\theta}}{r},\;\; \frac{\pd
A_{\phi}}{\pd r}=-\frac{A_{\phi}}{r} \quad (r=R),\quad\\
&&\!\!\!
A_r=\frac{\pd A_\theta}{\pd\theta}=A_\phi=0 \quad
(\theta=\theta_0,\pi-\theta_0).
\end{eqnarray}
We use small-scale low amplitude Gaussian noise as initial
condition for velocity and magnetic field.
On the latitudinal boundaries we assume that the density and
entropy have vanishing first derivatives, thus suppressing heat fluxes
through the boundaries.

On the upper radial boundary we apply a black body condition
\begin{equation}
\sigma T^4  = -K\nabla_r T - \chit \rho T \nabla_r s,
\label{eq:bbb}
\end{equation}
where $\sigma$ is the Stefan--Boltzmann constant.
We use a modified value for $\sigma$ that takes into account that both
surface temperature and energy flux through the domain are larger than in
the Sun.
The value of $\sigma$ can be chosen so that the flux at the surface carries 
the total luminosity through the boundary in the initial non-convecting 
state. However, in many cases we have changed the value of $\sigma$
during runtime to speed up thermal relaxation.

\subsection{Dimensionless parameters}
\label{sec:dimless}

To facilitate comparison with other work using different normalizations,
we present our results by normalizing with
physically meaningful quantities.
We note, however, that in the code we used non-dimensional quantities
by choosing
\begin{equation}
R = GM = \rho_0 = c_{\rm P} = \mu_0 = 1,
\end{equation}
where $\rho_0$ is the initial density at $r=r_0$.
The units of length, time, velocity, density, entropy, and magnetic field
are therefore
\begin{eqnarray}
&[x] = R,\,\,\, [t]=\sqrt{R^3/GM},\,\,\, [u] = \sqrt{GM/R},& \nonumber
\\ &[\rho] =\rho_0,\,\,\, [s] = c_{\rm P},\,\,\, [B]= \sqrt{\rho_0
  \mu_0 GM/R}.&
\end{eqnarray}
The radiative conductivity is proportional to
$K_0=(\mathcal{L}/4\pi)c_{\rm V}(\gamma-1)(n_{\rm ad}+1)\rho_0\sqrt{GMR}$,
where $\mathcal{L}$ is the non-dimensional luminosity, given below.
The corresponding nondimensional input parameters are the luminosity parameter
\begin{equation}
\mathcal{L} = \frac{L_0}{\rho_0 (GM)^{3/2} R^{1/2}},
\end{equation}
the normalized pressure scale height at the surface,
\begin{equation}
\xi = \frac{(\gamma-1) c_{\rm V}T_1}{GM/R},
\end{equation}
with $T_1$ being the temperature at the surface,
the Taylor number
\begin{equation}
\Ta=(2\Omega_0 R^2/\nu)^2,
\end{equation}
the fluid and magnetic Prandtl numbers
\begin{equation}
\Pra=\frac{\nu}{\chi_{\rm m}},\quad \Pra_{\rm SGS}=\frac{\nu}{\chitm},\quad \Pm=\frac{\nu}{\eta},
\end{equation}
where $\chi_{\rm m}=K/c_{\rm P} \rho_{\rm m}$ and $\rho_{\rm m}$ are
the thermal diffusivity and density at $r=r_{\rm m}$,
respectively.
Finally, we have the non-dimensional viscosity
\begin{equation}
\tilde{\nu}=\frac{\nu}{\sqrt{GMR}}.
\end{equation}
Instead of $\xi$, we often quote the initial density contrast,
$\Gamma_\rho^{(0)}\equiv\rho(r_0)/\rho(R)$.
The density contrast can change during the run. We list the 
final values of $\Gamma_\rho$ from the thermally saturated stage in 
Table~\ref{tab:runs}.

Other useful diagnostic parameters are the fluid and magnetic Reynolds numbers
\begin{equation}
\Rey=\frac{\urms}{\nu \kef},\quad \Rm=\frac{\urms}{\eta \kef},
\end{equation}
where $\kef=2\pi/\Delta r\approx21 R^{-1}$ is an estimate of the wavenumber of the largest eddies,
and $\Delta r=R-r_0=0.3\,R$ is the thickness of the layer.
The Coriolis number is defined as
\begin{equation}
\Co=\frac{2\Omega_0}{\urms \kef},
\end{equation}
where $\urms=\sqrt{(3/2)\brac{u_r^2+u_\theta^2}_{r\theta\phi t}}$ is
the rms velocity and the subscripts indicate averaging over $r$,
$\theta$,
$\phi$, and a time interval during which the run is thermally
relaxed and which covers several magnetic diffusion times.
The averaging procedures employ the correct volume or surface
elements of spherical polar coordinates.
Note that for $\urms$ we omit the contribution from the azimuthal
velocity, because its value is dominated by effects from the
differential rotation \citep{KMGBC11} and compensate for this with
the $3/2$ factor.
The Taylor number can also be
written as $\Ta=\Co^2\Rey^2(\kef R)^4$. Due
to the fact that the initial stratification is isentropic, we quote
the turbulent Rayleigh number $\Rat$ from the thermally
relaxed state of the run,
\begin{eqnarray}
\Rat\!=\!\frac{GM(\Delta r)^4}{\nu \chitm R^2} \bigg(-\frac{1}{c_{\rm P}}\frac{{\rm d}\brac{s}_{\theta \phi t}}{{\rm d}r} \bigg)_{r_{\rm m}}.
\label{equ:Co}
\end{eqnarray}
We also quote the value of $k_\omega=\omega_{\rm rms}/\urms$, where
$\bm\omega=\bm\nabla \times {\bm u}$, and $\omega_{\rm rms}$ is the
volume averaged rms value of $\bm\omega$.
The magnetic field is expressed in equipartition field strengths,
$\Beq(r)=\langle \mu_0 \rho \bm{u}^2 \rangle^{1/2}_{\theta\phi t}$, where
all three components of $\bm{u}$ are included.
We define mean quantities as averages over the $\phi$-coordinate and
denote them by overbars.
However, as we will see, there can also be significant power in non-axisymmetric
spherical harmonic modes with low azimuthal degree $m=1$ and 2, which will be discussed
at the end of the paper.

The simulations were performed with the {\sc Pencil
  Code}\footnote{http://pencil-code.googlecode.com/}, which uses a
high-order finite difference method for solving the compressible
equations of magnetohydrodynamics.

\begin{deluxetable*}{ccccccccccccccccc}
\tabletypesize{\scriptsize}
\tablecaption{Summary of the runs.}
\tablecomments{Columns 2--7 and 9--11 show 
quantities that are input parameters to the models whereas the quantities 
in the eight and the last four columns are results of the simulations
computed from the
saturated state. Here we use $\phi_0=\pi/2$ in Sets~A--D. In Set~E we 
use $\phi_0=\pi/4$ (Run~E1), $\phi_0=\pi/2$ (E2), $\phi_0=\pi$ (E3), 
and $\phi_0=2\pi$ (E4). Runs~C1 and E2 are the same model, which is also
the same as Run~B4m of \cite{KMB12a}. Here $\Gamma_\rho$ is the
density stratification in the final saturated state and
$\tilde\sigma=\sigma R^2 T_0^4/L_0$, where $T_0$ is the temperature at 
the base of the convection zone.}

\tablewidth{0pt}
\tablehead{
\colhead{Run} & \colhead{grid} & \colhead{$\Pra$} & \colhead{$\Pra_{\rm SGS}$} & \colhead{$\Pm$} & 
\colhead{$\mbox{Ta} [10^{10}]$} & \colhead{$\xi$} & \colhead{$\Gamma_\rho^{(0)}$} & \colhead{$\Gamma_\rho$} & \colhead{$\tilde\nu [10^{-5}]$} & \colhead{$\mathcal{L} [10^{-5}]$} & \colhead{$\tilde\sigma$} & \colhead{${\rm Ra}_{\rm t} [10^6]$} & \colhead{$\Rey$} & \colhead{$\Rm$} & \colhead{$\Co$} & 
}
\startdata
A1   & $128\times256\times128$  & 71 & $1.5$ & $1.0$ & $1.0$  & 0.29 &  2.0 & 2.1 & $1.7$ & $3.8$ & $0.92$ & $0.83$ & $26$ & $26$  & $8.6$ \\ 
A2   & $128\times256\times128$  & 71 & $1.5$ & $1.0$ & $1.8$ & 0.29 &  2.0 & 2.1 & $1.7$ & $3.8$ & $0.92$ & $0.11$ & $24$ & $24$  & $12.8$ \\ 
           \hline
B1   & $128\times256\times128$  & 82 & $2.5$ & $1.0$ &  $0.64$ & 0.09 &  5.0 & 5.3 & $2.9$ & $3.8$ & $10.9$ & $1.1$ & $22$ & $22$  & $8.1$  \\ 
B2   & $128\times256\times128$  & 82 & $2.5$ & $1.0$ & $1.4$ & 0.09 &  5.0 & 5.2 & $2.9$ & $3.8$ & $10.9$ & $1.1$ & $20$ & $20$  & $13.7$ \\ 
           \hline
C1   & $128\times256\times128$  & 56 & $2.5$ & $1.0$ & $1.4$ & 0.02 & 30 & 22 & $2.9$ & $3.8$ & $1.4\cdot10^{3}$ & $2.1$  & $35$ & $35$  &  $7.8$ \\ 
C2   & $128\times256\times128$  & 56 & $2.5$ & $1.0$ & $4.0$ & 0.02 & 30 & 21 & $2.9$ & $3.8$ & $1.4\cdot10^{3}$ & $2.7$  & $31$ & $31$  & $14.8$ \\ 
           \hline
D1   & $128\times256\times128$  & 503 & $7.5$ & $3.0$ & $0.16$ & 0.008 & 100 & 85 & $4.7$ & $0.63$ & $3.9\cdot10^{4}$ & $1.2$ & $11$ & $34$  &  $8.0$ \\ 
D2   & $256\times512\times256$  & 269 & $4.0$ & $2.0$ & $1.0$  & 0.008 & 100 & 74 & $2.5$ & $0.63$ & $3.9\cdot10^{4}$ & $2.4$ & $25$ & $50$  &  $9.1$ \\ 
           \hline
E1   & $128\times256\times64$   & 56 & $2.5$ & $1.0$ & $1.4$ & 0.02 &  30 & 22 & $2.9$ & $3.8$ & $1.4\cdot10^{3}$ & $2.1$ & $34$ & $34$  &  $7.9$ \\ 
E2   & $128\times256\times128$  & 56 & $2.5$ & $1.0$ & $1.4$ & 0.02 & 30 & 22 & $2.9$ & $3.8$ & $1.4\cdot10^{3}$ & $2.1$  & $35$ & $35$  &  $7.8$ \\ 
E3   & $128\times256\times256$  & 56 & $2.5$ & $1.0$ & $1.4$ & 0.02 & 30 & 22 & $2.9$ & $3.8$ & $1.4\cdot10^{3}$ & $2.4$  & $35$ & $35$  &  $7.9$ \\ 
E4   & $128\times256\times512$  & 67 & $3.0$ & $1.0$ & $1.0$ & 0.02 & 30 & 23 & $3.5$ & $3.8$ & $1.4\cdot10^{3}$ & $2.2$  & $28$ & $28$  &  $8.1$ 
\enddata
\label{tab:runs}
\end{deluxetable*}

\begin{deluxetable*}{cccccccccccc}
\tabletypesize{\scriptsize}
\tablecaption{Summary of diagnostic variables.}
\tablecomments{Here $\tilde\lambda=\lambda/(\urms \kef)$ is the 
  normalized growth rate of the magnetic field
  and $\tilde\urms=\urms/\sqrt{GM/R}$ is the non-dimensional rms velocity.
  $E_{\rm kin}=\onehalf \langle\rho {\bm u}^2\rangle$
  is the volume averaged kinetic energy. $E_{\rm mer}=\onehalf
  \langle \rho (\mean{u}_r^2+\mean{u}_\theta^2) \rangle$ and $E_{\rm
    rot}=\onehalf \langle \rho \mean{u}_\phi^2 \rangle$ denote the
volume averaged energies of the azimuthally averaged meridional circulation and
  differential rotation. Analogously $E_{\rm mag}=\onehalf \langle
  {\bm B}^2 \rangle$ is the total volume averaged magnetic energy while
  $E_{\rm pol}=\onehalf
  \langle (\mean{B}_r^2+\mean{B}_\theta^2) \rangle$ and
    $E_{\rm tor}=\onehalf \langle \mean{B}_\phi^2 \rangle$ are the energies 
  in the axisymmetric part of the poloidal and toroidal magnetic fields.
} 
\tablewidth{0pt} \tablehead{
\colhead{Run} & $\tilde\lambda$ & $\tilde\urms$ & \colhead{$E_{\rm mer}/E_{\rm kin}$} & \colhead{$E_{\rm rot}/E_{\rm kin}$} & \colhead{$E_{\rm mag}/E_{\rm kin}$} &  \colhead{$E_{\rm pol}/E_{\rm mag}$} & \colhead{$E_{\rm tor}/E_{\rm mag}$} & \colhead{$\Delta_\Omega^{(r)}$} & \colhead{$\Delta_\Omega^{(\theta)}$} & \colhead{$k_\omega$} 
}
\startdata
A1   & 0.084 & 0.010 & 0.000 & 0.580 & 0.418 & 0.045 & 0.396 & 0.013 & 0.089 & 62 \\ 
A2   & 0.095 & 0.009 & 0.000 & 0.490 & 0.553 & 0.068 & 0.338 & 0.009 & 0.050 & 62 \\ 
           \hline
B1   & 0.028 & 0.013 & 0.000 & 0.705 & 0.345 & 0.038 & 0.487 & 0.034 & 0.142 & 68 \\ 
B2   & 0.098 & 0.012 & 0.000 & 0.757 & 0.222 & 0.056 & 0.427 & 0.023 & 0.072 & 72 \\ 
           \hline
C1   & 0.006 & 0.021 & 0.001 & 0.440 & 0.346 & 0.138 & 0.203 & 0.047 & 0.068 & 93 \\ 
C2   & 0.105 & 0.019 & 0.001 & 0.326 & 0.706 & 0.198 & 0.238 & 0.016 & 0.030 & 94 \\ 
           \hline
D1   & 0.003 & 0.011 & 0.002 & 0.222 & 0.472 & 0.166 & 0.135 & 0.011 &-0.000 & 89  \\ 
D2   & 0.003 & 0.013 & 0.000 & 0.617 & 0.222 & 0.133 & 0.190 & 0.045 & 0.058 & 116 \\ 
           \hline
E1   & 0.007 & 0.021 & 0.001 & 0.478 & 0.393 & 0.133 & 0.328 & 0.048 & 0.069 & 92  \\ 
E2   & 0.006 & 0.021 & 0.001 & 0.440 & 0.346 & 0.138 & 0.203 & 0.047 & 0.068 & 93 \\ 
E3   & 0.005 & 0.021 & 0.001 & 0.375 & 0.380 & 0.120 & 0.172 & 0.037 & 0.055 & 92 \\ 
E4   & 0.024 & 0.020 & 0.001 & 0.410 & 0.477 & 0.016 & 0.080 & 0.028 & 0.054 & 89 
\enddata
\label{tab:runs2}
\end{deluxetable*}

\subsection{Relation to reality}
\label{sec:reality}
In simulations, the maximum possible Rayleigh number is much smaller
than in real stars due to the higher diffusivities.
This implies higher energy fluxes and thus larger Mach numbers \citep{BCNS05}.
To have realistic Coriolis numbers, the angular velocity
in the Coriolis force has to be increased in proportion to one third power 
of the increase of the energy flux, but the
centrifugal acceleration is omitted, as it would otherwise be unrealistically
large \cite[cf.][]{KMGBC11}.
In the present models this would mean that the centrifugal acceleration is
of the same order of magnitude as gravity, thus significantly altering the
hydrostatic balance.

We note that we intend to use low values of $\mathcal{L}$ so that
the Mach number is sufficiently below unity.
This is particularly important when the stratification is strong.
In our current formulation the unresolved turbulent heat conductivity,
$\chit$, acts on the total entropy and thus contributes to the radial
heat flux.
In the current models with $\Pra_{\rm SGS}$ greater than unity, the SGS-flux
accounts for a few per cent of the total flux within the convection zone.
Using smaller values of $\Pra_{\rm SGS}$ at the same Reynolds number would
lead to a greater contribution due to the SGS-flux.
To minimize the effects of the SGS-flux within the convection zone,
we use the smallest possible value of $\chit$ that is still compatible
with numerical stability.

The span of time scales in our model is strongly compressed
so as to comprise the full range all the way to the viscous,
thermal, and resistive time scales.
In the following we define acoustic, convective, thermal,
resistive, and viscous time scales as follows:
\begin{equation}
\tau_{\rm ac}=\sqrt{R^3/GM},\quad
\tau_{\rm conv}=H_{\rm P0}/\urms^{\rm(ref)},
\end{equation}
\begin{equation}
\tau_{\rm th}=H_{\rm P0}^2/\chi_0,\quad
\tau_{\rm res}=H_{\rm P0}^2/\eta,\quad
\tau_{\rm visc}=H_{\rm P0}^2/\nu,
\end{equation}
where $H_{\rm P0}$ is the pressure scale height at $r_0$,
$\urms^{\rm(ref)}=(F_0/\rho_0)^{1/3}$ is a convective reference velocity
based on the luminosity of the model, $L=4\pi r_0^2 F_0$,
and $F_0$ is the total flux at $r_0$.

\begin{figure}[t]
\centering
\includegraphics[width=0.46\textwidth]{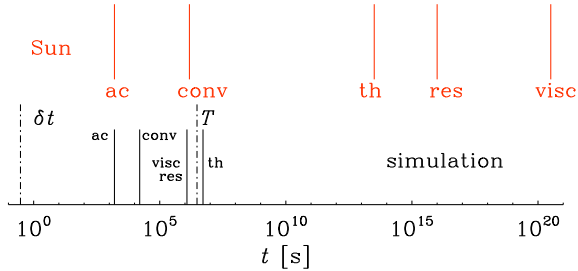}
\caption{
Visual comparison of acoustic, convective, thermal, resistive, and viscous
time scales both in the Sun (upper part, in red) and in Run~C1 (lower part).
In our models, resistive and viscous time scales are often equal, and
the thermal time scale is 1.5--7.5 times longer; see
Columns 4 and 5 of Table~\ref{tab:runs}.
The simulation time scales are confined between the length
of the time step, $\delta t$, and the maximum run time, $T$.
}\label{fig:ptime_scales}
\end{figure}

A visual comparison of these different time scales for the Sun
and Run~C1 is given in Figure~\ref{fig:ptime_scales}.
In order to allow for slow thermal and resistive relaxation processes,
we require that their respective time scales are shorter than the run time
$T$ of the simulation.
As stated in Section~\ref{sec:dimless}, the acoustic time scale of
our model is equal to that of the Sun.
This implies that all the other time scales must be significantly
reduced: $\tau_{\rm conv}$ by a factor 70--100,
$\tau_{\rm th}$ and $\tau_{\rm res}$ by a factor $10^7$,
and $\tau_{\rm visc}$ by a factor $10^{14}$.
This is accomplished by taking values of ${\cal L}$ that are not as
small as in the Sun (where ${\cal L}\approx5\times10^{-11}$),
but typically $3.8\times10^{-5}$ for Run~C1.
This just corresponds to taking values of the Rayleigh number that are
on the order of $10^6$ rather than solar values (in excess of $10^{24}$).
Likewise, shorter thermal, resistive, and viscous time scales are obtained
by choosing values of the magnetic and fluid Reynolds number that are
not as large as in the Sun and by choosing magnetic and fluid Prandtl
numbers that are not as small as in the Sun.

For the purpose of comparing dynamo time scales of the model with the Sun,
it is useful to rescale them such that $\tau_{\rm conv}$
coincides with that of the Sun.
We can then compare the rotation rates of our models
in Table~\ref{tab:runs} with that of the Sun: Runs~A1 and A2 are
2 and 3 times solar, B1 and B2 are 3 and 4.4 times solar,
C1 (including all of Set~E) and C2 are 4.4 and 7 times solar, and
D1 and D2 are 4.3 and 6 times solar.

\section{Results}
\label{sec:results}

We perform runs for four values of $\xi$, corresponding to initial
density contrasts $\Gamma_\rho^{(0)}=2, 5, 30$, and $100$.
These runs are referred to as Sets A--D.
In Set E we use $\Gamma_\rho^{(0)}=30$ and vary $\phi_0$ with all other
parameters being kept the same as in Run~C1, except in Run~E4 where
we use 20\% higher viscosity and magnetic diffusivity than
in the other runs in Set~E.
For each series, we consider different values of $\Ta$ and, as a
consequence, of Co and Re.
The hydrodynamic progenitors of
the Runs~B1, C1, and D1 correspond to Runs~A4, B4, and C4,
respectively, from
\cite{KMB11}. The rest of the simulations were run from the
initial conditions described in Section~\ref{sec:initcond}.

Earlier studies applying fully spherical simulations have shown that
organized large-scale magnetic fields appear provided the rotation of
the star is rapid enough \citep{BBBMT10} and that at even higher
rotation rates, cyclic solutions with poleward migration of the
activity belts are obtained
\citep{BMBBT11}. A similar transition has been observed in the spherical
wedge models of \cite{KKBMT10} and \cite{KMB12a}. However, in the
former case the oscillatory mode showed poleward migration, whereas in
the latter an equatorward branch appears
near the equator. Furthermore, in these runs the dynamo
mode changes from one showing a high frequency cycle with poleward
migration near the equator to another mode with lower frequency and
equatorward migration when the magnetic
field becomes dynamically important. 

\begin{figure}[t]
\centering
\includegraphics[width=0.46\textwidth]{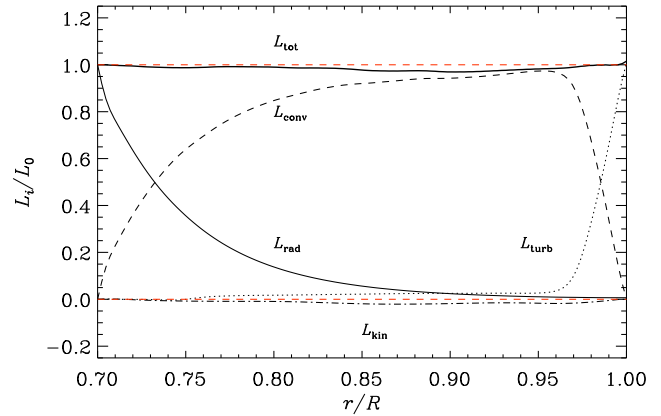}
\caption{Luminosity of the energy fluxes from Run~E4: radiative conduction
(thin solid line), enthalpy (dashed), kinetic energy (dot-dashed),
and unresolved subgrid scale (dotted) fluxes.
The thick solid line is the sum of all contributions.
The two dashed red lines indicate the zero and unity lines.
}\label{fig:pflu}
\end{figure}

There are several differences between the models of \cite{KKBMT10} and
\cite{KMB12a}: the amount of density stratification
(a density contrast of 3 in comparison to
30), the efficiency of convective energy transport (20\% versus
close to 100\% in the majority of the domain achieved by the
use of $\chit$;
see also Figure~\ref{fig:pflu}), and the top
boundary condition for entropy (constant temperature versus black body
radiation). Here we concentrate on studying the influence of the
density stratification on models similar to those presented in
\cite{KMB12a}.

\subsection{Thermal boundary effects and energy balance}

\begin{figure}[t]
\centering
\includegraphics[width=0.46\textwidth]{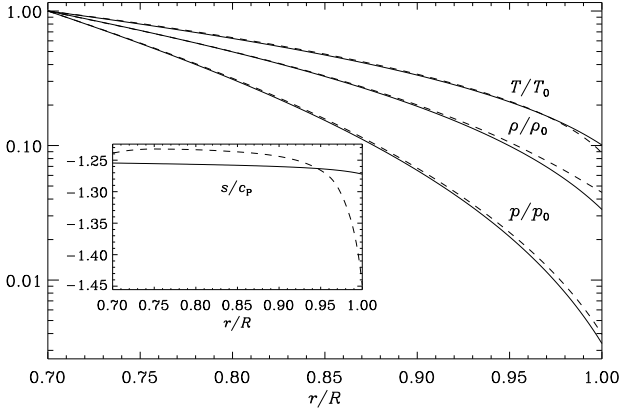}
\caption{Initial (solid lines) and saturated (dashed) radial profiles
  of temperature $T$, density $\rho$, and pressure $p$, normalized by
  their respective values at the bottom of the domain (indicated by 
  the subscript zero) from Run~C1. The
  inset shows the specific entropy $s/c_{\rm P}$ from the same
  run.}\label{fig:pstrat}
\end{figure}

In \cite{KMB11} we started to apply the black--body boundary condition,
Equation~(\ref{eq:bbb}), that has previously been used in mean-field models
with thermodynamics \citep{R89,BMT92,KM00}. Instead of using the physical
value for the Stefan--Boltzmann constant, we estimate the value of
$\sigma$ so that the flux at the upper boundary is approximately that
needed to transport the total luminosity of the star through
the surface; see Table~\ref{tab:runs}. However, the final thermally
relaxed state of the
simulation can significantly deviate from the initial state. In
combination with the nonlinearity of Equation~(\ref{eq:bbb}), the final
stratification is usually somewhat different from the initial one; see
Figure~\ref{fig:pstrat} for an illustrative example from Run~C1. The
final density stratification in this case is around 22, down from 30
in the initial state.

The main advantage of the black--body condition is that it allows the
temperature at the surface more freedom than in our previous models
where a constant temperature was imposed
\citep{KKBMT10,KMGBC11}. In particular, as the temperature is allowed
to
vary at the surface, this can be used as a diagnostic for possible
irradiance variations.
These issues are discussed further in Section~\ref{sec:irra}.

Considering the energy balance, we show the averaged radial energy
fluxes for Run~E4
in Figure~\ref{fig:pflu}. We find that the simulation is thermally
relaxed and that the total luminosity is close to the input
luminosity, i.e., $L_{\rm tot}-L_0\approx0$.
The fluxes are defined as:
\begin{eqnarray}
\mathcal{F}_{\rm rad} &=& -K  \langle \nabla_r T \rangle,\\
\mathcal{F}_{\rm conv} &=& c_{\rm P}\langle(\rho u_r)^{\prime}T^{\prime} \rangle,\\
\mathcal{F}_{\rm kin} &=& \onehalf \left\langle\rho u_r{\bm u}^2\right\rangle,\\
\mathcal{F}_{\rm visc} &=& -2\nu \left\langle \rho u_i S_{ir} \right\rangle,\\
\mathcal{F}_{\rm turb} &=& -\chit \langle\rho T
\nabla_r s\rangle,\\
\mathcal{F}_{\rm Poyn} &=& \langle E_\theta B_\phi - E_\phi B_\theta
\rangle/\mu_0,
\end{eqnarray}
where ${\bm E}=\eta \mu_0 {\bm J}-{\bm u}\times{\bm B}$, the primes
denote fluctuations, and angle brackets abbreviate 
$\langle\cdot\rangle_{\theta,\phi,t}$.
The radiative flux carries energy into the convection zone and drops
steeply as a function of radius so that it contributes only a few per
cent in the middle of the convection zone. The resolved convection is
responsible for transporting the energy through the majority of the
layer, whereas the unresolved turbulent transport carries energy
through the outer surface. 
The viscous and Poynting fluxes are much smaller, and are thus omitted in
this figure.
The flux of kinetic energy is also very small in the rapid
rotation regime considered here \citep[see also][]{ABBMT12}. 

\subsection{Dynamo excitation and large-scale magnetic fields}

\begin{figure}[t]
\centering
\includegraphics[width=0.48\textwidth]{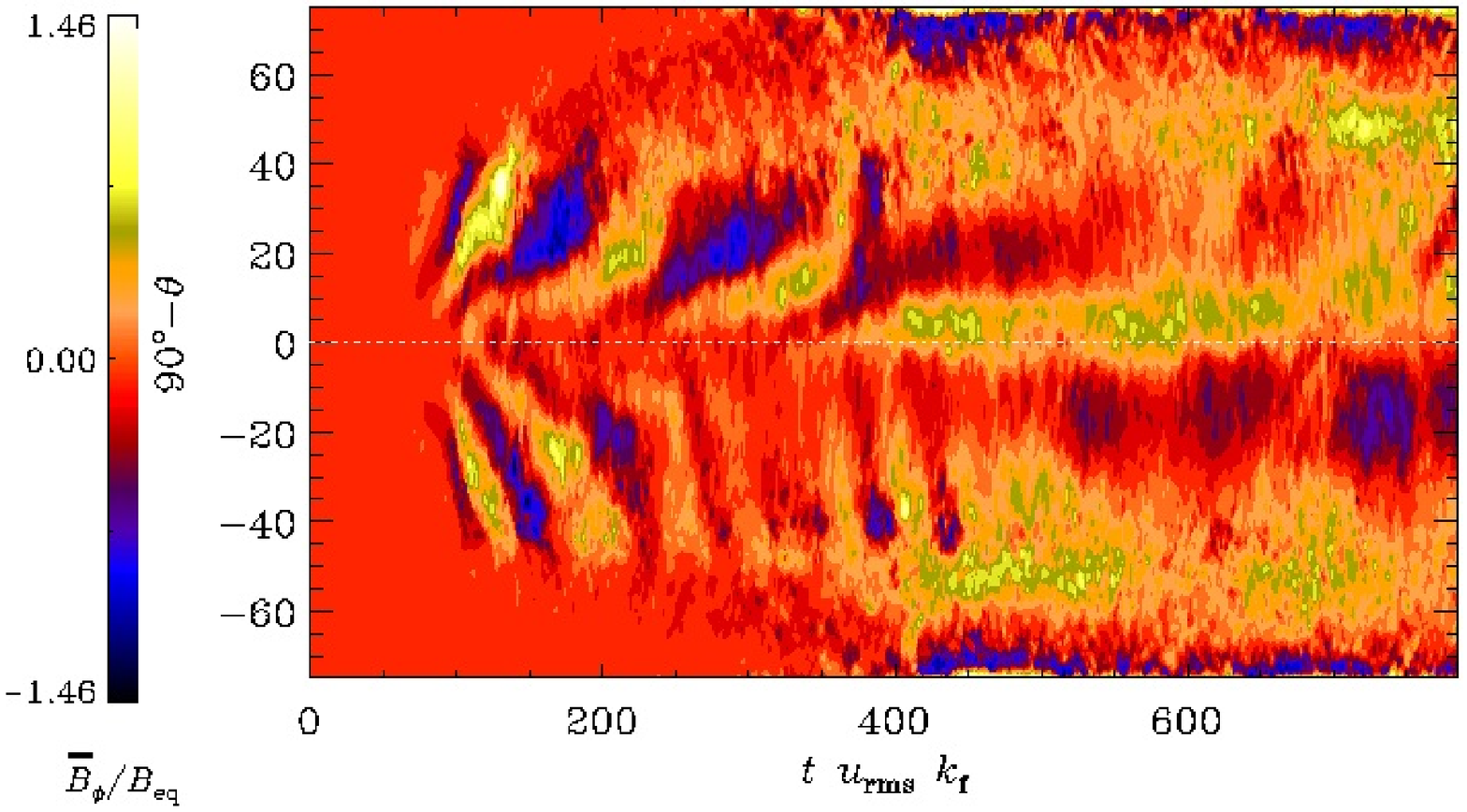}
\includegraphics[width=0.48\textwidth]{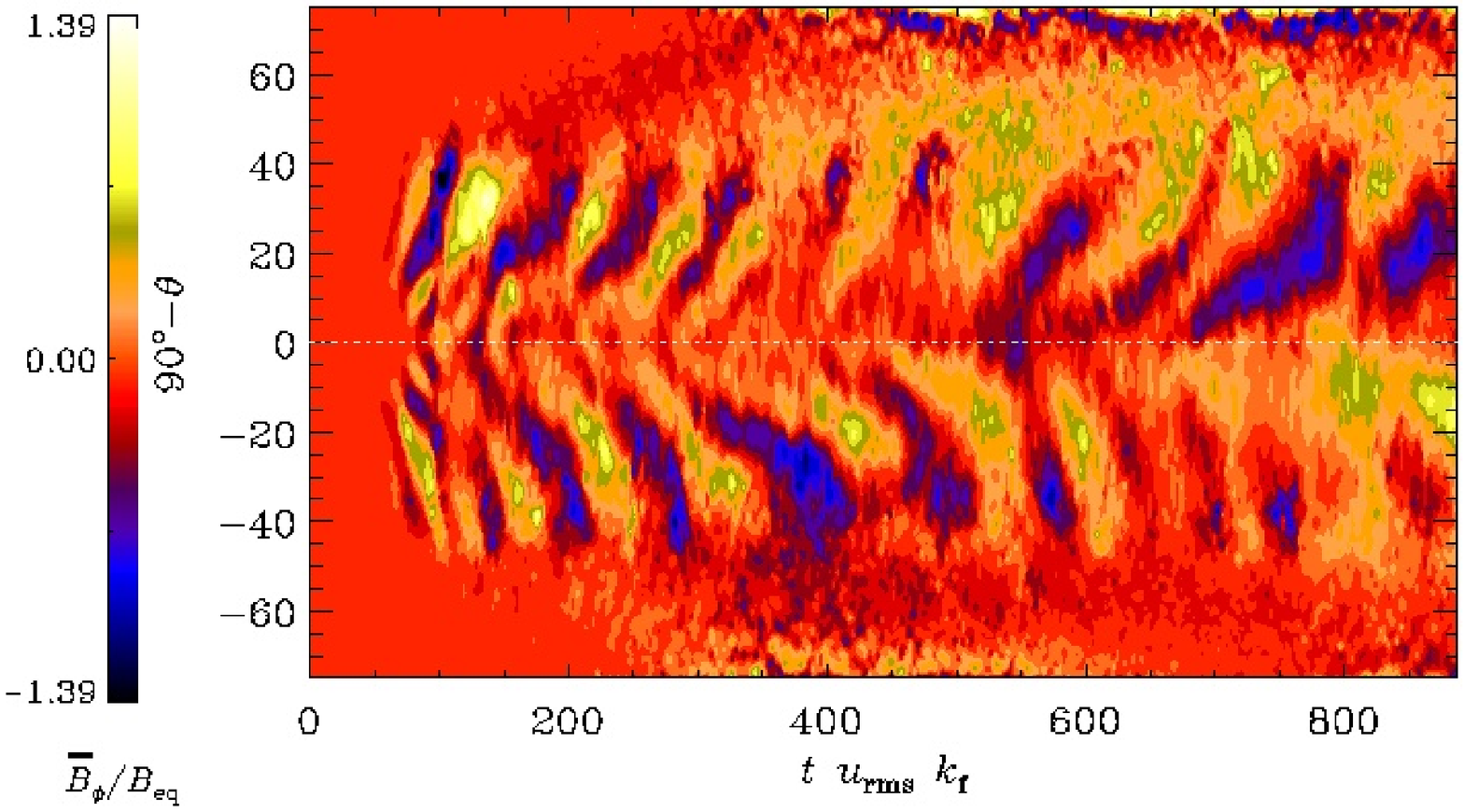}
\caption{$\mean{B}_\phi$ near the surface of the star at $r=0.98\,R$ as
 a function of latitude ($=90\degr-\theta$) and time for Runs~A1 (top) and
  A2 (bottom). The white dotted line denotes the equator
  $90\degr-\theta=0$.  }\label{fig:butterflyA}
\end{figure}

\begin{figure}[t]
\centering
\includegraphics[width=0.48\textwidth]{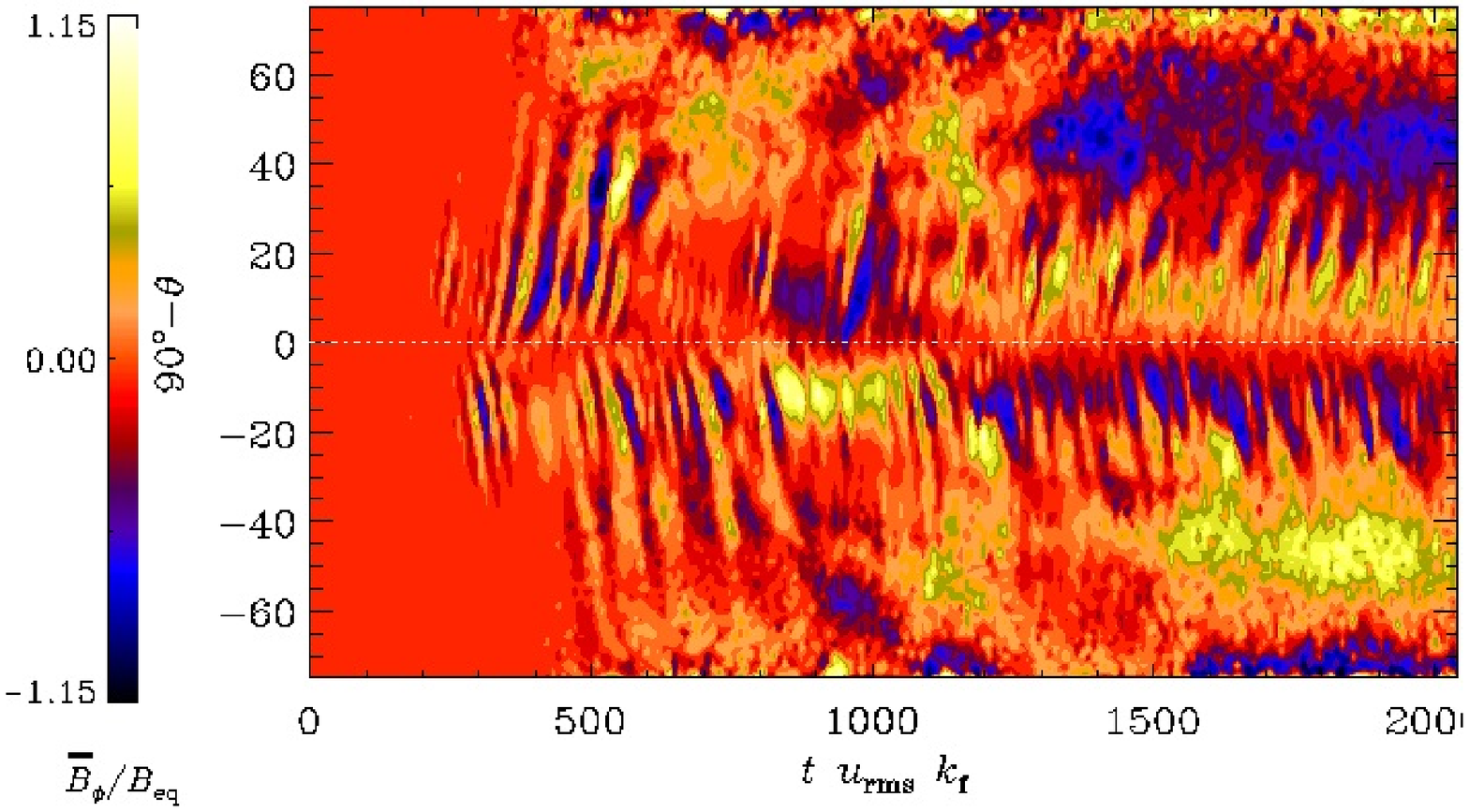}
\includegraphics[width=0.48\textwidth]{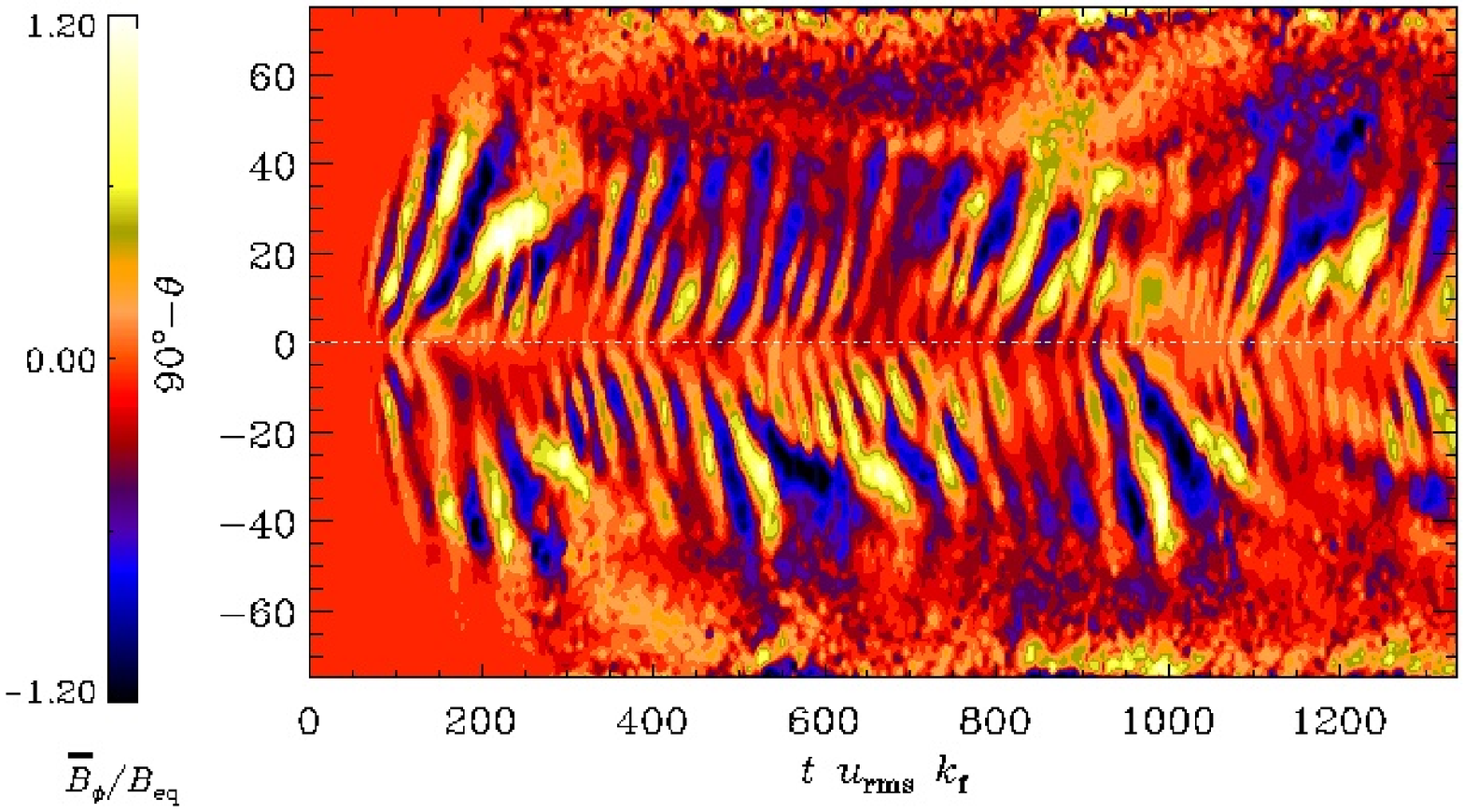}
\caption{Same as Figure~\ref{fig:butterflyA}, but for Runs B1 (top)
  and B2 (bottom).}\label{fig:butterflyB}
\end{figure}

\begin{figure}[t]
\centering
\includegraphics[width=0.48\textwidth]{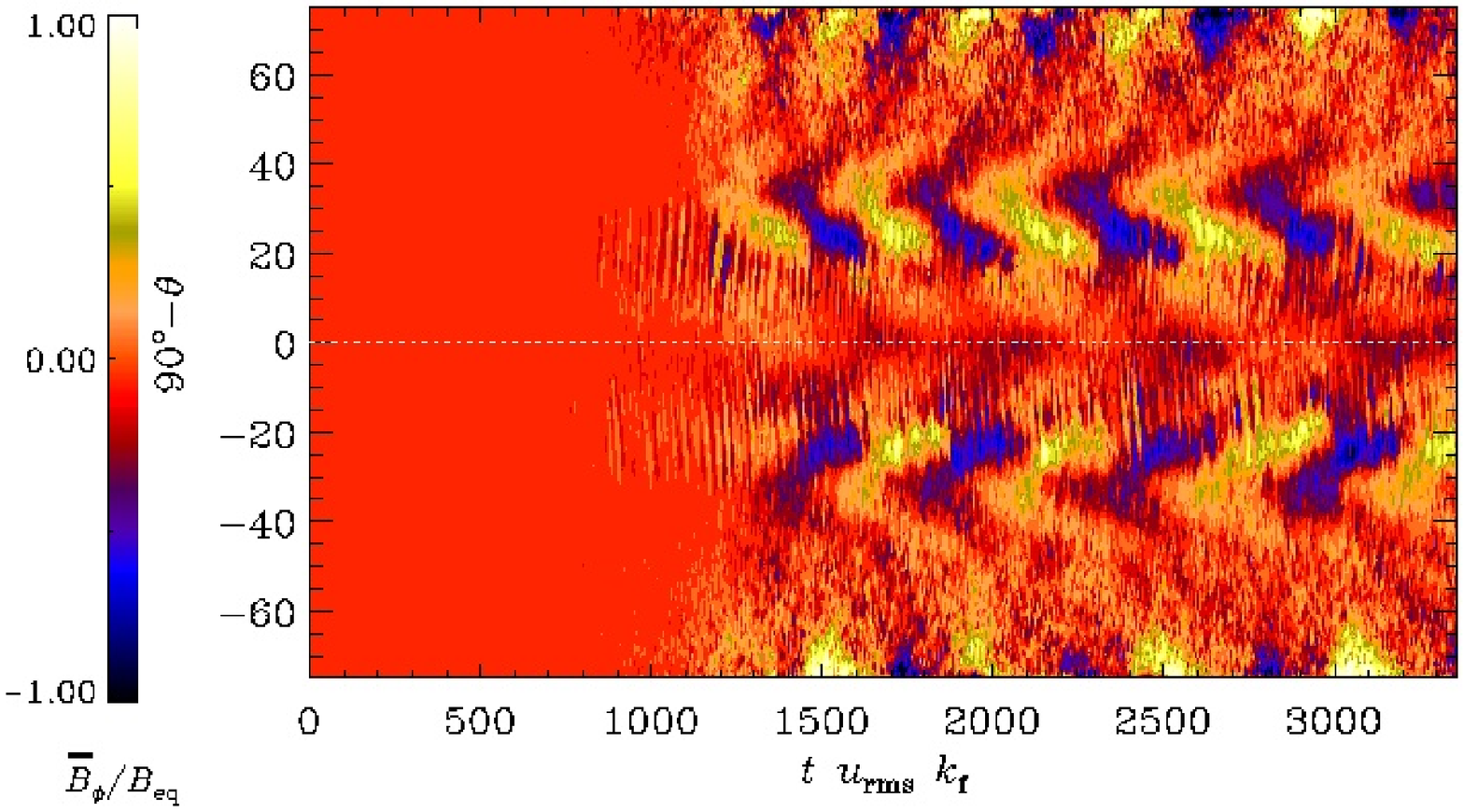}
\includegraphics[width=0.48\textwidth]{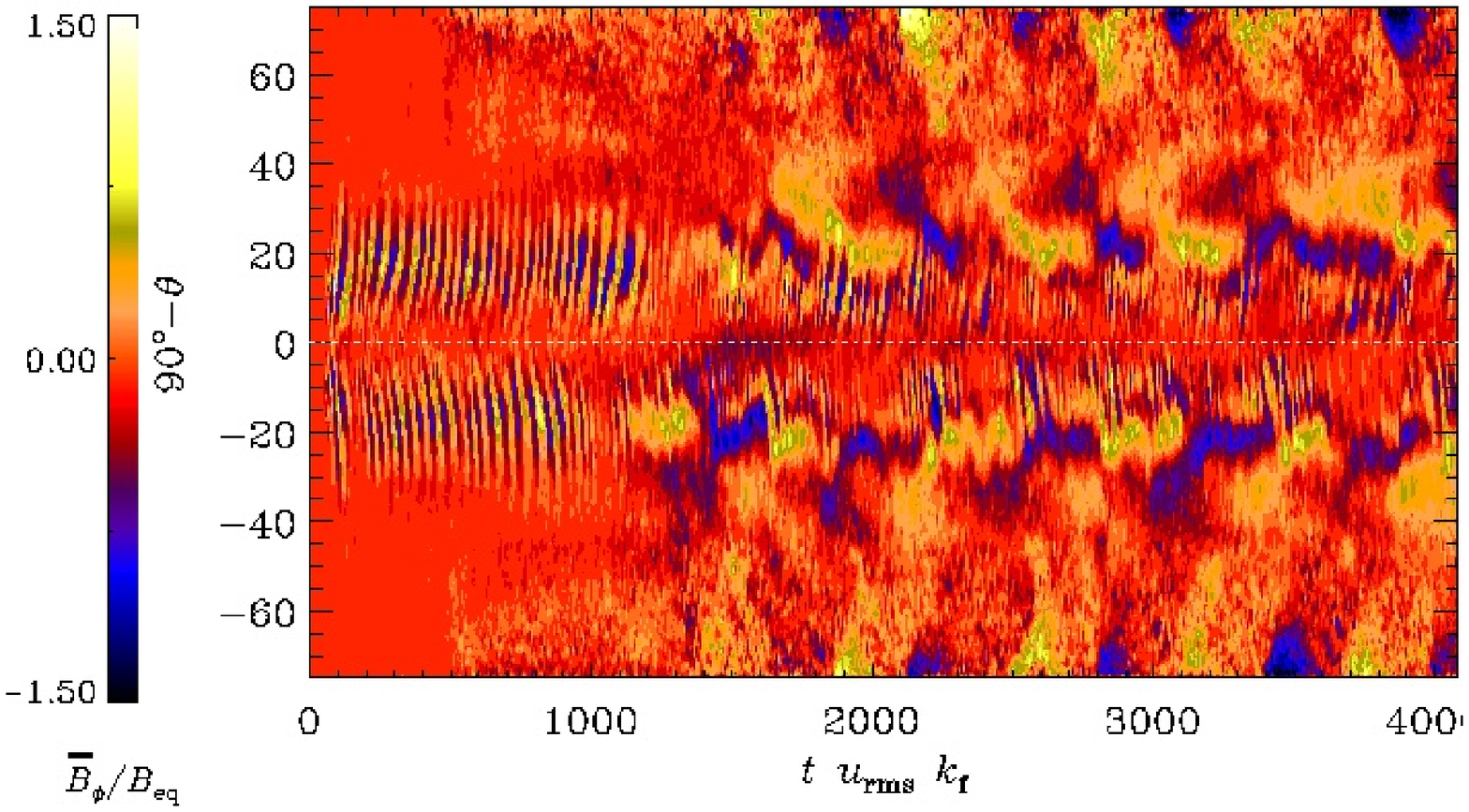}
\caption{Same as Figure~\ref{fig:butterflyA}, but for Runs C1 (top)
and C2 (bottom).
Note the difference in cycle frequency between the early times when the
frequency is similar to that of Run~B2 (Figure~\ref{fig:butterflyB})
and late times.
}\label{fig:butterflyC}
\end{figure}

\begin{figure}[t]
\centering
\includegraphics[width=0.48\textwidth]{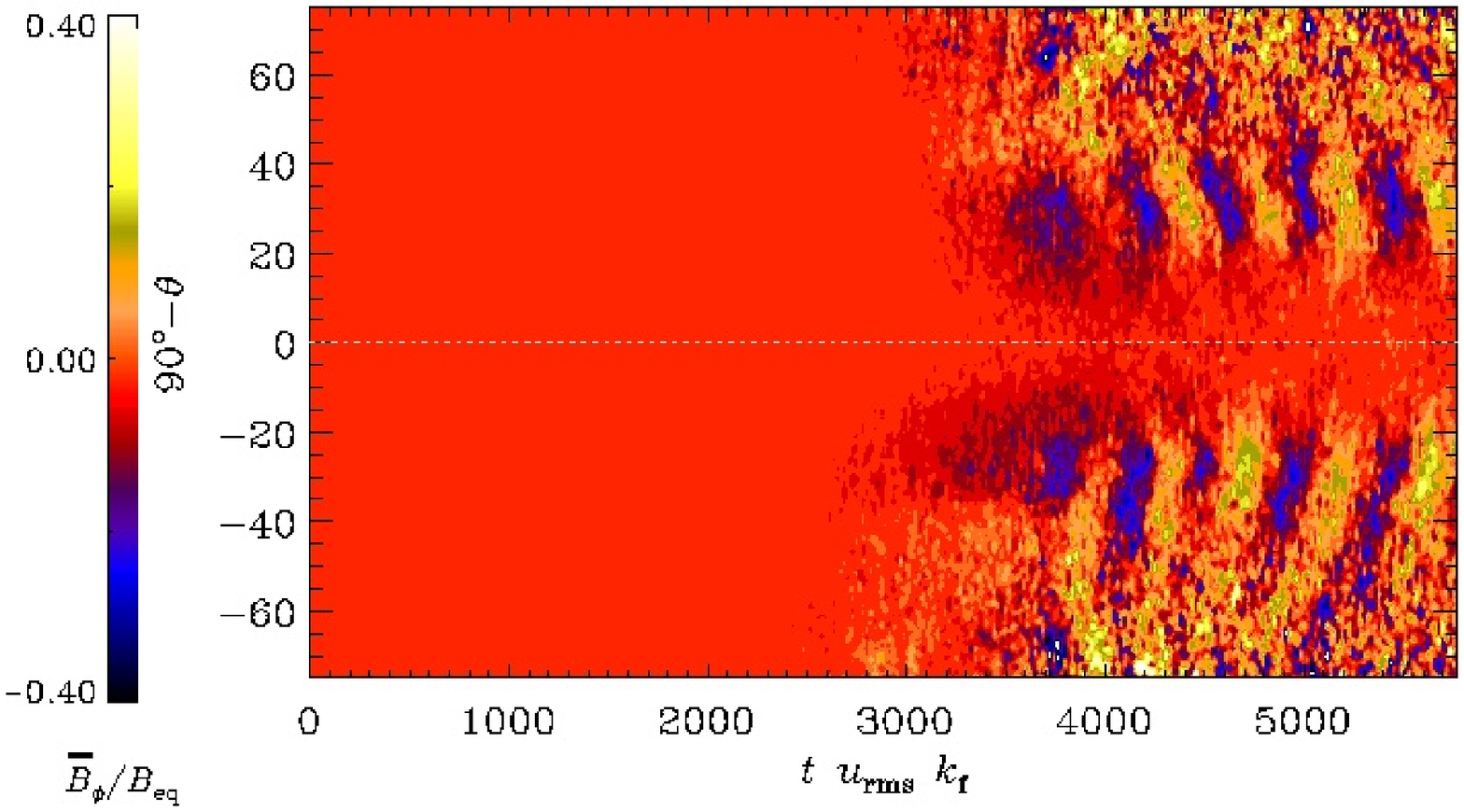}
\includegraphics[width=0.48\textwidth]{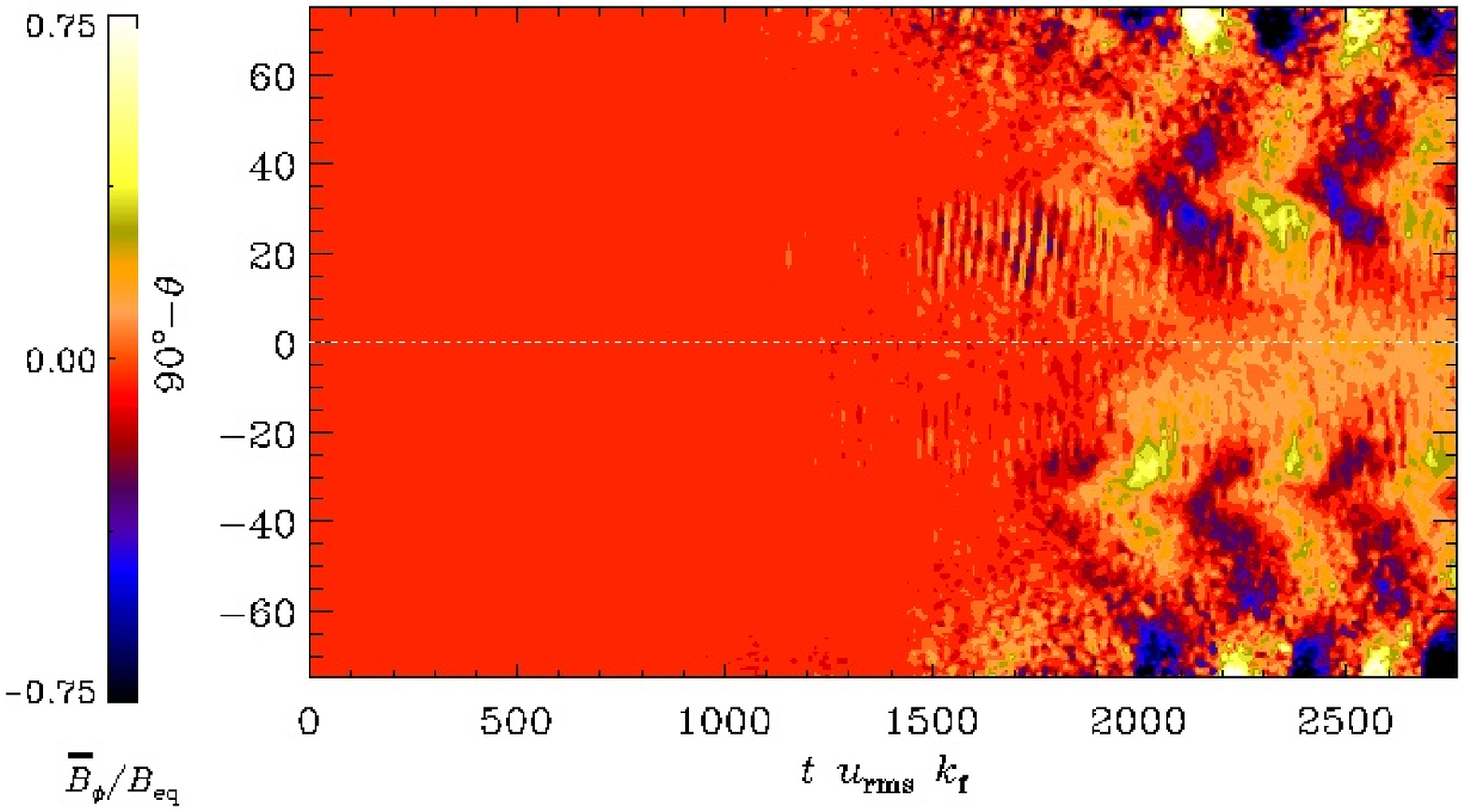}
\caption{Same as Figure~\ref{fig:butterflyA}, but for Runs D1 (top)
  and D2 (bottom).}\label{fig:butterflyD}
\end{figure}

The azimuthally averaged toroidal magnetic fields from Sets~A--D listed
in Tables~\ref{tab:runs} and \ref{tab:runs2} are shown in
Figures~\ref{fig:butterflyA}--\ref{fig:butterflyD}. The full time
evolution from the introduction of the seed magnetic field to the final
saturated state is shown for each run. Note that the magnitude of the
seed field in terms of the equipartition strength is different in each
set so direct comparisons between different sets are not possible. We
measure the
average growth rates during the kinematic stage,
\begin{equation}
\lambda=\langle d\ln B_{\rm rms}/dt \rangle_t,
\end{equation}
and find that $\lambda$ is greater for smaller stratification; see 
Column 2 of
Table~\ref{tab:runs2} for $\tilde\lambda=\lambda/(\urms \kef)$.
Comparing Runs~A1, B1, C1, and D2 with roughly
comparable Reynolds and Coriolis numbers shows that the normalized
growth rate decreases monotonically from 0.084 in Run~A1 to just 0.003
in Run~D2. 
Another striking feature is that $\tilde\lambda$ increases by a
factor of nearly 20 from Run~C1 to C2 whose only difference is that
the latter has a roughly two times higher Coriolis number.
It turns out that in all of the cases (Runs~A1, A2,
B1, B2, C2, and E4) with the highest growth rates, a dynamo mode with
poleward migration at low latitudes, is excited first.
In some of
the runs this mode is later overcome by another one that can be
quasi-stationary (Runs~A1 and B1) or oscillatory with equatorward
migration and a much longer cycle period (Runs~C2 and E4).

Table~\ref{tab:runs2} shows that, even though the growth rates decrease
dramatically with increasing stratification, many properties of the
saturated stages are similar.
In particular, the ratio of magnetic to kinetic energies does not seem
to systematically depend on stratification, but rather on the Coriolis
number, which varies only little between different runs.

In Figure~\ref{fig:butterflyA} we show the azimuthally averaged toroidal
magnetic field $\mean{B}_\phi$ near the surface of the computational
domain ($r=0.98\,R$) for two runs (A1 and A2; see Table~\ref{tab:runs})
with $\Gamma_\rho^{(0)}=2$. We find that in Run~A1 with
$\Co\approx8.7$ the mean magnetic field is initially oscillatory with
poleward propagation of the activity belts. At $t\urms\kef\approx400$
the dynamo mode changes to a quasi-steady configuration.
In Run~A2 a poleward mode persists throughout the simulation, although
the oscillation period is irregular and significant hemispherical
asymmetry exists. 
This behavior is similar to Run~A4 presented in
\cite{KKBMT10} with comparable stratification ($\Gamma\approx 3$)
and Reynolds ($\approx 20$) numbers, but a somewhat lower Coriolis
number\footnote{Note that the values of $\Rey$ and $\Co$ have been
  recalculated with the same definition of $\urms$ as in the current
  paper.} ($\approx 4.7$). The transition to oscillatory solutions
thus occurs at a lower $\Co$ in the models of \cite{KKBMT10}. A 
possible explanation is that in the present models we lack a lower 
overshoot layer which could affect the dominant dynamo mode.

In Set~B with $\Gamma_\rho^{(0)}=5$ the situation is similar: in Run~B1 
with $\Co\approx8.1$ there is a
poleward mode near the equator with a short cycle period which is
visible from early times; see Figure~\ref{fig:butterflyB}. However,
after around $t \urms \kef=1200$ there is a dominating non-oscillatory
mode that is especially clear at high latitudes. There are still hints
of the poleward mode near the equator. In Run~B2 with
$\Co\approx13.7$, however, the poleward mode
also prevails at late times. As in Run~A2, the cycles show significant
variability and hemispheric asymmetry.
The runs in Sets~A and B also show signs of non-axisymmetric `nests'
of convection \citep[cf.][]{Bu02,BBBMT08} in the hydrodynamical and
kinematic stages. Once the magnetic field becomes dynamically
important, these modes
either vanish or they are significantly damped.

Increasing the stratification further to $\Gamma_\rho^{(0)}=30$ (Set~C) the
dynamo solutions at lower rotation rates, $\Co\lesssim 5$, are still
quasi-steady; see Figure~2 of \cite{KMB12a}. However, a watershed
regarding the oscillatory modes at higher $\Co$ seems to have been
reached so that the irregular poleward migration seen in Sets~A and B
is replaced by more regular equatorward patterns.
In Run~C1 with $\Co\approx8.7$ the poleward migration near the equator
is also visible in the kinematic stage where the equatorward mode
is not yet excited; see
Figure~\ref{fig:butterflyC}.
The poleward mode
near the equator is more prominent in the early stages of Run~C2 with
$\Co\approx14.7$, but subdominant at late times.

For $\Gamma_\rho^{(0)}=100$ (Set~D) the general picture is similar to that
in Set~C. Quasi-steady configurations at lower rotation rates change
into
equatorward migrating solutions at sufficiently high values of $\Co$. We find
that this transition occurs between $\Co=5$ and $8$, similar to
Set~C; see Fig.~2 of \cite{KMB12a}.
For Set~D the equatorward mode is visible for both of its runs; see
Figure~\ref{fig:butterflyD}.
In Run~D1 no poleward migration at low
latitudes is seen in the kinematic stage. Also, the poleward migrating
branch at high latitudes is missing in the non-linear stage. Both of
these features are present in Run~D2.
The apparently slower growth of the magnetic field in Run~D1 is due
to a two orders of magnitude lower seed magnetic field than in Run~D2.

\subsection{Diagnostic stellar activity diagrams}

To identify the possibility of different types of dynamos, it is useful
to classify them in diagrams relating their characteristic properties.
In the geodynamo literature it has become customary to consider
the Elsasser number as a measure of the magnetic energy.
It correlates well with $\Rm$ \citep{CA06}, but this is partially
explained by the fact that $\Rm$ itself enters in the definition of the
Elsasser number.
Geodynamo models are mostly dominated by a strong dipolar component.
\cite{GDW12} have shown that such solutions fall on a branch that is
distinct from the cyclic solutions studied here, and that the latter
solutions become favored once density stratification is large and rotation
is sufficiently rapid so that large-scale non-axisymmetric fields become
dominant \citep[see also][]{NBBMT13}.
However, this type of analysis is not well suited for the present work,
where $\Rm$ and $\Co$ vary only little.
Furthermore, these tools do not characterize the nature of magnetic cycles,
which is the focus of this section.

\begin{figure}[t]
\centering
\includegraphics[width=0.46\textwidth]{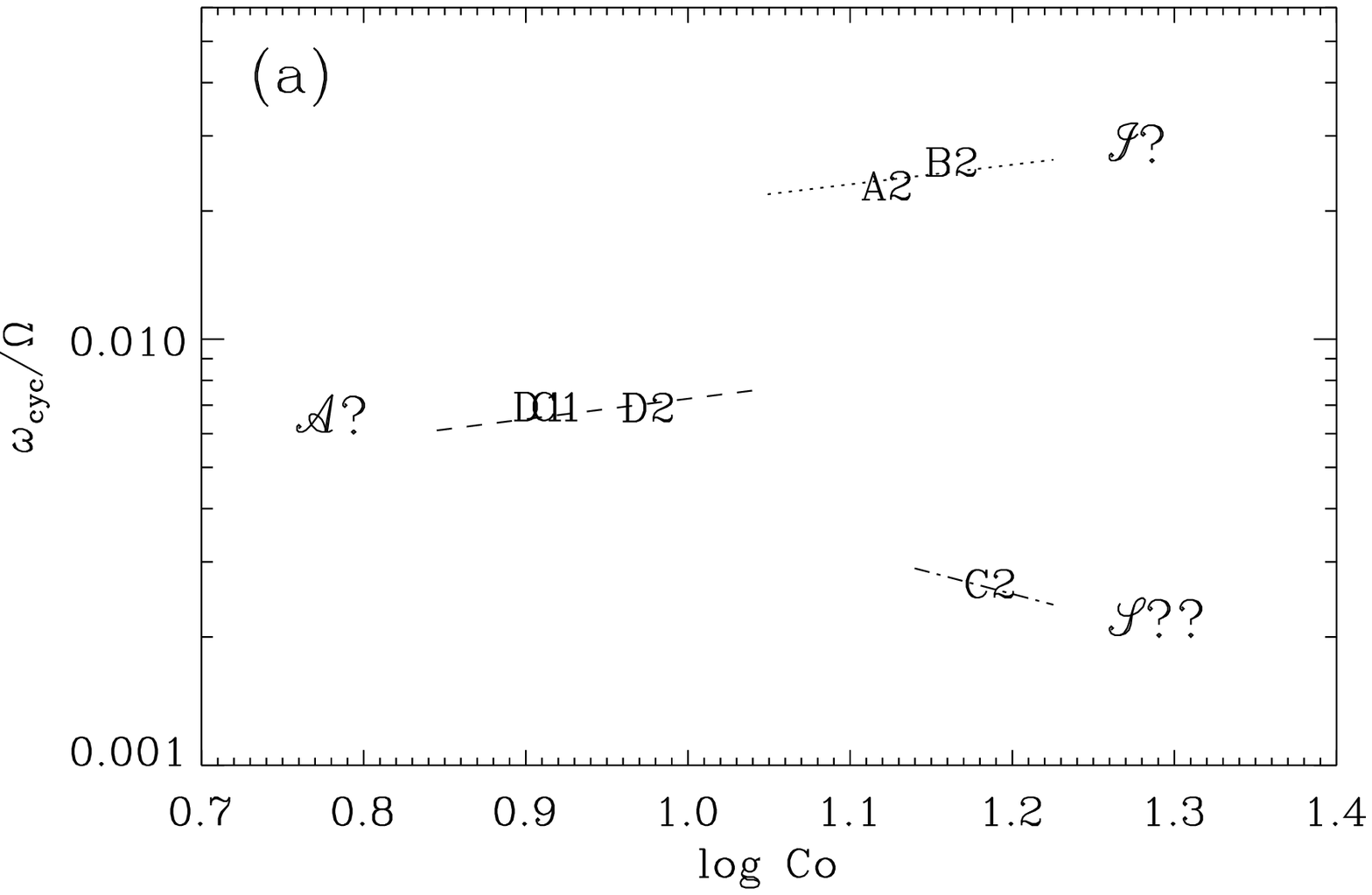}
\includegraphics[width=0.46\textwidth]{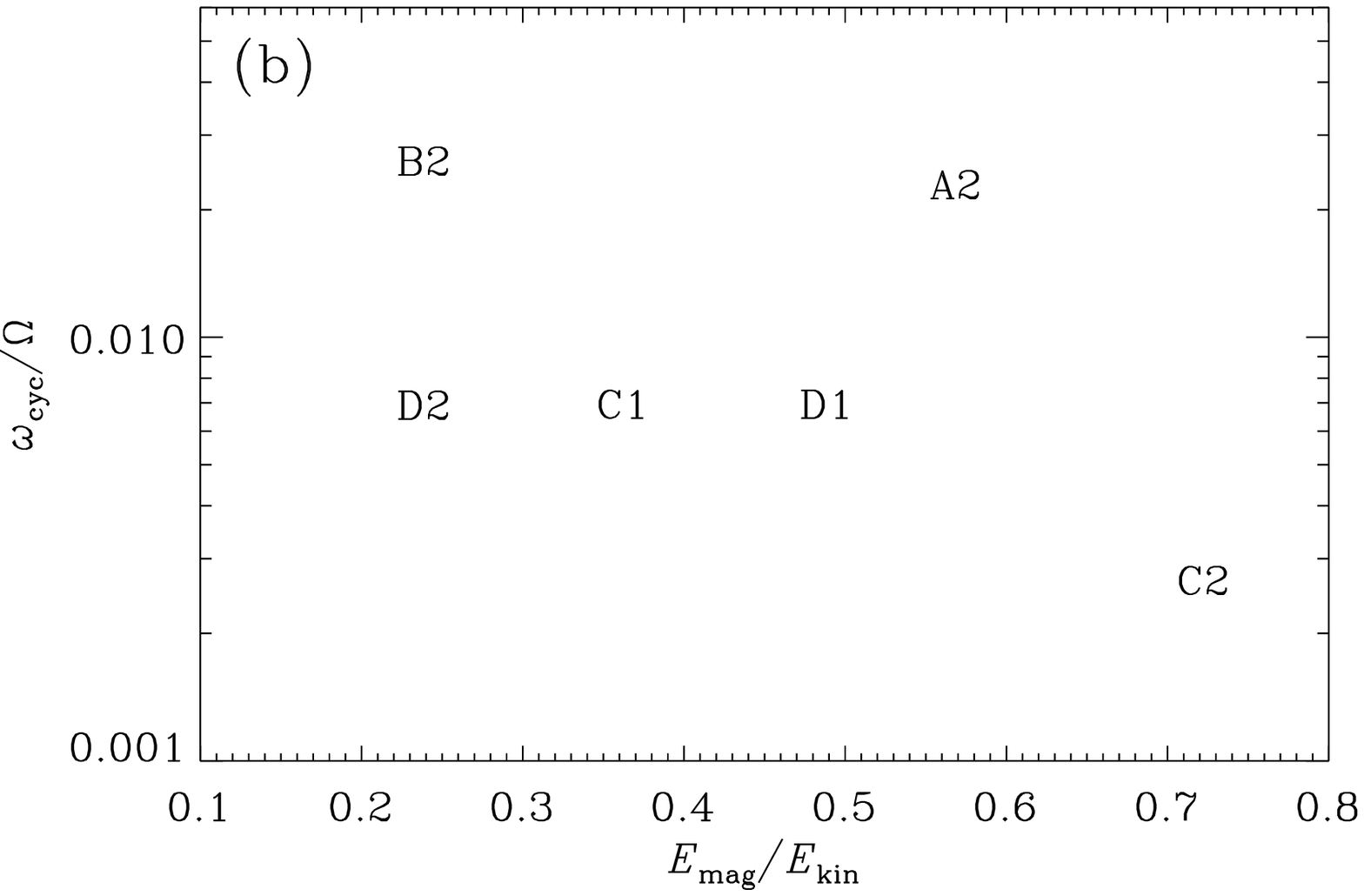}
\includegraphics[width=0.46\textwidth]{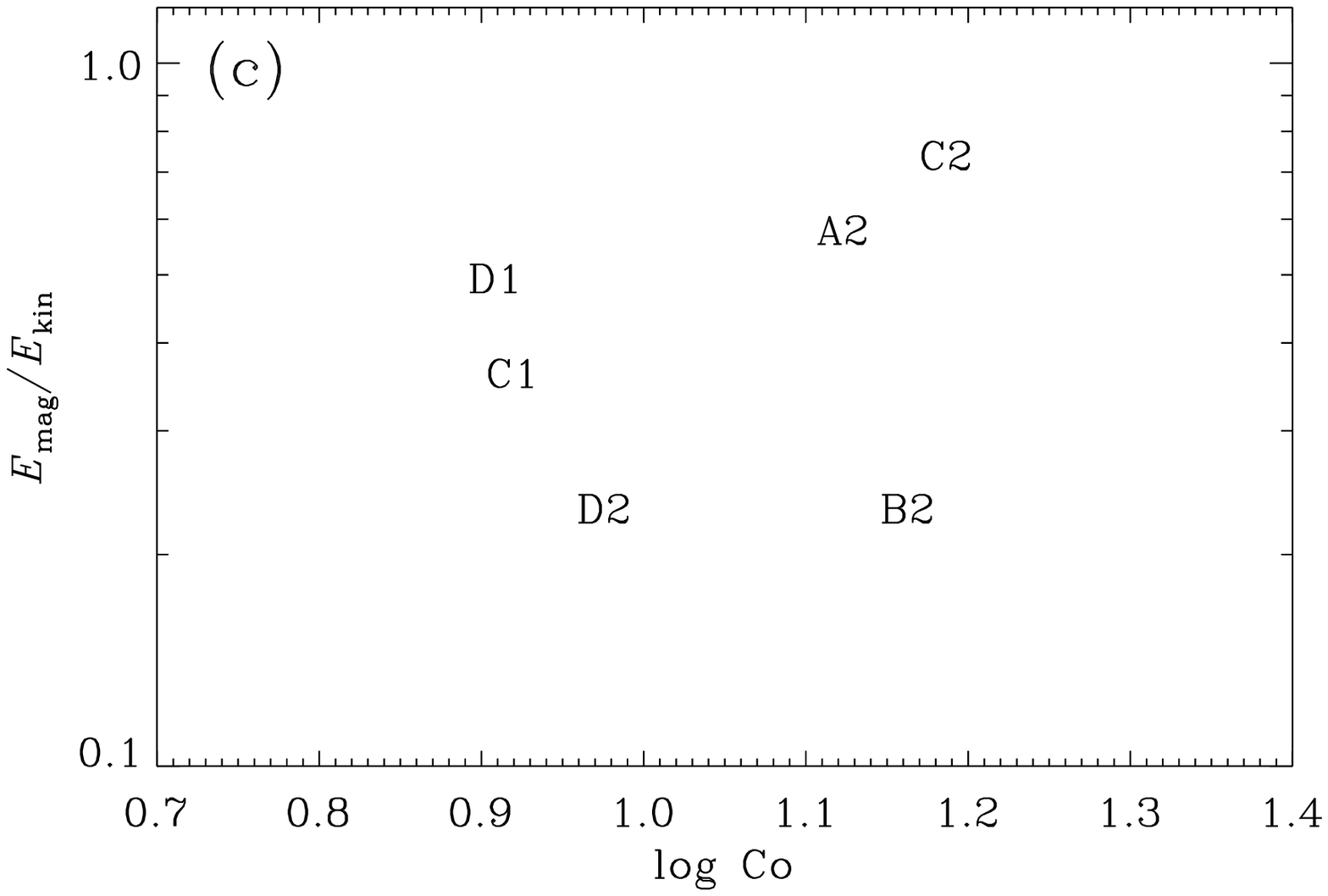}
\caption{Diagnostic diagrams from six runs that show cyclic activity.
   (a) Ratio of cycle and rotation frequencies vs.\ $\log\Co$.
   The dotted and dashed lines are given by $c_i \Co^{\sigma_i}$,
   where $\sigma_i$ correspond to those in \cite{BST98} for active
   (labeled `${\cal A}$?') and inactive (`${\cal I}$?') stars, while
   $c_i$ are used as fit parameters.
   The label `${\cal S}$??' indicates the possibility of the superactive branch
   in \cite{SB99}.
   (b) Ratio of cycle and rotation frequencies vs.\ magnetic to
   kinetic energy $E_{\rm mag}/E_{\rm kin}$.
   (c) Time averaged $E_{\rm mag}/E_{\rm kin}$ vs.\ $\log\Co$.}
\label{fig:pocycom}
\end{figure}

To connect our results with observations of magnetically active
stars we compute the ratio of cycle to rotation frequency
$\omega_{\rm cyc}/\Omega_0$, where $\omega_{\rm cyc}=2\pi/T_{\rm cyc}$
is the cycle frequency of magnetic energy of the mean field and
$T_{\rm cyc}$ its period.
Plotting this ratio as a
function of the Coriolis number for stars exhibiting chromospheric
activity has shown that stars tend to group along inactive and active
branches \citep{BST98}, and for higher Coriolis numbers along a
super-active branch \citep{SB99}.
Six of our simulations (Runs~A2,
B2, C1, C2, D1, and D2), excluding the runs in Set~E (which are very
similar to each other and to Run~C1), show cycles and can
thus be used in this analysis. We compute the cycle frequency from the
highest peak of a temporal Fourier transformation of the time series
for $\mean{B}_\phi$ averaged over a latitudinal strip of
$\pm10\degr\ldots30\degr$ near the surface. The results
are shown in Figure~\ref{fig:pocycom}(a).

Three of the models, Runs~C1, D1, and D2, fall on
a branch labeled `${\cal A}$?' for active stars, while Run~C2 might
be suggestive of the superactive stars of \cite{SB99}, labeled here
`${\cal S}$??'.
Runs~A2 and B2 show irregular cycles and group along the branch labeled
`${\cal I}$?' for inactive stars.
The question marks on these labels in Figure~\ref{fig:pocycom}(a)
indicate that the association with real
branches is quite uncertain and somewhat
premature, because there are too few models.
We cannot be sure that there are no models connecting
the group of Runs~C1, D1, D2 with that of A2 and B2 through a single
line with a steeper slope.
Nevertheless, this plot allows us to see that, while the separation
in the ratio $\omega_{\rm cyc}/\Omega_0$ is slightly less for the two
groups of runs compared with active and inactive stars,
their relative ordering in the value of $\Co$ is actually
the other way around.
One would therefore not have referred to Runs~A2 and B2 as inactive
just because their $\omega_{\rm cyc}/\Omega_0$ ratio agrees with that
of inactive stars.
In fact, their $E_{\rm mag}/E_{\rm kin}$ ratios (a measure of stellar activity)
in Table~\ref{tab:runs2}
are typically larger than for Runs~C1, D1, and D2.

As is visible from Figure~\ref{fig:pocycom}(c),
there is no clear relation between $\Co$ and $E_{\rm mag}/E_{\rm kin}$,
which is different from stars for which there is a clear relation
between $\Co$ (referred to as the inverse Rossby number in that context)
and
stellar activity;
see \cite{BST98} for details and references.
Furthermore, there are also no indications of branches in the
graph of $\omega_{\rm cyc}/\Omega_0$ versus $E_{\rm mag}/E_{\rm kin}$;
see Figure~\ref{fig:pocycom}(b).
Instead, there might just be one group in it, possibly with a positive
correlation, i.e., $\omega_{\rm cyc}/\Omega_0$ might increase with
$E_{\rm mag}/E_{\rm kin}$.
Such a possibility does indeed arise when considering the frequency ratio
versus the dimensional rotation rate \citep{OKS00}.
However, as discussed by \cite{BST98}, a positive slope is not easily
explained in the
framework of standard mean-field dynamo theory, where the frequency
ratio is usually a decreasing function of normalized rotation
rate and activity parameter \citep{To98,SB99}.

In conclusion, we reiterate that the quantity $\omega_{\rm cyc}/\Omega_0$ is an
important and robust property of cyclic dynamo models and its
dependence on other properties of the model should therefore be
a useful characteristics that can be compared with other models and
ultimately with actual stars.
Here we have made a first attempt in classifying model results in this way.

\begin{figure}[t]
\centering
\includegraphics[width=0.9\columnwidth]{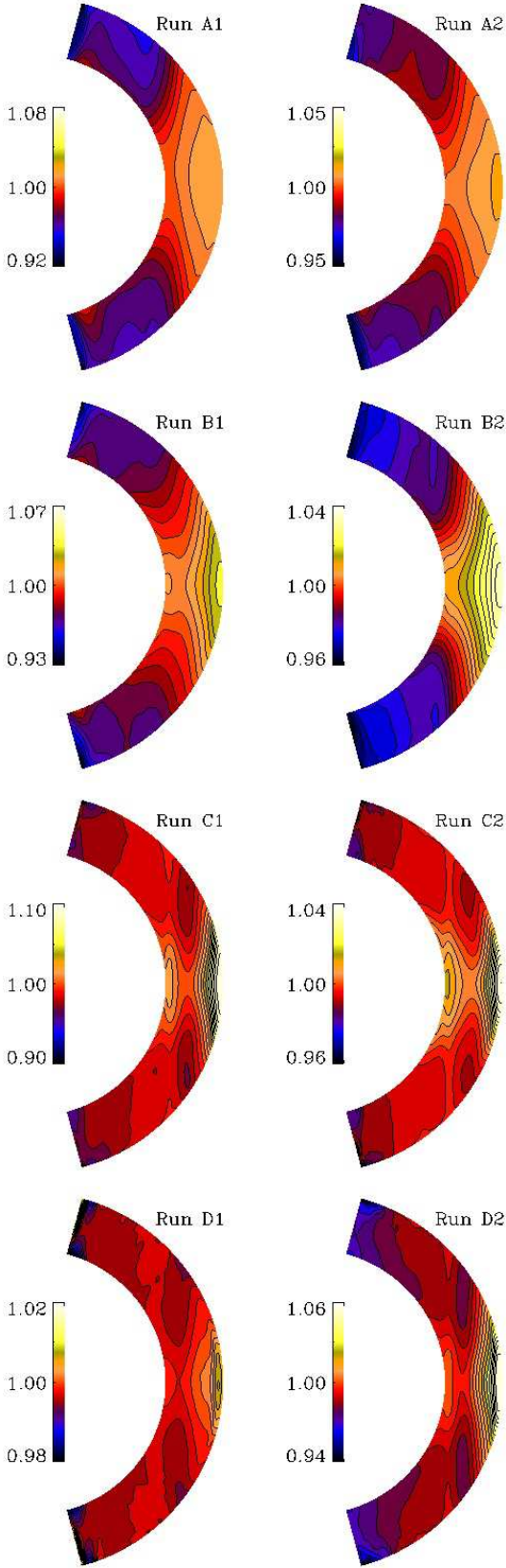}
\caption{Time averaged mean rotation profiles
  $\mOmega/\Omega_0$ (gray/color scale and line contours) from
  Sets~A, B, C, and D.}\label{fig:pOm}
\end{figure}

\subsection{Differential rotation and meridional circulation}
\begin{figure}[t]
\centering
\includegraphics[width=0.49\columnwidth]{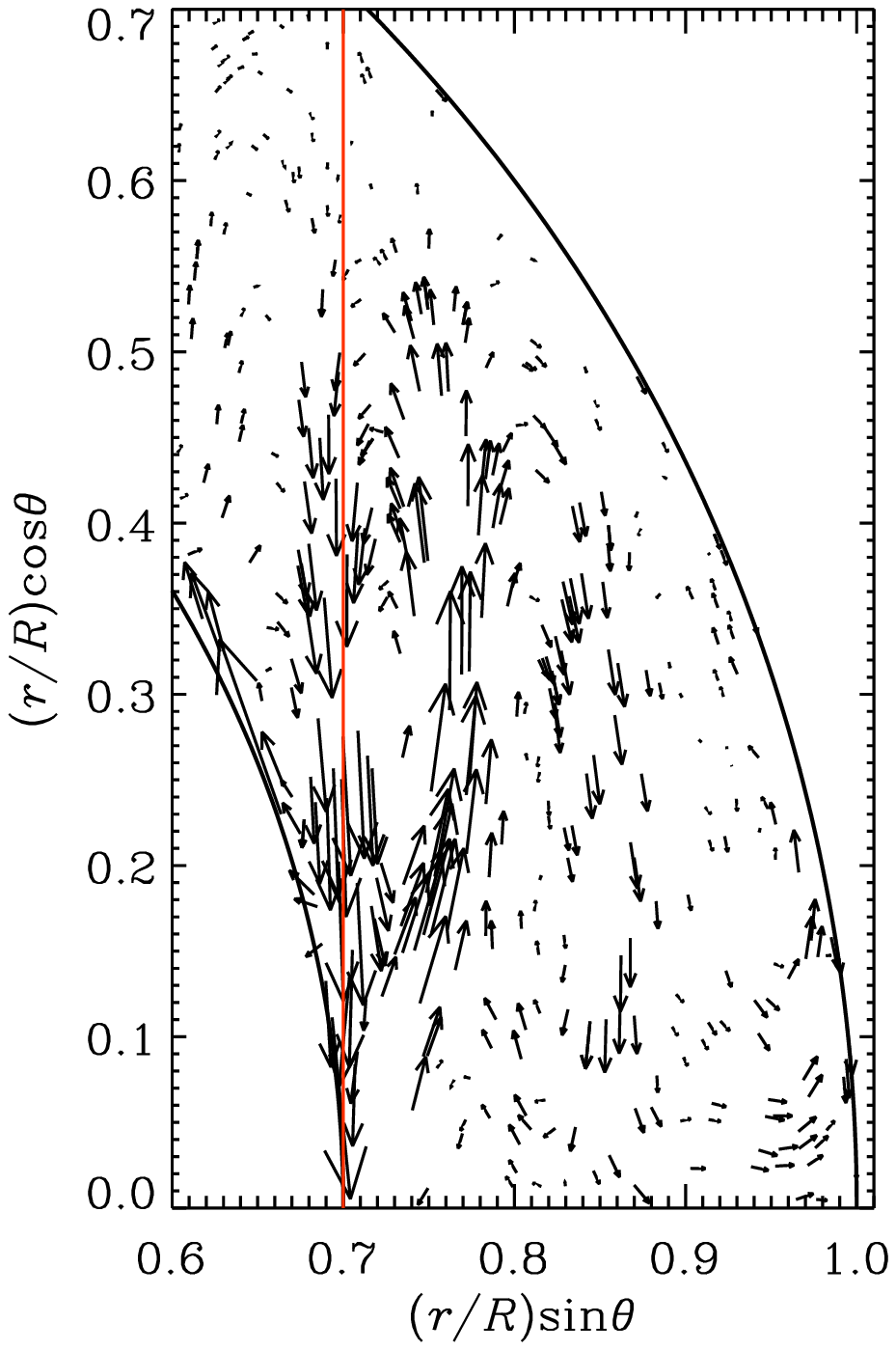}
\includegraphics[width=0.49\columnwidth]{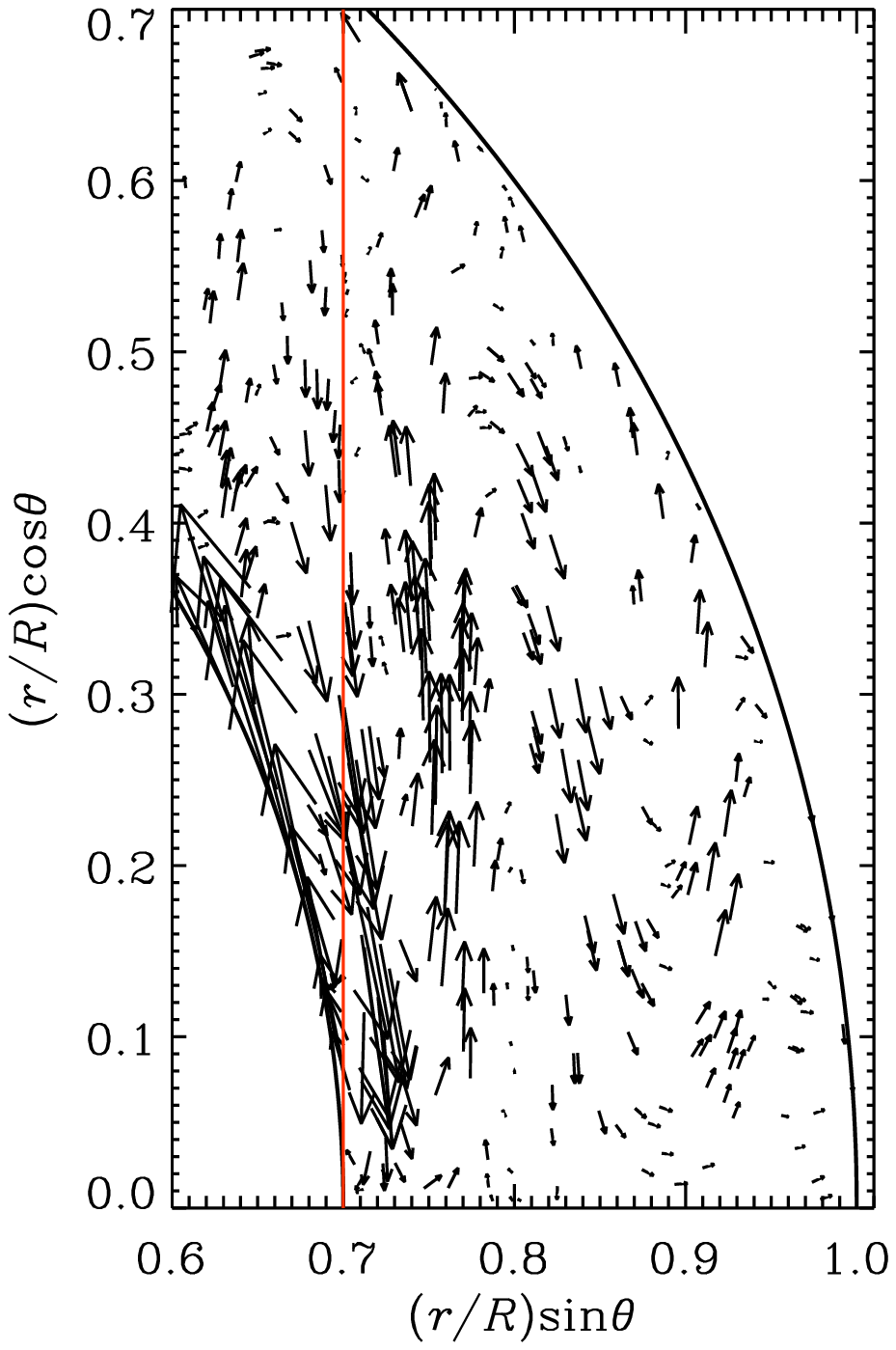}
\caption{
Meridional circulation in the
northern hemisphere of the convection zone of Run~C1 (left panel) and
Run~D1 (right)
shown as vectors of the mass flux
$\mean{\rho}(\mean{u}_r,\mean{u}_\theta,0)$, which is also averaged
over a time span of around 250 turnover times in the saturated state.
The black solid lines indicate the surface ($r=R$) and the bottom of
the convection zone ($r=0.7\,R$), and the red solid line indicates the
position of the inner tangent cylinder.
Note, that for Run~D1 (right), the mass flux have been multiplied by a factor
of 5 to emphasize the structure.
}
\label{fig:meri}
\end{figure}
Non-uniform rotation of the convection zone of the Sun is an important
ingredient in maintaining the large-scale magnetic field. Furthermore,
the sign of the radial gradient of the mean angular velocity plays a crucial role
in deciding whether the dynamo wave propagates toward the pole or the
equator in $\alpha$--$\Omega$ mean-field models
\citep[e.g.,][]{Pa55b,Pa87a}.
In the following we use the local angular velocity defined as
$\mOmega=\Omega_0+\mean{u}_\phi/r\sin\theta$.
Azimuthally averaged rotation profiles from the runs in Sets~A to D
are shown in Figure~\ref{fig:pOm}. The rotation profiles of Runs~E1,
E3, and E4 are very similar to that of Run~C1.
We quantify the radial and latitudinal differential rotation by
\begin{eqnarray}
\Delta_\Omega^{(r)} &=& \frac{\mOmega_{\rm eq}-\mOmega_{\rm bot}}{\mOmega_{\rm eq}},\\
\Delta_\Omega^{(\theta)} &=& \frac{\mOmega_{\rm eq}-\mOmega_{\rm pole}}{\mOmega_{\rm eq}},
\end{eqnarray}
where $\mOmega_{\rm eq}=\mOmega(R,\pi/2)$ and $\mOmega_{\rm
  bot}=\mOmega(r_0,\pi/2)$ are the angular velocities at the top and
bottom at the equator, respectively, and $\mOmega_{\rm
pole}=[\mOmega(R,\theta_0) + \mOmega(R,\pi-\theta_0)]/2$.
It has long been recognized that dynamo-generated magnetic fields can have
an important effect on the angular velocity \citep{Gi83,Gl85,Gl87}.
Indeed, magnetic fields affect the turbulence that gives rise to
Reynolds stress and turbulent convective heat flux
\citep[e.g.,][]{KPR94,KKT04}. Furthermore, the large-scale flows are
directly influenced by the Lorentz force when the
magnetic field is strong enough \citep[e.g.,][]{MP75}.
A magnetically caused decrease of $\Delta_\Omega^{(\theta)}$ has also
been observed in LES models \citep[e.g.,][]{BMT04}. 
Comparing the latitudinal differential
rotation in Run~B1 with that of the otherwise identical hydrodynamic
Run~A4 of \cite{KMB11}, we find that $\Delta_\Omega^{(\theta)}$
decreases only slightly from 0.15 to 0.14. For $\Delta_\Omega^{(r)}$
the change is more dramatic---from 0.079 to 0.034. The fraction of
kinetic energy contained in the differential rotation, $E_{\rm
  rot}/E_{\rm kin}$, drops from 0.91 to 0.71. A
similar decrease is observed in Run~C1 in comparison to its
hydrodynamical parent Run~B4 of \cite{KMB11} with
$\Delta_\Omega^{(\theta)}$ changing from 0.08 to 0.07,
$\Delta_\Omega^{(r)}$ from 0.066 to 0.047, and $E_{\rm rot}/E_{\rm
  kin}$ dropping from 0.58 to 0.44.
Similar changes have also been seen in dynamos from forced
turbulence in Cartesian domains \citep{B01}, in addition to those from
convective turbulence in spherical shells \citep{BMT04}.

In all cases in Figure~\ref{fig:pOm}, we see a rapidly spinning equator with a positive radial
gradient of $\mOmega$.
The latitudinal variation of angular velocity is, however, not always
monotonic and there can be local minima at mid-latitudes, as is seen,
for example, in Run~C1.
Similar features have previously been seen \citep[see,
e.g.,][]{METCGG00,KMGBC11}
and might be related to the lack of small-scale turbulence.
Especially at larger stratification one would expect smaller-scale
turbulent structures to emerge, but this means large Reynolds numbers
and thus requires sufficient resolution, which is not currently
possible.

The amount of latitudinal differential rotation (here 0--0.09;
see Table~\ref{tab:runs2}) is clearly less than in the Sun where
$\Delta_\Omega^{(\theta)}\approx0.2$ between the equator and latitude
$60\degr$ \citep[e.g.,][]{Schouea98}. Furthermore, $\Delta_\Omega^{(\theta)}$
generally decreases within each set of runs as $\Co$ increases, except for
Runs~D1 and D2 where the value increases; see Table~\ref{tab:runs2}.
However, in Run~D1 the lower Reynolds number possibly
contributes to the weak differential rotation in comparison to Run~D2
with comparable $\Co$.
The rotation profiles appear to be dominated by the Taylor-Proudman
balance, except at very low latitudes where the baroclinic term is
significant; see Figure 9 of \cite{WKMB13}.
In this companion paper, we show that an outer coronal
layer seems to favor a
solar-like rotation, which shows even radially orientated contours
of constant rotation.
Such `spoke-like' rotation profiles have thus far only been obtained in
mean-field models involving anisotropic heat transport
\citep[e.g.,][]{BMT92,KR95} or a subadiabatic tachocline \citep{Re05}, and
in purely hydrodynamic LES models where a latitudinal entropy gradient
is enforced at the lower boundary \citep{MBT06}, or where a stably stratified
layer is included below the convection zone \citep{BMT11}.

The meridional circulation is weak in all cases and typically shows
multiple cells in the radial direction.
In Figure~\ref{fig:meri}, we plot the mean mass flux,
$\mean{\rho}(\mean{u}_r,\mean{u}_\theta,0)$, of the meridional
circulation for Runs~C1 and D1.
In Run~C1 the circulation pattern is mostly concentrated in
the equatorial region outside the inner tangent cylinder, where we
find a solar-like anti-clockwise cell at low latitudes
($<30^\circ$) in the upper third of the convection zone.
There are additional cells deeper down and also at higher latitudes.
Only the cell near the surface seems to have the same curvature as the
surface,
while the others, in particular the strong one above the inner tangent
cylinder, seem to be parallel to the rotation axis.
This is similar to earlier results by \cite{KMB12a} where the
meridional circulation pattern was shown in terms of the velocity.
The circulation pattern in Run~D1 is qualitatively quite similar, but
the velocity is smaller by roughly a factor of five.
Similar patterns of multi-cellular meridional circulation have also
been seen in anelastic simulations using
spherical harmonics \citep[see, e.g.,][]{NBBMT13}
and in models with an outer coronal layer \citep{WKMB13}.
In addition, as we will show in the next section,
the importance of meridional circulation
relative to the turbulent magnetic diffusivity is rather low, which is another
reason why it cannot play an important role in our models.

\begin{figure}[t]
\centering
\includegraphics[width=0.9\columnwidth]{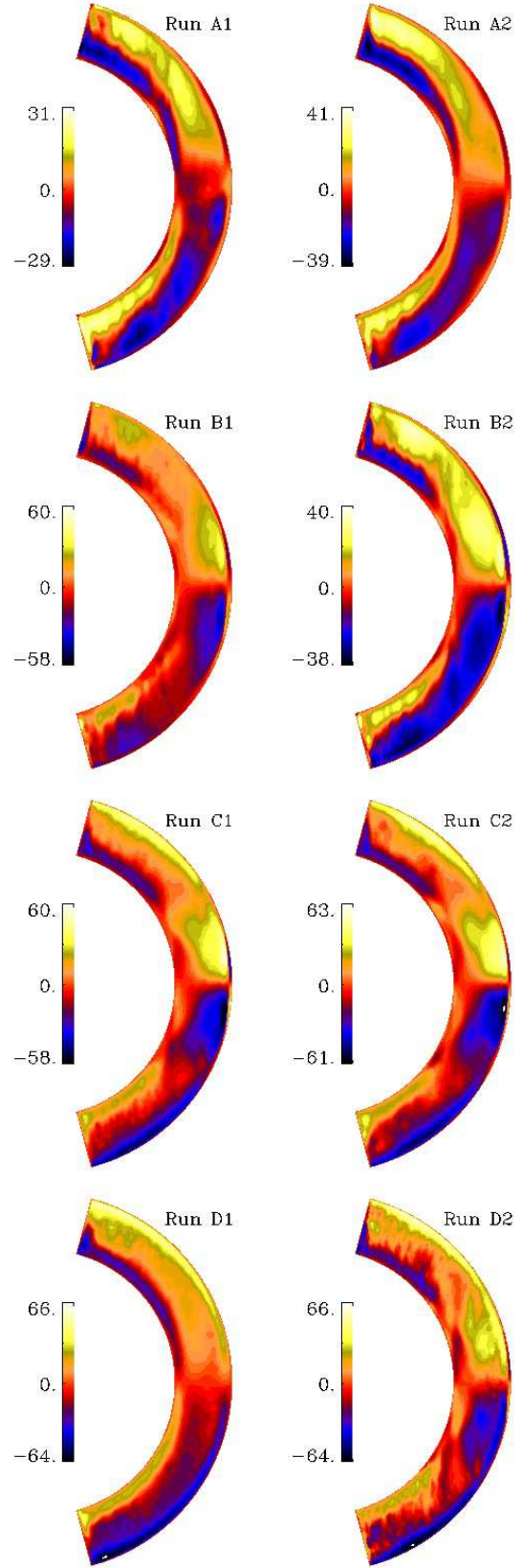}
\caption{Local dynamo parameter $c_\alpha$ from
  Sets~A, B, C, and D.}\label{fig:pCalp_sat}
\end{figure}

\begin{figure}[t]
\centering
\includegraphics[width=0.895\columnwidth]{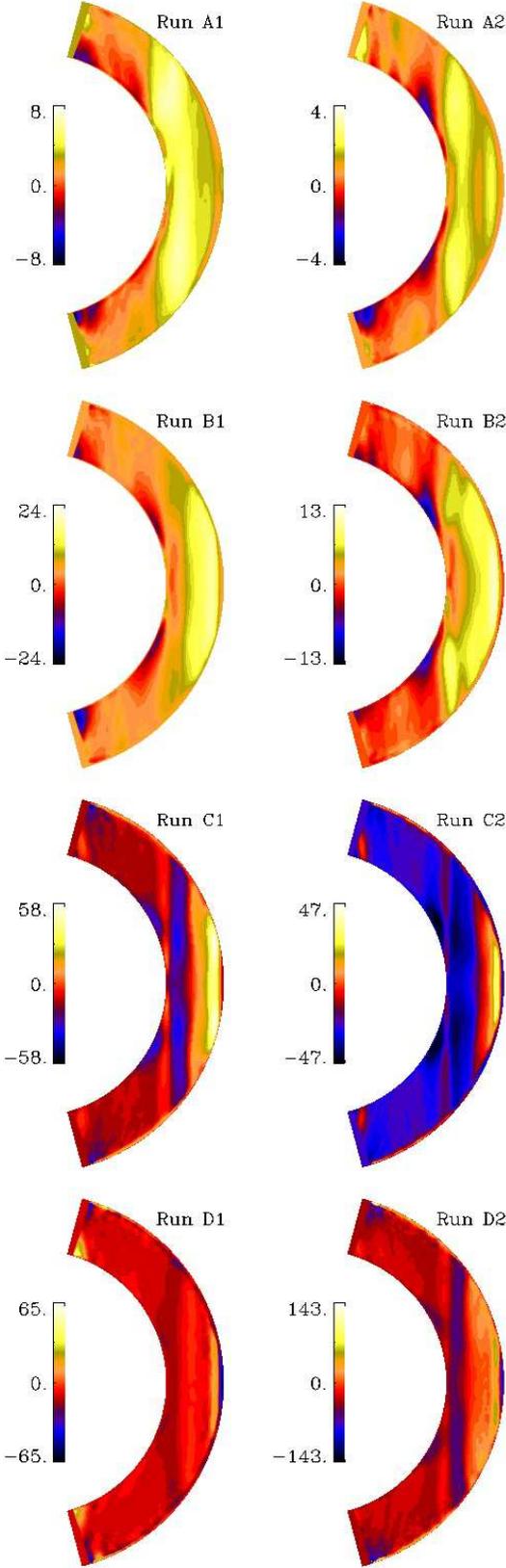}
\caption{Local dynamo parameter $c_\Omega$ from
  Sets~A, B, C, and D. We omit regions closer to $2.5\degr$ from the
  latitudinal boundaries.}\label{fig:pCOm}
\end{figure}

\begin{figure}[t]
\centering
\includegraphics[width=0.9\columnwidth]{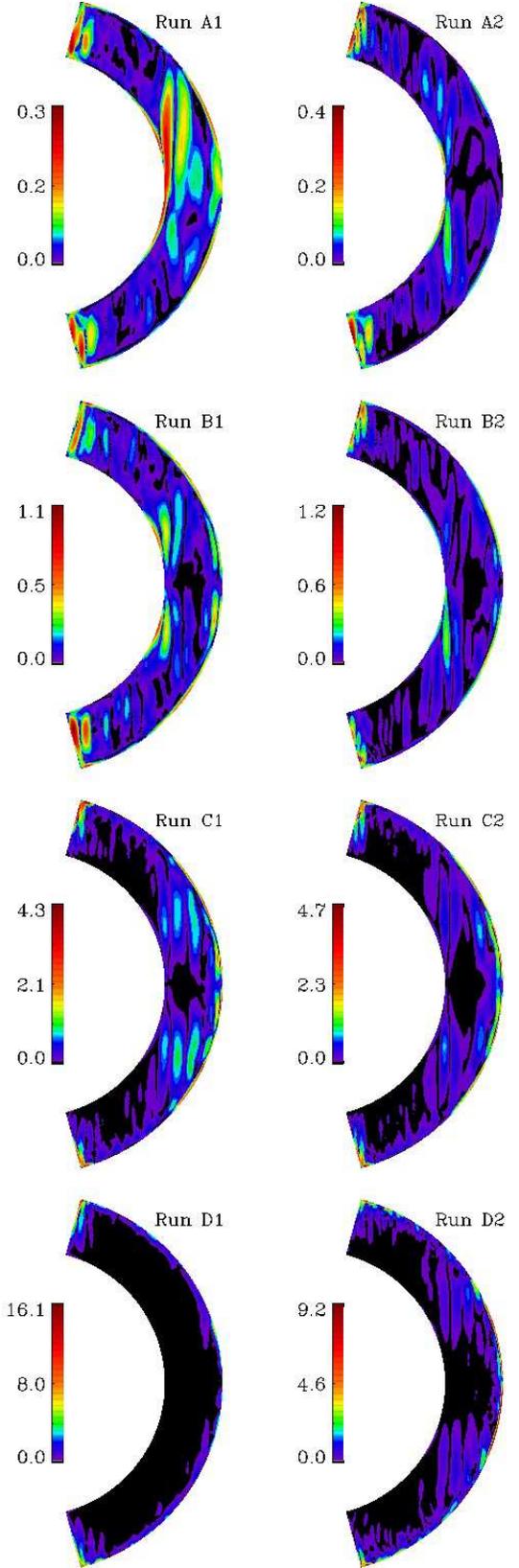}
\caption{Local dynamo parameter $c_U$ from Sets~A, B, C, and
  D.}\label{fig:pCU}
\end{figure}

\subsection{Estimates of local dynamo parameters}
\label{DynamoParameters}

To estimate the dynamo parameters related to $\alpha$-effect, radial
differential rotation, and meridional circulation, we consider
local ($r$- and $\theta$-dependent) versions of dynamo numbers, referred to as
local dynamo parameters that are defined by
\begin{equation}
c_\alpha=\frac{\alpha \Delta r}{\etatz},\quad
c_\Omega=\frac{\pd \mOmega/\pd r (\Delta r)^3}{\etatz},\quad
c_U=\frac{\mean{u}_{\rm mer}^{\rm rms} \Delta r}{\etatz},
\end{equation}
where $\pd \mOmega/\pd r$ is the $r$- and $\theta$-dependent radial 
gradient of $\mOmega$, $\Delta r=R-r_0$ is the thickness of the layer,
and $\alpha$ is a proxy of the $\alpha$-effect \citep{PFL76},
\begin{equation}
\alpha = -\onethird \tau (\mean{\mbox{\boldmath $\omega$}\cdot{\bm u}}
-\mean{{\bm j}\cdot{\bm b}}/\mean{\rho}),
\end{equation}
with $\tau=\alpha_{\rm MLT}H_{\rm P}/\urms(r,\theta)$ being the
local convective turnover time and $\alpha_{\rm MLT}$ the mixing length
parameter. We use $\alpha_{\rm MLT}=5/3$ in this work. We estimate
the turbulent diffusivity by
$\etatz=\tau\urms^2(r,\theta)/3$.
Furthermore, $\mean{u}_{\rm mer}^{\rm rms}=\sqrt{\mean{u}_r^2 +
  \mean{u}_\theta^2}$ is the rms value of the meridional circulation.

The results for the local dynamo parameters are shown in
Figures~\ref{fig:pCalp_sat}--\ref{fig:pCU}.
Generally, the values of $c_\alpha$ are fairly large,
and those of $c_\Omega$ surprisingly small, suggesting that
the dynamos might mainly be of $\alpha^2$ type.
In the following, however, we focus on relative changes between different runs.
It turns out that there is a weak tendency
for $c_\alpha$ to increase as a function of $\Gamma_\rho$
(from Sets~A to D) and $\Co$ (from subsets~1 to 2).
In Set~B, however, $c_\alpha$ decreases by a third from Run~B1
to B2. The spatial distribution of $c_\alpha$ becomes more
concentrated near the radial boundaries as $\Gamma_\rho$ increases.

We find that differential rotation is strongest near
the equator in all cases.
Sets~A and B have extended regions outside the inner tangent cylinder
and at low latitudes where $c_\Omega$ is large, but in all cases
$c_\Omega$ is clearly smaller than $c_\alpha$.
This is surprising given the fact that the energy of the mean toroidal field
is greater than that of the mean poloidal field by a significant
factor
(see $E_{\rm pol}$ and $E_{\rm tor}$ in Table~\ref{tab:runs2})
which would be expected if differential rotation dominates over the
$\alpha$-effect in maintaining the field. In Runs~C1, C2, and D1,
$c_\alpha$ and $c_\Omega$ have comparable magnitudes whereas in Run~D2
the maximum of $c_\Omega$ is roughly twice that of $c_\alpha$.
However, in these cases the toroidal and poloidal field
energies are roughly comparable (see Table~\ref{tab:runs2}).
For Set~C (and especially for Run~C2) there are broad regions
where $c_\Omega$ is negative.
In this connection we recall that in the diagnostic diagrams
(Figure~\ref{fig:pocycom}), C2 appears
as an outlier and far away from the ${\cal A?}$ and ${\cal I?}$
branches.
Furthermore, in the more strongly stratified models, $c_\alpha$ shows
enhanced values at low latitudes.
However, for the most strongly stratified models this is only true
of Run~D2, which is rotating slightly faster than Run~D1.
This is interesting in view of the fact that many mean-field dynamos
produce too strong fields at high latitudes, which is then `artificially'
reduced by an ad-hoc factor proportional to $\sin^2\theta$ \citep{RB95}
or other such variants \citep{DdTGAW04,PK11} for $\alpha$.
We note that in local convection simulations, the $\alpha$-effect has
been found to peak at mid-latitudes for rapid rotation \citep{KKOS06}.

We find that $c_U$ is always small in comparison to both
$c_\alpha$ and $c_\Omega$.
Note, however, that the range of $c_U$ does increase as we
go from Set~A to Set~D.
Figure~\ref{fig:pCU} also shows the
concentration of coherent meridional circulation cells in the
equatorial regions with a multi-cell structure.

In flux transport dynamos, $c_U$ has values of several hundreds \citep{KRS01}.
This is a consequence of choosing a small value of the turbulent
magnetic diffusivity.
In our simulations, on the other hand, $c_U$ is much smaller.
This is a consequence of faster turbulent motions,
making the turbulent diffusivity large and therefore $c_U$ small.
Whether or not this also applies to more realistic models remains to be seen.

\subsection{Phase relation and nature of the dynamo}

\begin{figure}[t]
\centering
\includegraphics[width=\columnwidth]{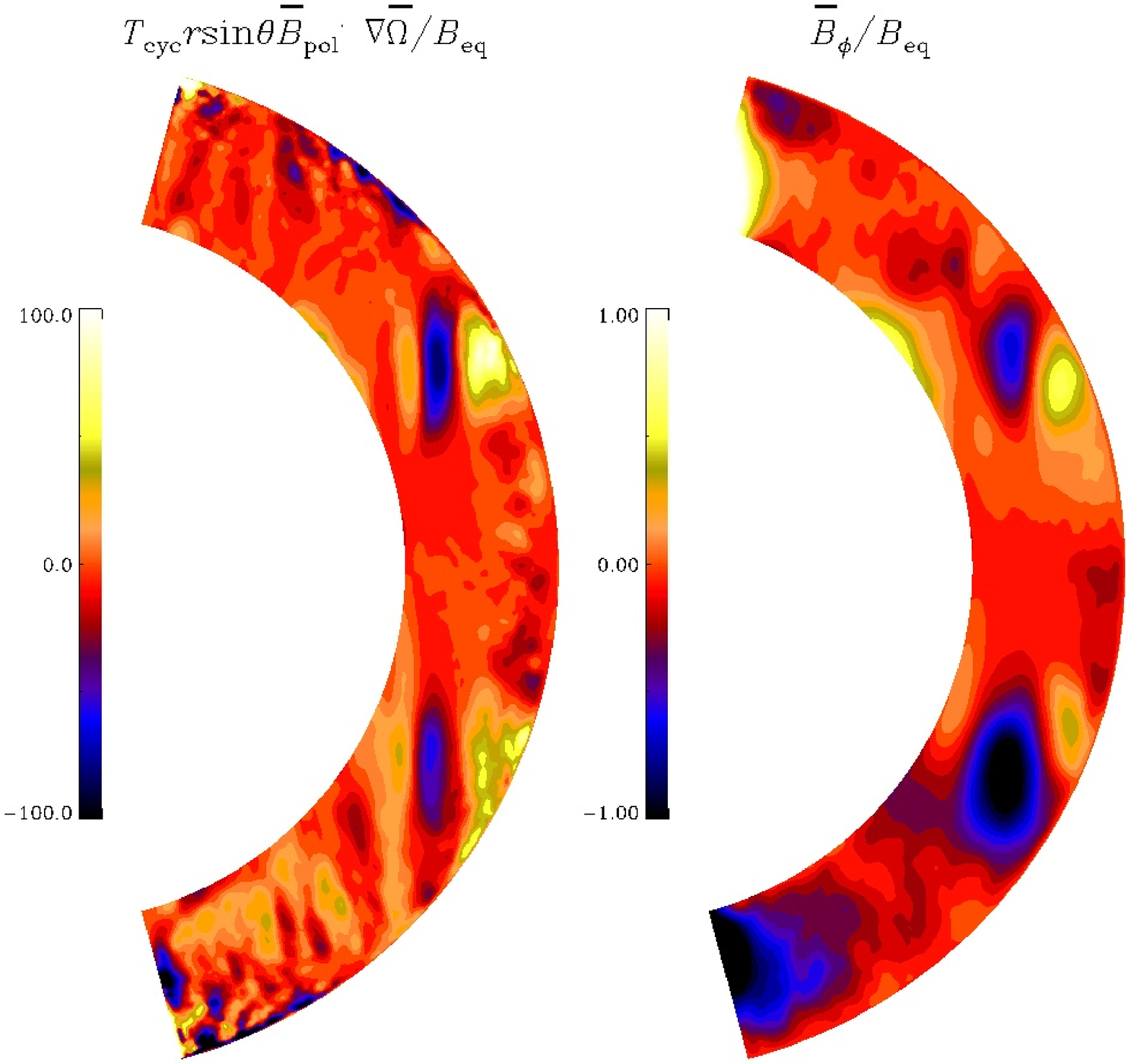}
\includegraphics[width=\columnwidth]{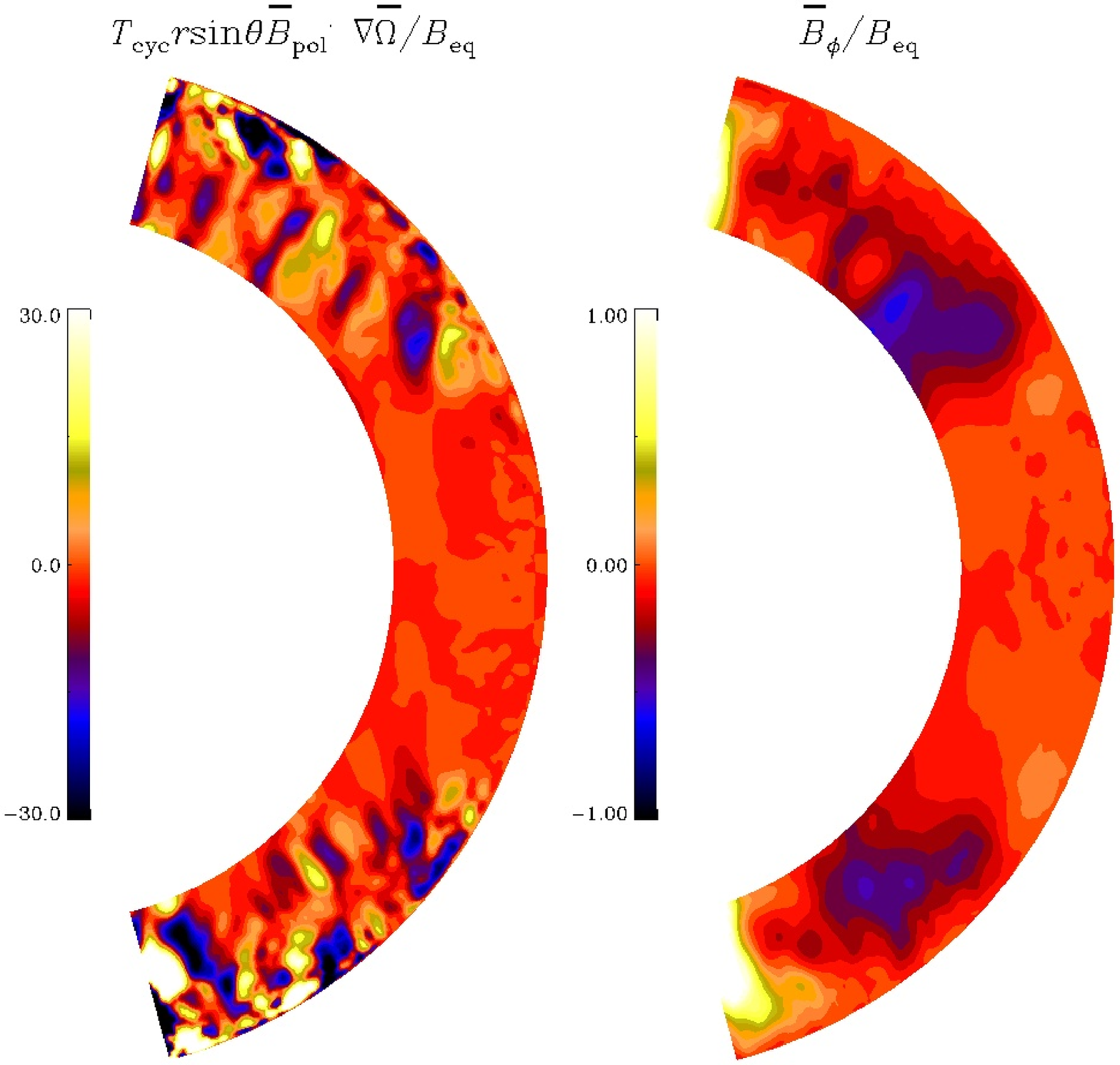}
\caption{$\Omega$-effect, as quantified by
$T_{\rm cyc} r \sin \theta \mean{\bm B}_{\rm pol} \cdot
  \bm\nabla \mean{\Omega}$, where $\mean{\bm B}_{\rm
    pol}=(\mean{B}_r,\mean{B}_\theta)$ (left panels), and the mean
  toroidal magnetic field $\mean{B}_\phi$ (right panels) normalized by
  $\Beq$ from the saturated states of Runs~C1 (upper panels) and D1
  (lower panel). The data is averaged over the longitude and
  approximately 60 convective turnover times in both cases.}
\label{fig:BpgOm}
\end{figure}

\begin{figure}[t]
\centering
\includegraphics[width=\columnwidth]{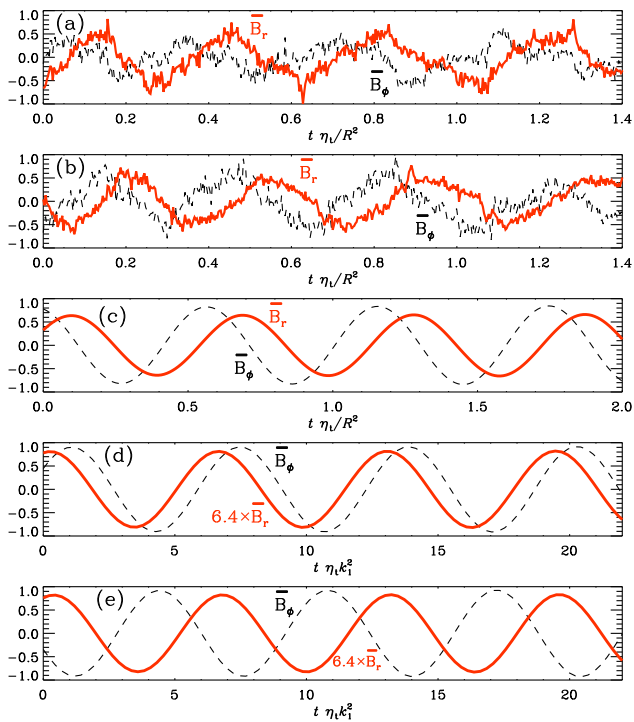}
\caption{
Phase relations of $\mean{B}_r$ (thick red lines) and $\mean{B}_\phi$ (black dashed lines) for Run~C1 at
(a) $70^\circ$ and (b) $30^\circ$ latitude, compared with results of
mean-field dynamos of (c) $\alpha^2$ type and (d), (e) $\alpha^2\Omega$ type,
with positive and negative shear, respectively.
The amplitudes have been rescaled to unity.
Note that only the $\alpha^2$ dynamo has approximately the phase relation
seen in the simulations.
}\label{fig:pphase}
\end{figure}

The relative magnitudes of the estimated values of $c_\alpha$ and
$c_\Omega$, and also the comparable amplitudes of $\mean{B}_r$ and
$\mean{B}_\phi$, shown in Figure~4(a) of \cite{KMB12a},
strongly suggest that the dynamos of this study are not of
$\alpha^2\Omega$ type, as is usually expected for the Sun.
This can be motivated further through direct inspection of the
$\Omega$ term in the equation for the mean toroidal field.
Following \cite{SPD12}, we compare the $\Omega$-effect,
$r\sin\theta \mean{\bm B}_{\rm pol}\cdot\bm\nabla \mean{\Omega}$,
with the mean toroidal field.
The results for Runs~C1 and D1 are shown in Figure~\ref{fig:BpgOm},
where we have scaled
the $\Omega$ term by the magnetic cycle period, $T_{\rm cyc}$.
A fraction of this would be responsible for the production of
mean toroidal field for the next cycle.
For Run~C1, the magnitude of this term is actually large compared with
$\mean{B}_\phi$, and the two are clearly correlated at latitudes below
$\pm35\degr$, which is also where equatorward migration is seen.
For Run~D1, however, no clear correlation is seen even at low latitudes.
The possibility of $\alpha^2\Omega$ type dynamo action therefore remains
unclear, and especially for Run~D1 it may not be the dominant mechanism.
To explore the possibility that our dynamo is of $\alpha^2$ type,
we now consider the phase relation between $\mean{B}_r$
and $\mean{B}_\phi$; see Figure~\ref{fig:pphase}.

\begin{figure*}[t]
\centering
\includegraphics[height=5cm]{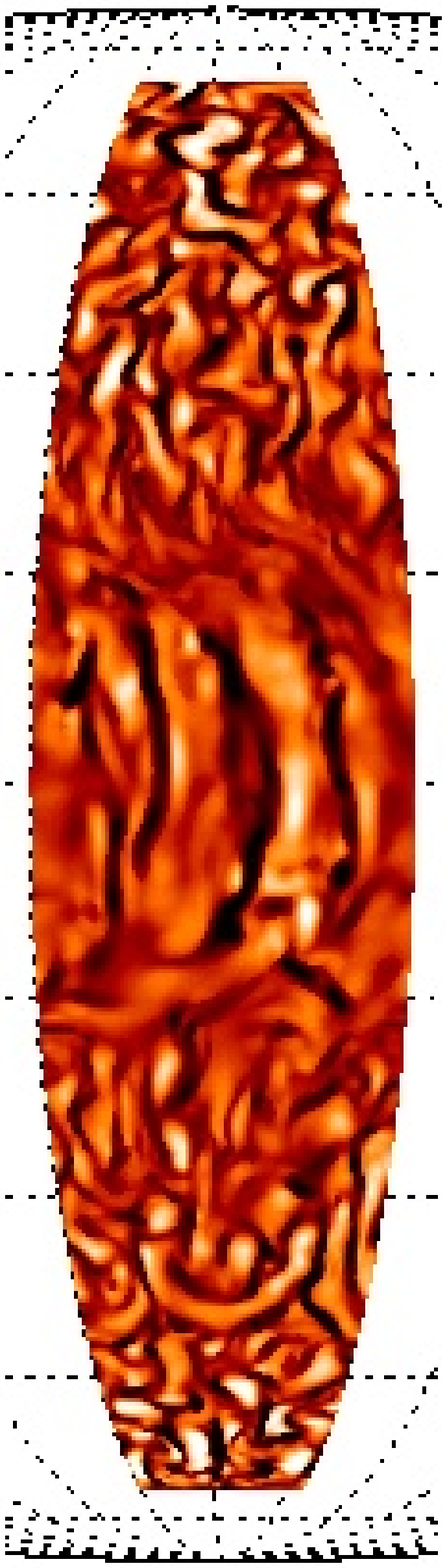}\includegraphics[height=5cm]{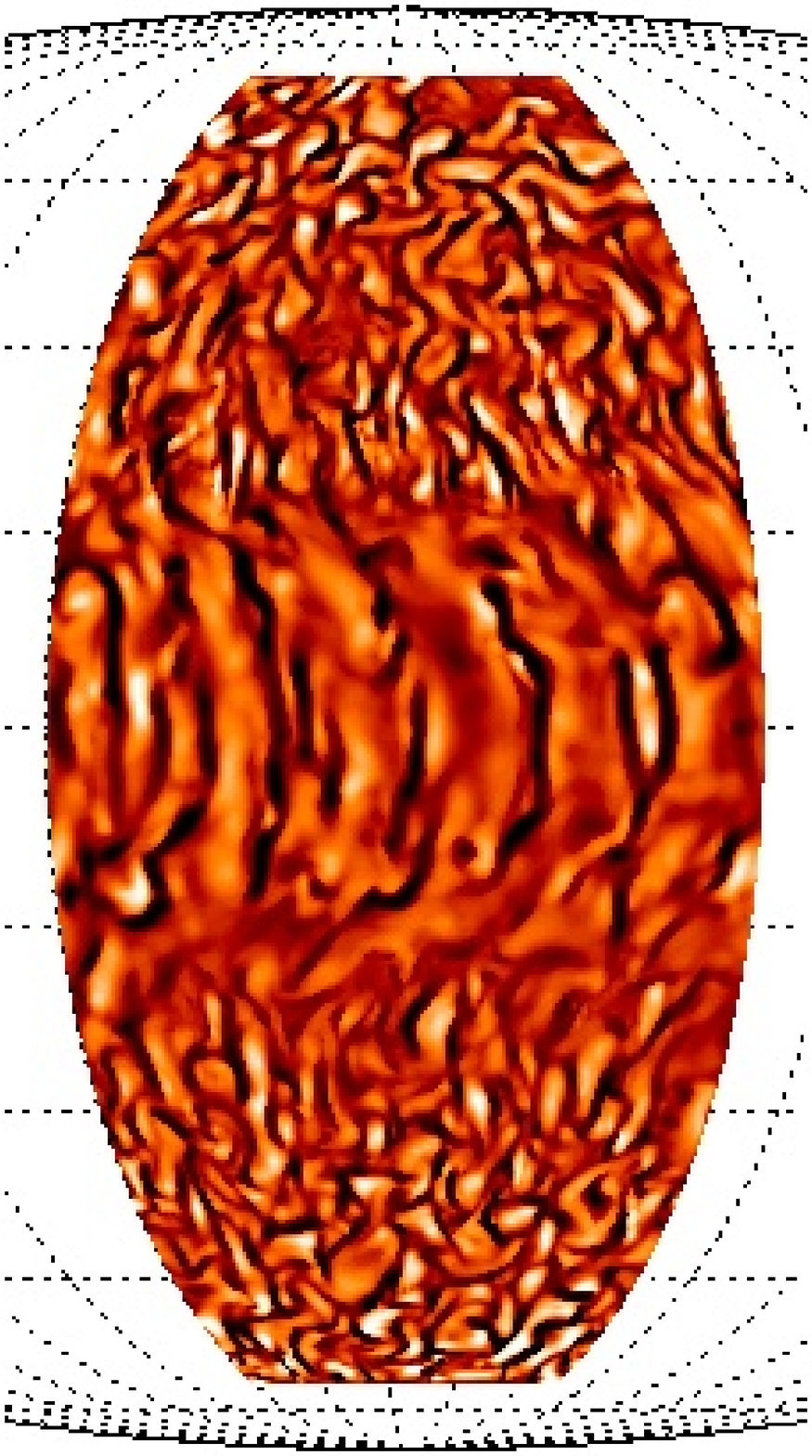}\includegraphics[height=5cm]{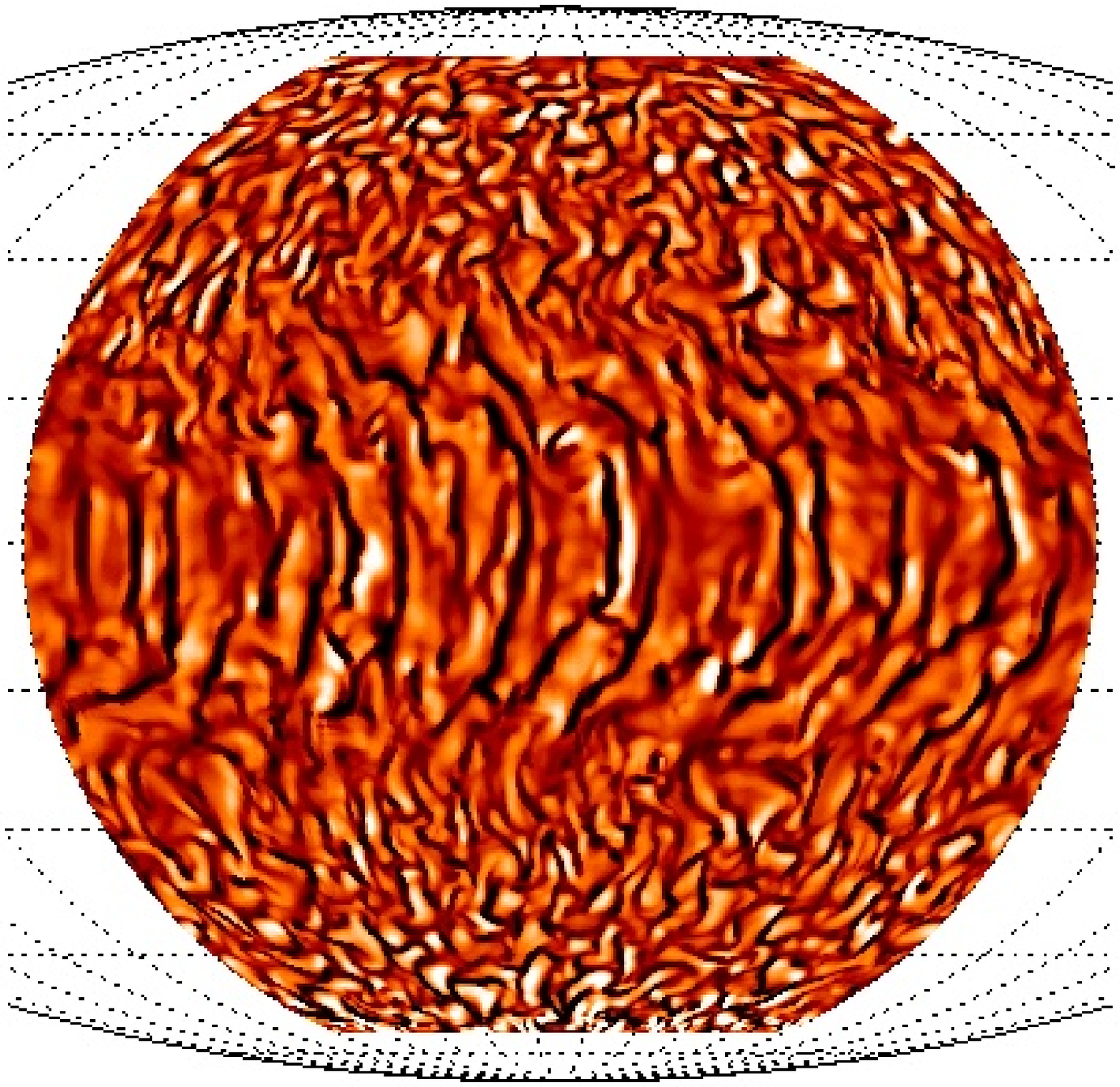}\includegraphics[height=5cm]{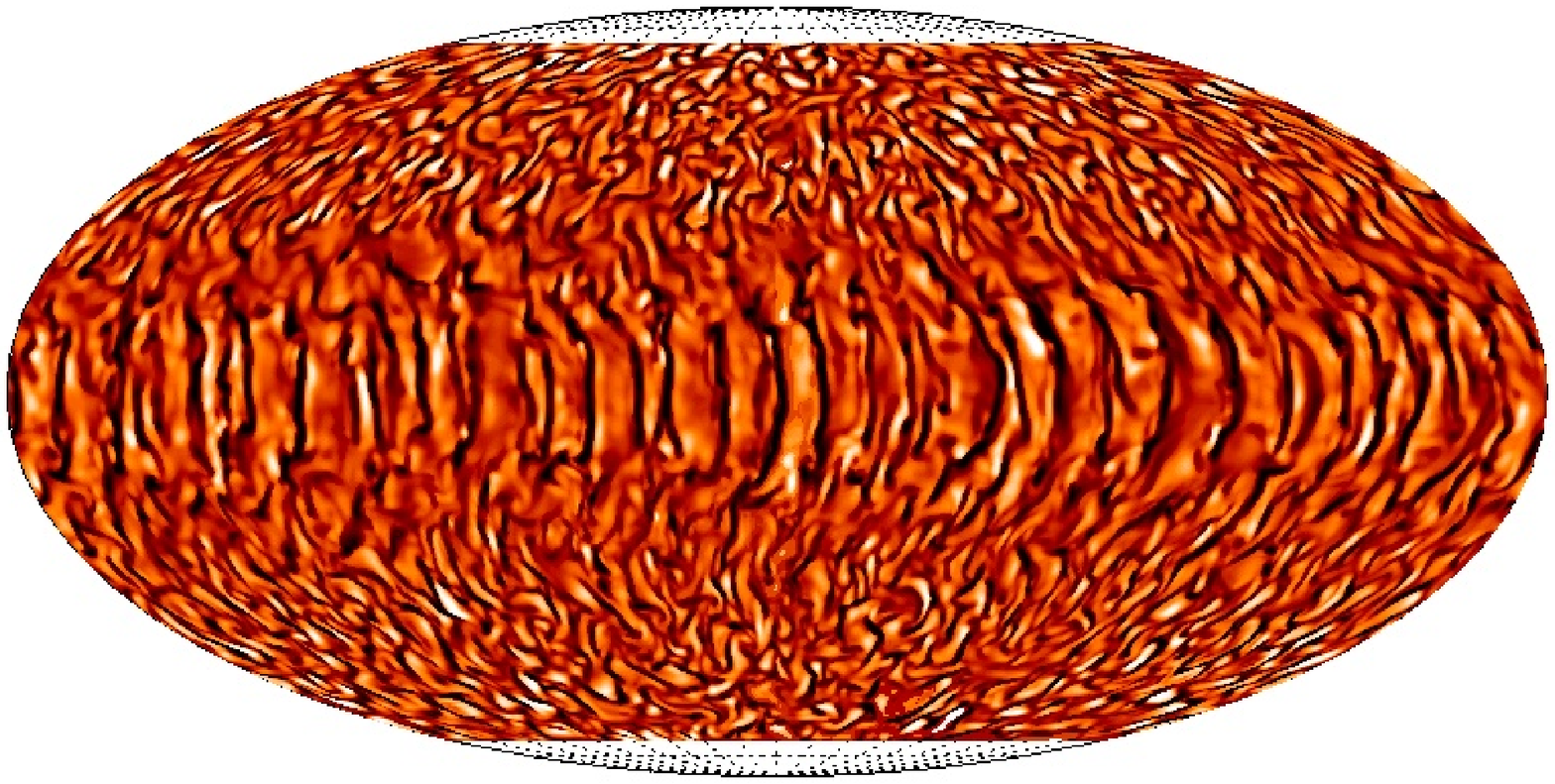}
\includegraphics[height=5cm]{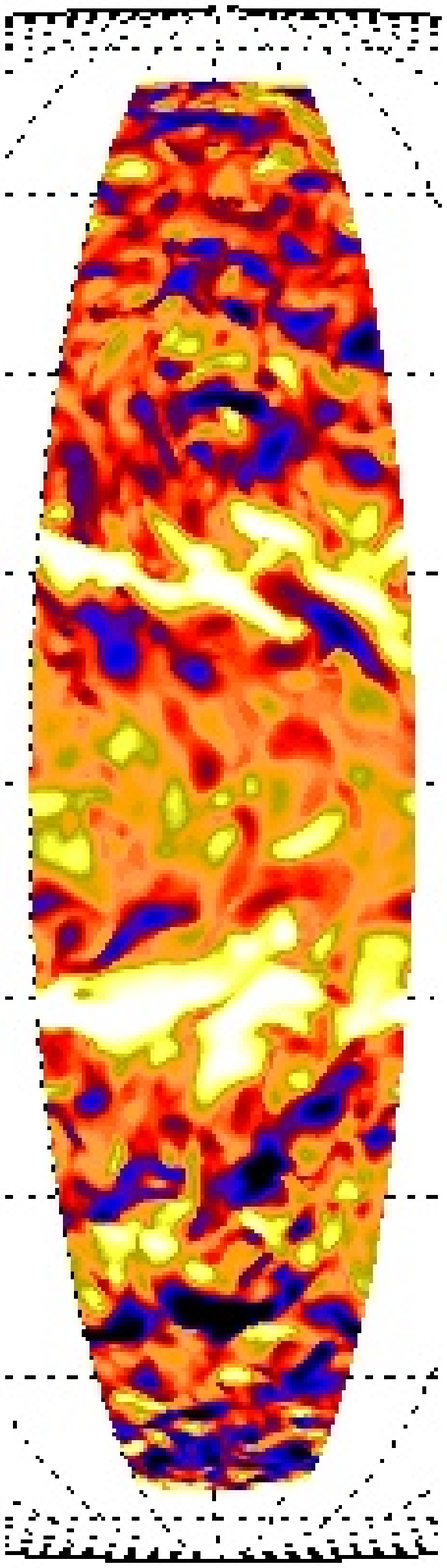}\includegraphics[height=5cm]{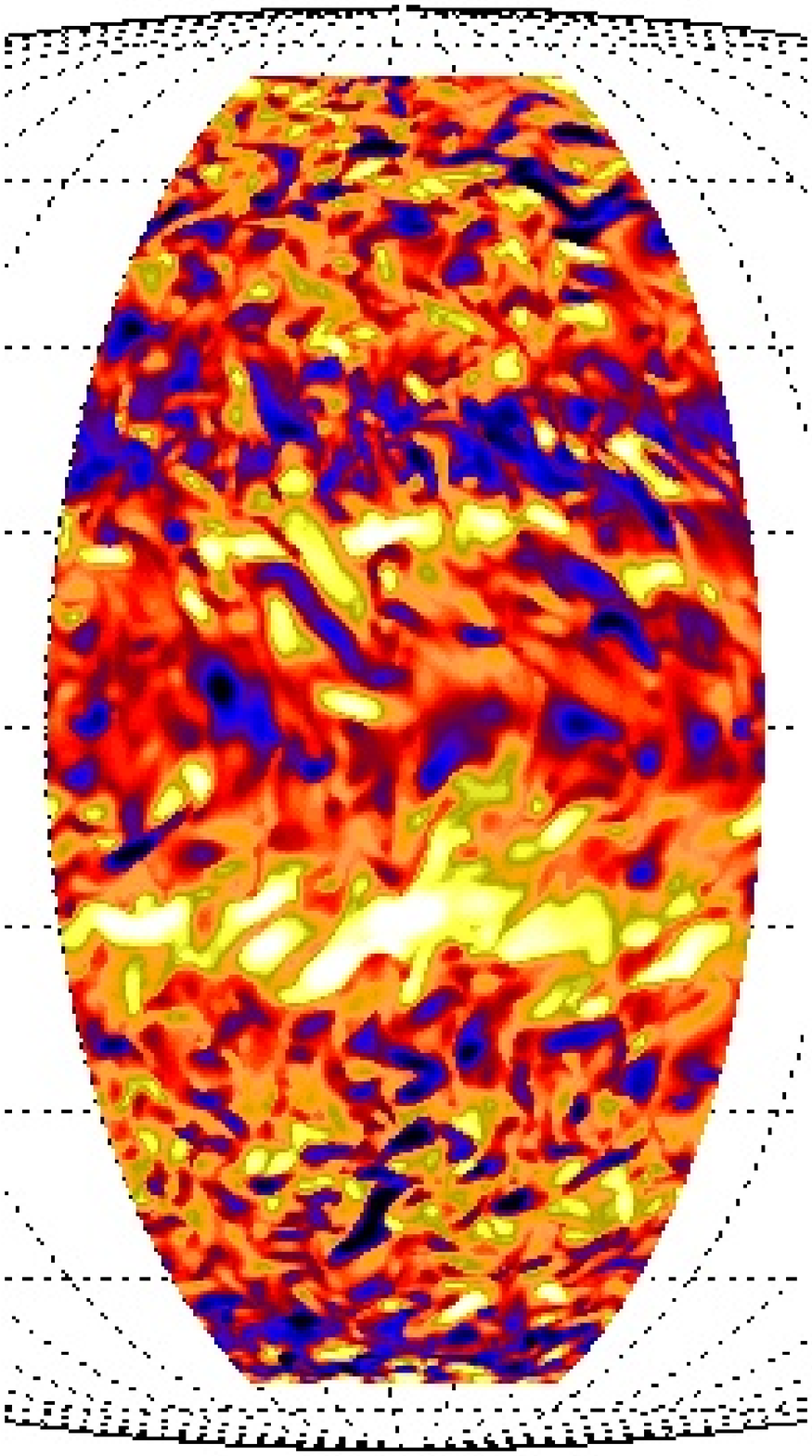}\includegraphics[height=5cm]{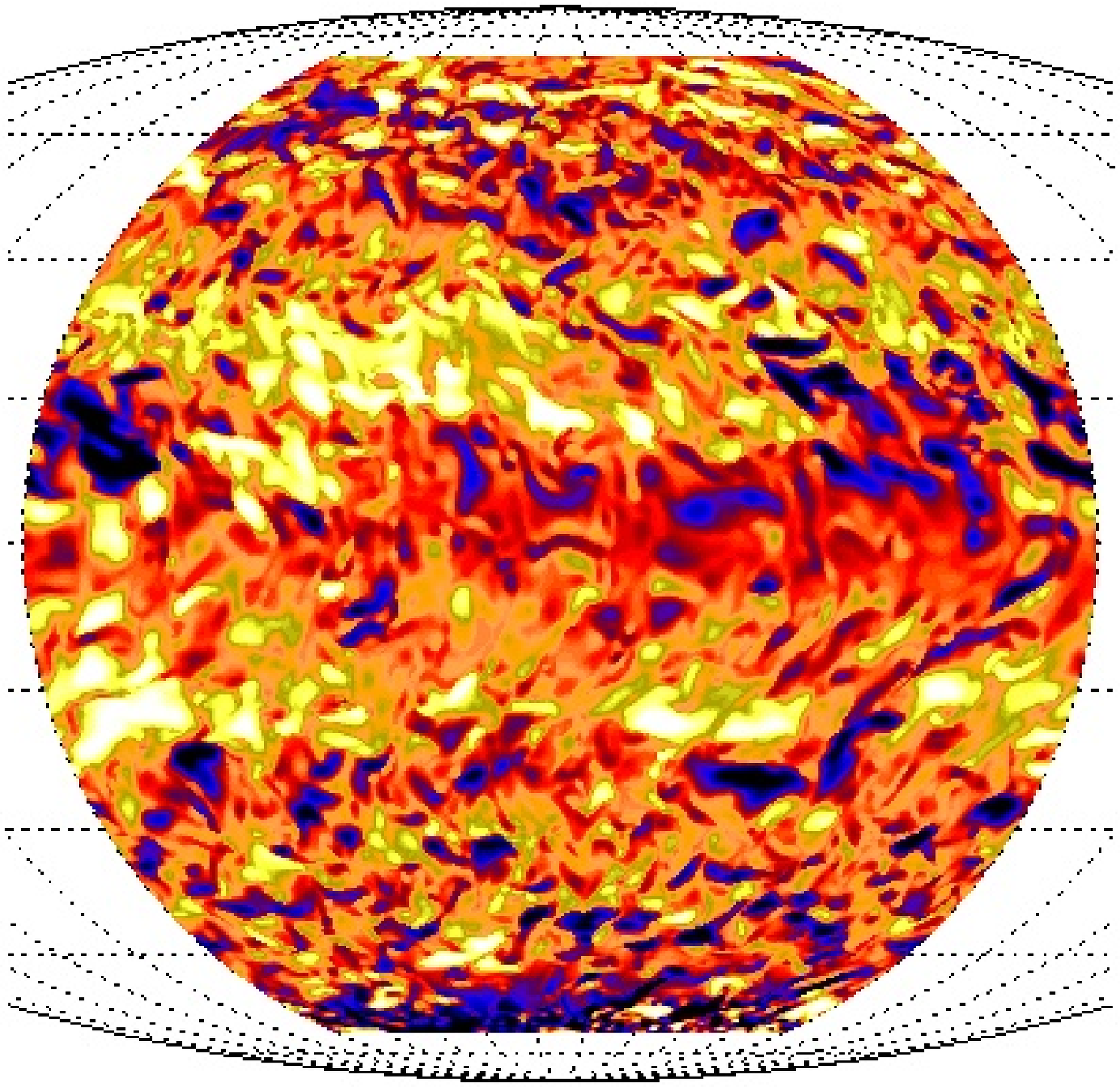}\includegraphics[height=5cm]{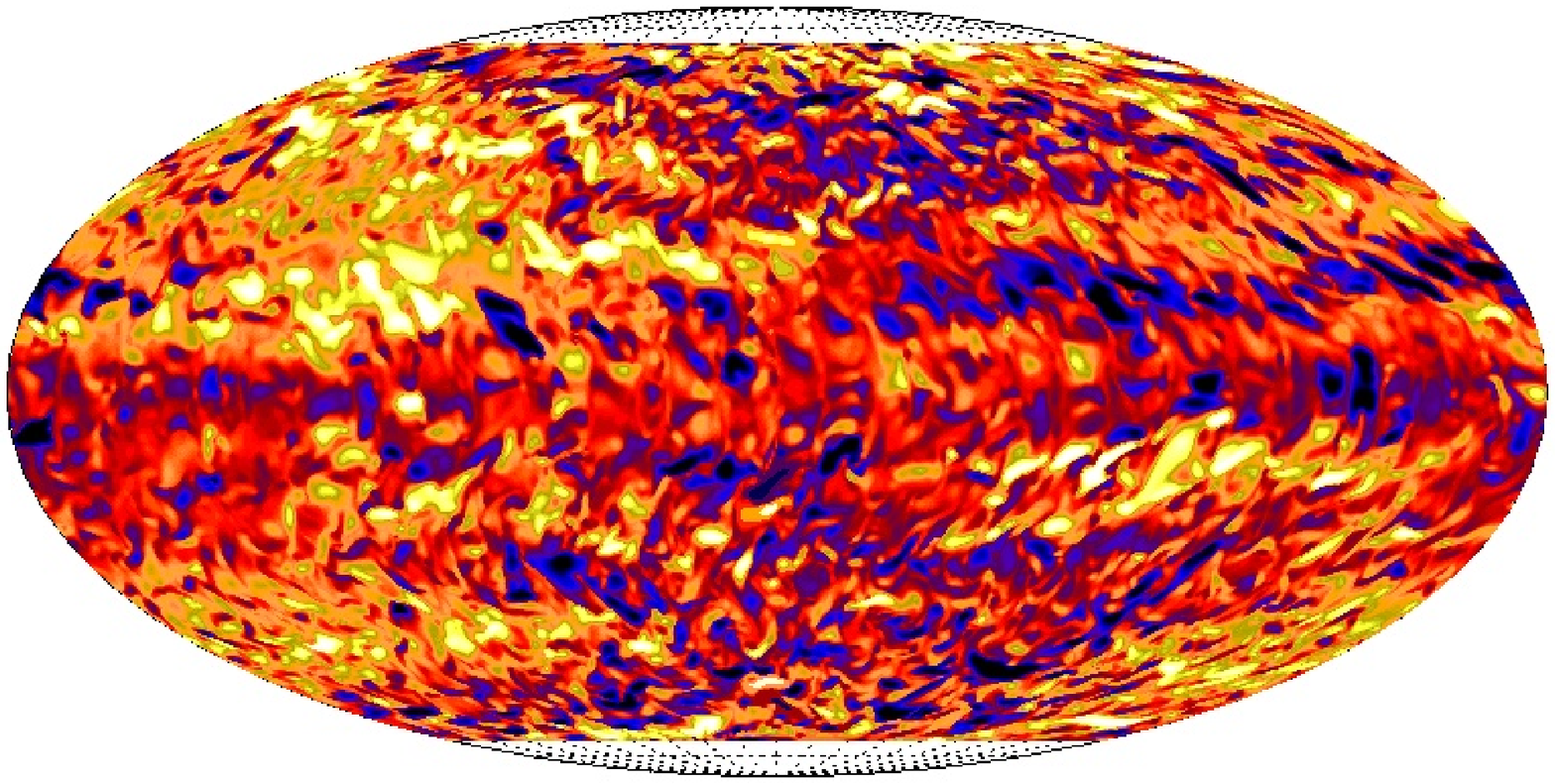}
\caption{Radial velocity $u_r$ (top row) and azimuthal magnetic field $B_\phi$
  (bottom) near the surface of the star $r=0.98\,R$ in Mollweide
  projection from Runs~E1 (left), E2, E3, and E4
  (right). See http://youtu.be/u55sAtN2Fqs for an animation of the
  magnetic field in Run~E4.}\label{fig:pmoll_uu1bb3}
\end{figure*}

\begin{figure}[t]
\centering
\includegraphics[width=0.48\textwidth]{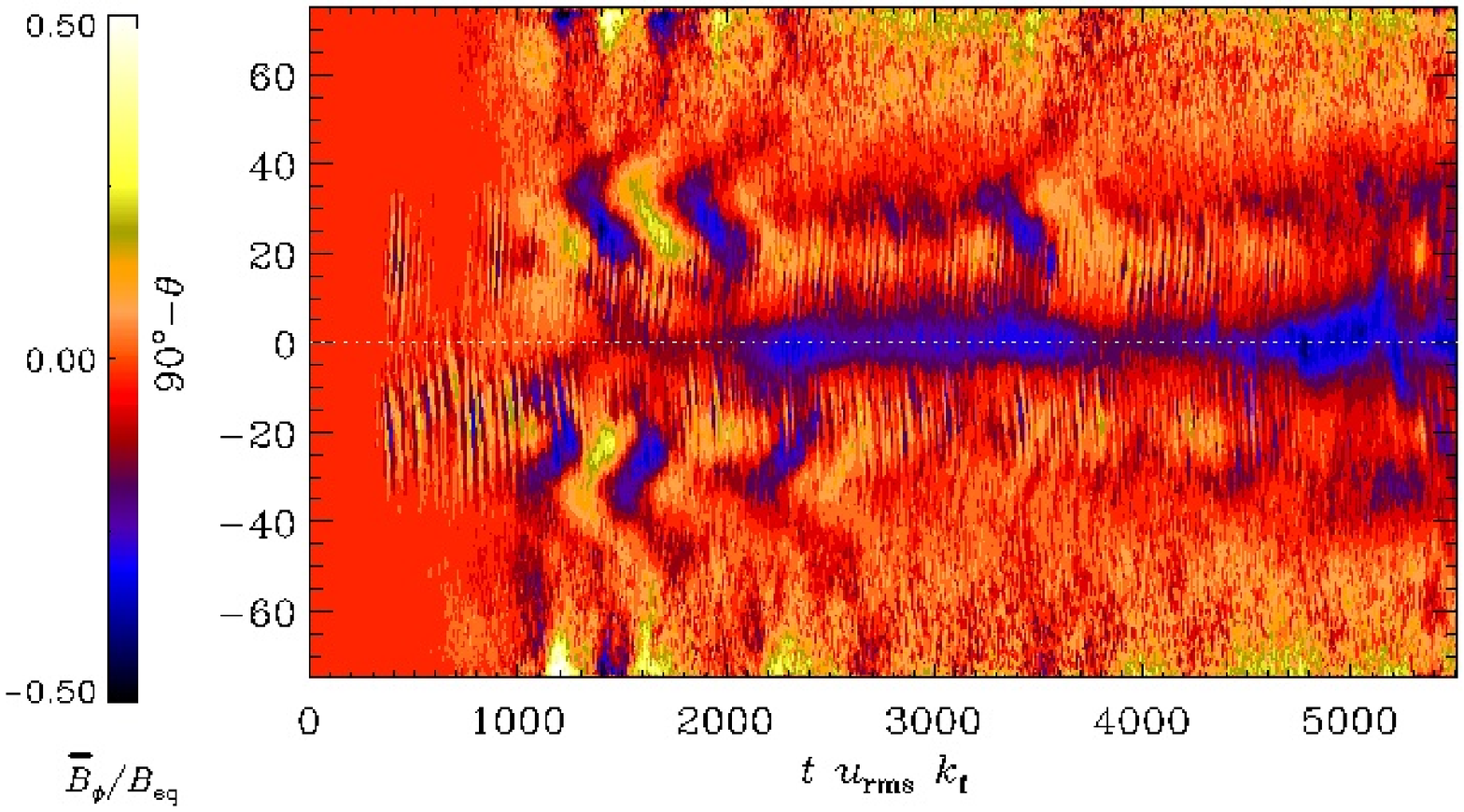}
\caption{Same as Figure~\ref{fig:butterflyA} but for
  Run~E4.}\label{fig:butterflyE4}
\end{figure}

For $\alpha^2\Omega$ dynamos, the phase relation between $\mean{B}_r$ and
$\mean{B}_\phi$ is commonly used to determine the sign of the radial differential
rotation \citep{St76,Yo76}.
By contrast, the sign of $\alpha$ is determined by the sense of
migration of the dynamo wave.
For negative radial shear, $\mean{B}_r$ and $\mean{B}_\phi$ are approximately in
antiphase with $\mean{B}_\phi$ preceding $\mean{B}_r$ by $\approx3\pi/4$.
For positive radial shear, $\mean{B}_r$ and $\mean{B}_\phi$ are approximately
in phase with $\mean{B}_\phi$ lagging $\mean{B}_r$ by $\approx\pi/4$.
In our simulations, radial shear is indeed positive, but $\mean{B}_\phi$
precedes $\mean{B}_r$ by a certain amount; see Figures~\ref{fig:pphase}(a) and (b).
This cannot be explained by an $\alpha^2\Omega$ dynamo where (for positive
radial shear) $\mean{B}_\phi$ {\it lags} $\mean{B}_r$ by $\pi/4$.

Another possibility are oscillatory $\alpha^2$ dynamos of the type
recently found by \cite{MTKB10} using direct numerical simulations of forced
turbulence in a spherical wedge.
Those models have also been used to study the effects
of an outer coronal layer to shed magnetic helicity \citep{WBM11}.
Oscillatory $\alpha^2$ dynamos were first studied by
\cite{BS87} and \cite{Ra87}; see also the monograph of \cite{RH04}.
Such solutions have also been studied in connection with the geodynamo,
where the $\alpha$-effect might change sign in the middle of the outer
liquid iron core \citep{SG03}.
By contrast, in the simulations of \cite{MTKB10} and \cite{WBM11},
$\alpha$ changes sign about the equator.
They used a perfect conductor boundary condition at high latitudes
and found equatorward migrating dynamo waves.
With a vacuum condition, on the other hand, mean-field simulations
have predicted poleward migration \citep{BCC09}.
Those simulations were done in Cartesian geometry, where $(x,y,z)$
can be identified with $(r,\phi,-\theta)$.
Looking at their Figure~2, it is clear that $\mean{B}_y$ lags $\mean{B}_x$
by $\pi/2$.

We have verified the phase relations of the Cartesian model of \cite{BCC09}
with a one-dimensional spherical model\footnote{
\url{http://www.nordita.org/$^\ensuremath{\sim}$brandenb/PencilCode/MeanFieldSpherical.html}
}, where $\alpha=\alpha_0\cos\theta$ has been assumed, which changes sign
about the equator at $\theta=\pi/2$.
The dynamo number for the marginally excited case is
$\alpha_0 R/\etatz\approx23.63$ and, as expected,
$\mean{B}_\phi$ lags $\mean{B}_r$ by $\pi/2$; see Figure~\ref{fig:pphase}(c).
The amplitudes have been rescaled to unity.
The corresponding behavior for an $\alpha^2\Omega$ dynamo is shown in
Figures~\ref{fig:pphase}(d) and (e), where $\mean{B}_\phi$ either precedes
$\mean{B}_r$ by $\pi/4$ or lags $\mean{B}_r$ by $3\pi/4$.
In this case, we have used a Cartesian model with constant $\alpha$,
constant shear, $S=du_y/dx={\rm const}$, and periodic boundaries
in a domain $0<z<L$,
where the critical dynamo number is $\alpha S L^3/\etatz^2\approx8\pi^2$.
In this model, the Cartesian coordinates $(x,y,z)$ correspond to
$(-r,\phi,\theta)$, so positive (negative) values of $S$ correspond
to negative (positive) radial angular velocity gradients.
Neither of the phase relations of these two models agrees with those
of the DNS.

Another hint pointing toward an $\alpha^2$ dynamo in Run~C1 is, that the
magnetic field is particularly strong in the middle of the convection
zone ($0.8\,R<r<0.9\,R$), from where dynamo waves seem to propagate toward the
surface and the bottom of the convection zone; see Figure~3(a) of
\cite{KMB12a}.
Even though there exists no tachocline at the bottom or a near-surface
shear layer at the top of our convection zone, the $\Omega$-effect
appears to be larger toward the bottom and top of the
convection zone; see Figure~\ref{fig:pCOm}.
Therefore, an $\Omega$-effect would produce the magnetic field mainly
at the bottom and the top of the convection zone, which is not the
case in our simulation.
The case of Run~D1 is more clear because the $\Omega$-effect is weak
except near the boundaries (Figure~\ref{fig:pCOm}) and the toroidal
field shows no correlation with it; see Figure~\ref{fig:BpgOm}.
We therefore suggest that oscillatory $\alpha^2$ dynamos of the type found
by \cite{MTKB10} might explain the origin of equatorward migrating
dynamo waves in the spherical wedge simulations of \cite{KMB12a}.
It is also possible that this mechanism explains the poleward migration at
high latitudes, but detailed comparisons must await a proper determination
of $\alpha$-effect and turbulent diffusivity tensors.
A first step toward this has recently been attempted by
\cite{RCGBS11} who estimated the tensor components of $\alpha$ by
correlating the electromotive force with the mean magnetic field using
singular value decomposition. These results were applied in mean-field
models of \cite{SCB13}, in an effort to explain the dynamos seen by
\cite{GCS10}.
However, this analysis is flawed in the sense that
the diffusive part of the electromotive force cannot be separated from the one
related to the $\alpha$-effect. This has been shown to lead to
erroneous estimates of $\alpha$ \citep{KKB10}. The only reliable way
to compute the turbulent diffusion tensor is currently possible with the
test-field method \citep{SRSRC05,SRSRC07}. We postpone such analysis
to a future publication.

\subsection{Effect of domain size}

We recently reported equatorward migration of activity belts in a
spherical wedge simulation \citep{KMB12a}. There we gave results
from simulations with a $\phi$-extent of $\pi/2$.
However, at large values of the Coriolis number, the $\alpha$-effect
becomes sufficiently anisotropic and differential rotation weak so
that non-axisymmetric solutions become possible; see \cite{MB95} for
corresponding mean-field models with dominant $m=1$ modes in the
limit of rapid rotation.
To allow for such modes, we now choose a $\phi$-extent of up to $2\pi$
for the same model as in \cite{KMB12a}.
In the present case, we find that for $\Co\approx7.8$ it is possible that
non-axisymmetric dynamo modes of low azimuthal order ($m=1$ or 2)
can be dominant.
This was not possible in the simulations of \cite{KMB12a}.
The same applies
to non-axisymmetric modes excited in hydrodynamic convection
\citep[e.g.,][]{Bu02,BBBMT08,KMGBC11,ABBMT12}.

We test the robustness of the equatorward migration by performing runs
with $\phi_0=\pi/4$, $\pi/2$, $\pi$, and $2\pi$ with otherwise similar
parameters; see~Table \ref{tab:runs}. We find that the same dynamo
mode producing equatorward
migration is ultimately excited in all of these runs. The only
qualitatively different run is that with $\phi_0=2\pi$ where the
poleward mode near the equator grows much faster than in the other
cases. However, after $t \urms \kef\approx1500$ the equatorward
mode takes over similarly as in the runs with a smaller
$\phi_0$.

The velocity field shows no marked evidence of low-degree
non-axisymmetric constituents, but there are indications of $m=1$
structures in the instantaneous magnetic field
(Figure~\ref{fig:pmoll_uu1bb3}); see also
http://youtu.be/u55sAtN2Fqs for an animation
of the toroidal magnetic field.
This is also reflected by the fraction of the axisymmetric part
of the total magnetic energy; see Columns 5 and 6 of
Table~\ref{tab:runs2}. We find that the energy of the mean toroidal
field decreases monotonically when $\phi_0$ is increased so that there
is a factor of three in $E_{\rm tor}/E_{\rm mag}$ between the extreme cases of Runs~E1
and E4. The axisymmetric part still exhibits an oscillatory mode
with equatorward migration in all runs in Set~E.
The most prominent exception is visible in Figure~\ref{fig:butterflyE4}, 
where we show the butterfly diagram of the $m=0$ contribution for Run~E4.
Clearly, equatorward migratory events are now rare and superimposed
on a background of small-scale, high-frequency poleward migratory field.

\begin{figure*}[t]
\centering
\includegraphics[width=\textwidth]{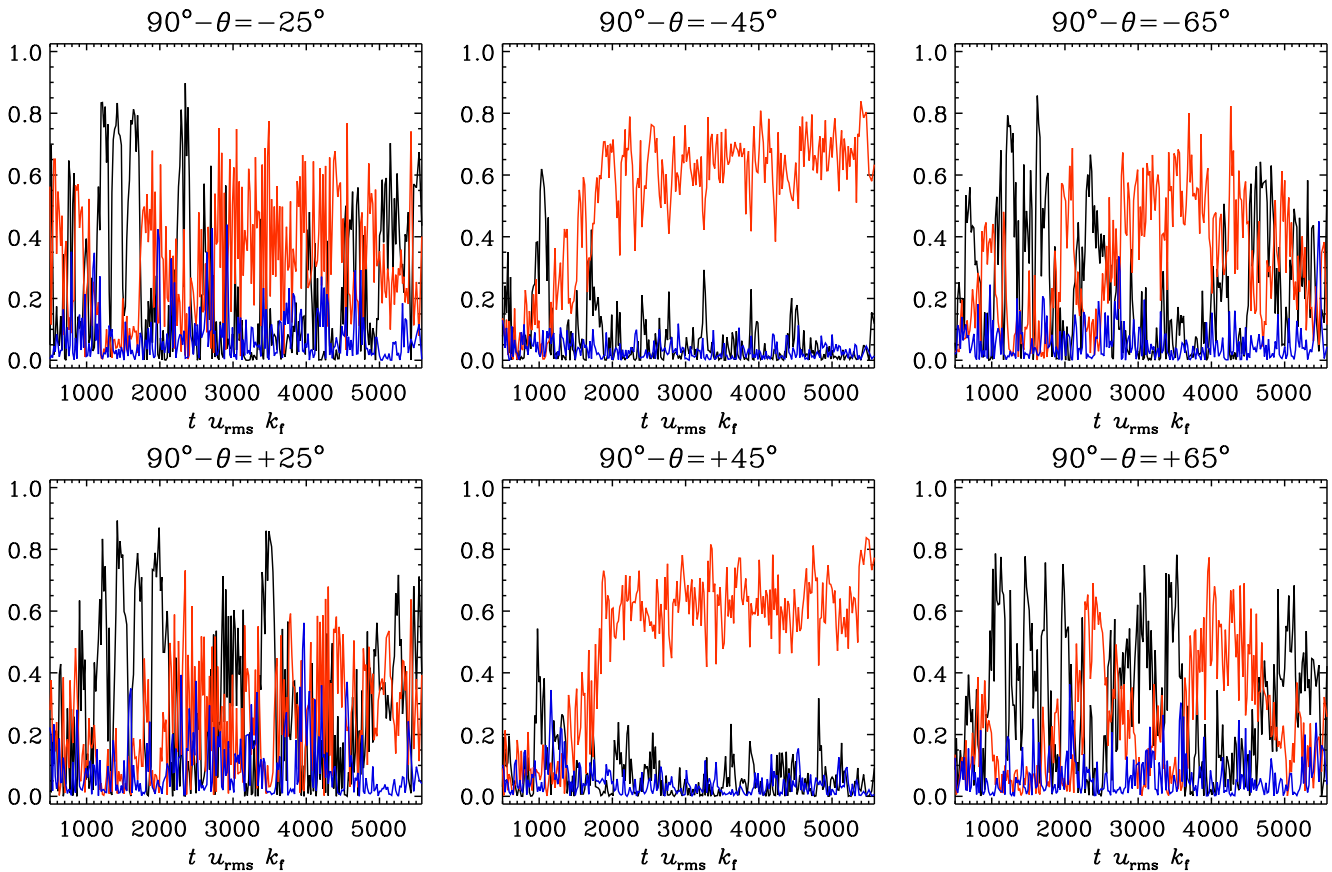}
\caption{Energies of the $m=0$ (black lines), 1 (red), and 2 (blue)
modes of the azimuthal magnetic field as functions of time near the
 surface of the star ($r=0.98\,R$) in Run~E4. The data is averaged over
$10\degr$ latitude strips centered at latitudes $90\degr-\theta=\pm25\degr$
  (left panels), $\pm45\degr$ (middle), and $\pm65\degr$ (right) and
  normalized by the total energy within each strip.
  The top and bottom rows refer to negative and positive latitudes,
  respectively.
}\label{fig:pps}
\end{figure*}

We compute power spectra of the azimuthal component of the magnetic
field from the Run~E4 over three $10\degr$ latitude strips from each hemisphere,
centered around latitudes of $\pm25\degr$, $\pm45\degr$,
and $\pm65\degr$. The results for the three lowest degrees $m=0, 1, 2$ are shown in
Figure~\ref{fig:pps}. We find that at low ($\pm25\degr$) and high ($\pm65\degr$)
latitudes the axisymmetric ($m=0$) mode begins to dominate after
around $1000$ turnover times and shows a cyclic pattern consistent
with that seen in the time-latitude diagram of the azimuthally
averaged field. After $t \urms \kef \approx 1600$, however, the $m=1$
mode becomes stronger in the southern hemisphere, coinciding with
the growth of the $m=1$ mode at mid-latitudes ($\pm45\degr$) where it
dominates
earlier in both hemispheres.
This is in rough agreement with some observational results of rapid
rotators, which show the most prominent non-axisymmetric temperature
\citep[e.g.,][]{Thomas01,Heidi07,Marjaana11} and magnetic structures
\citep{Oleg11} at the latitudinal range around 60$\degr$--80$\degr$, while
the equatorial and polar regions are more axisymmetric; some
temperature inversions even show almost completely axisymmetric
distributions in the polar regions and rings of azimuthal
field at low latitudes \citep[e.g.,][]{Donati03}. The strength of the
axisymmetric versus the non-axisymmetric part in such objects has also been
reported to vary over time with a time scale of a few years
\citep{Oleg11}.

\subsection{Irradiance variations}
\label{sec:irra}
\begin{figure}[t]
\centering
\includegraphics[width=\columnwidth]{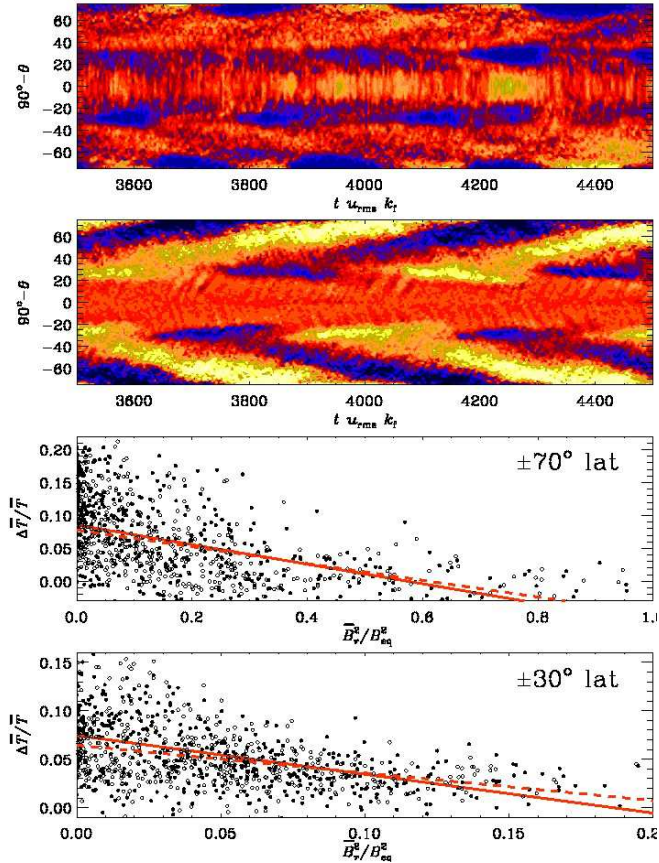}
\caption{
Top panel: azimuthally averaged temperature fluctuations, normalized
with their temporal average, $\Delta\mean{T}/\mean{T}$,
at the surface as a function of time.
Second panel: azimuthally averaged radial field at the surface,
$\mean{B}_r/B_{\rm eq}$, as function of time.
Lower panels: scatter plots of $\Delta\mean{T}/\mean{T}$ vs.\
$\mean{B}_r/B_{\rm eq}$ at $\pm70^\circ$ and $\pm30^\circ$ latitude.
Here, solid (dashed) red lines and filled (open) symbols refer to
northern (southern) latitudes.
The slopes are $Q_T\approx0.14$ and 0.33 at $\pm70$ and $\pm30$ latitude.
All plots show quantities from the saturated stage of Run~C1.
}\label{fig:pTTmxy}
\end{figure}

In contrast to the constant temperature condition used earlier, the
black body boundary condition (\ref{eq:bbb}) allows the temperature to
vary
at the surface of the star and thus enables the study of irradiance
variations due to the magnetic cycle \citep{Spr00}.
Such variations might even be responsible for driving torsional
oscillations in the Sun \citep{Spr03,Re06}.
In Figure~\ref{fig:pTTmxy} we compare time--latitude
surface representations ($r=R$) of azimuthally averaged
temperature variations relative to its temporal average,
$\Delta\mean{T}(\theta,t)=\mean{T}(\theta,t)-\brac{\mean{T}}_t(\theta)$,
with those of the azimuthally averaged radial magnetic field,
$\mean{B}_r(\theta,t)$, for Run~C1 in the saturated state of the dynamo.
We also show scatter plots of $\Delta\mean{T}/\mean{T}$ versus
$\mean{B}_r/B_{\rm eq}$ at $\pm70^\circ$ and $\pm30^\circ$ latitude
to demonstrate that there are many instances where enhanced surface
magnetic activity leads to a local decrease in surface temperature.
We see that
\begin{equation}
\Delta\mean{T}/\mean{T} \approx -Q_T \mean{B}_r^2/B_{\rm eq}^2
\end{equation}
with `quenching' coefficients $Q_T$ of $\approx0.14$ at high latitudes
and $\approx0.33$ at low latitudes.
However, there is also considerable scatter, even though our data
is already longitudinally averaged.
Without such averaging, the correlation between individual
structures on the surface would be rather poor.
The temperature modulation is best seen near the poles; see
Figure~\ref{fig:pTTmxy}.
This could be a consequence of a strong radial magnetic field
that builds up some 50--100 turnover times earlier
and thus precedes the temperature signal.
A weaker modulation is also seen near the equator.
The peak values of $\Delta\mean{T}/\mean{T}$ at high latitudes are
15\%--20\% of the
surface temperature; see the last two panels of Figure~\ref{fig:pTTmxy}.
This is relatively large compared with earlier work using mean-field
models \citep{BMT92}, which showed remarkably little relative variation
of the order of $10^{-3}$ in the bulk of the convection zone and
even less at the surface.
This difference in the modulation amplitude
is probably related to the importance of latitudinal
variations that were also present in the mean-field model of
\cite{BMT92} and referred to as thermal shadows \citep{Pa87b}.

\section{Conclusions}
\label{sec:conclusions}

We have studied the effects of density stratification on the dynamo
solutions found in simulations of rotating turbulent convection in
spherical wedge geometry for four values of $\Gamma_\rho$,
which is the ratio of the densities at the bottom and at the surface
of the convection zone. In addition, we vary the rotation rate for
each value of $\Gamma_\rho$. For all stratifications we find
quasi-steady large-scale dynamos for lower rotation and oscillatory
solutions when rotation is rapid enough. The transition from
quasi-steady to oscillatory modes seems to occur at a lower $\Co$ for
higher stratification. Furthermore, for low values of $\Gamma_\rho$ the
oscillatory solutions show only poleward propagation of the activity
belts whereas at higher $\Gamma_\rho$ an equatorward branch appears
at low latitudes.

The equatorward branch was first noted by \cite{KMB12a} using a
wedge with $\phi_0=90\degr$ longitude extent. Here we test the
robustness of this result by varying $\phi_0$ from $45\degr$ to
full $360\degr$. We find a very similar pattern of the axisymmetric
part of the field in all cases. However, the energy of the
axisymmetric magnetic field decreases with increasing $\phi_0$. In
the simulation with the full $\phi$-extent of $2\pi$ we observe an
$m=1$
mode which is visible even by visual inspection
(see Figure~\ref{fig:pmoll_uu1bb3}). Such field configurations have
been observed in rapidly rotating late-type stars \citep[see
e.g.,][]{Oleg11} and our simulation is one of the first to
reproduce such features \citep[see also][]{GD08,GDW12}.
We are currently investigating
the rapid rotation regime with more targeted runs which will be
reported in a separate publication \citep{CKMB13}.

The ratio between cycle to rotation frequency, $\omega_{\rm cyc}/\Omega_0$,
is argued to be an important non-dimensional output parameter of a
cyclic dynamo.
For the Sun and other relatively inactive stars, this ratio is around 0.01,
while for the more active stars it is around 0.002.
For our models we find values in the range 0.002--0.01, but for most of the
runs it is around 0.004.
Although it is premature to make detailed comparisons with other stars
and even the Sun, it is important to emphasize that kinematic mean-field dynamos
produce the correct cycle frequency only for values of the turbulent magnetic
diffusivity that are at least 10 times smaller than what is suggested by
standard estimates \citep{Ch90}.
In our case, these longer cycle periods (or smaller cycle frequencies)
might be a result of nonlinearity as they are only obtained in the
saturated
regime of the dynamo.
The detailed reason for this is unclear, but it has been speculated that
it is connected with a slow magnetic helicity evolution \citep{Br05}.
On the other hand, magnetic helicity effects are expected to become
important only at values of $\Rm$ between 100 and 1000 \citep{DSGB13},
which is much larger than what has been reached in the present work.
Equally unclear is the reason for equatorward migration, which, as we have
seen, might be a consequence of nonlinearity, as well.
It will therefore be important to provide an accurate determination of all
the relevant turbulent transport coefficients.
The explanation favored in the present paper is that the dynamo wave is
that expected for an oscillatory $\alpha^2$ dynamo caused by the change
of sign of $\alpha$ about the equator.
This is evidenced by our finding that $\mean{B}_\phi$ lags $\mean{B}_r$
by about $\pi/2$, which cannot be explained by an $\alpha^2\Omega$ dynamo.

\acknowledgements
We thank Matthew Browning, Martin Schrinner, L$\acute{\rm u}$cia
Duarte, and Matthias Rheinhardt,
as well as the referee for making a number of useful suggestions.
The simulations were performed using the supercomputers hosted by CSC
-- IT Center for Science Ltd.\ in Espoo, Finland, who are administered
by the Finnish Ministry of Education. Financial support from the
Academy of Finland grants No.\ 136189, 140970 (PJK) and 218159, 141017
(MJM), as well as the Swedish Research Council grant 621-2007-4064,
and the European Research Council under the AstroDyn Research Project
227952 are acknowledged as well as the HPC-Europa2 project, funded by
the European Commission - DG Research in the Seventh Framework
Programme under grant agreement No.\ 228398.
The authors thank NORDITA for hospitality during their visits.

\bibliography{paper}

\begin{thebibliography}{99}
\expandafter\ifx\csname natexlab\endcsname\relax\def\natexlab#1{#1}\fi

\bibitem[{{Arlt} {et~al.}(2005){Arlt}, {Sule}, \& {R{\"u}diger}}]{ASR05}
{Arlt}, R., {Sule}, A., \& {R{\"u}diger}, G. 2005, \aap, 441, 1171

\bibitem[{{Augustson} {et~al.}(2012){Augustson}, {Brown}, {Brun}, {Miesch}, \&
  {Toomre}}]{ABBMT12}
{Augustson}, K.~C., {Brown}, B.~P., {Brun}, A.~S., {Miesch}, M.~S., \&
  {Toomre}, J. 2012, \apj, 756, 169

\bibitem[{{Babcock}(1961)}]{Bab61}
{Babcock}, H.~W. 1961, \apj, 133, 572

\bibitem[{{Baryshnikova} \& {Shukurov}(1987)}]{BS87}
{Baryshnikova}, I., \& {Shukurov}, A. 1987, AN, 308, 89

\bibitem[{{Bonanno} {et~al.}(2002){Bonanno}, {Elstner}, {R{\"u}diger}, \&
  {Belvedere}}]{BERB02}
{Bonanno}, A., {Elstner}, D., {R{\"u}diger}, G., \& {Belvedere}, G. 2002, \aap,
  390, 673

\bibitem[{{Brandenburg}(2001)}]{B01}
{Brandenburg}, A. 2001, \apj, 550, 824

\bibitem[{{Brandenburg}(2005)}]{Br05}
{Brandenburg}, A. 2005, \apj, 625, 539

\bibitem[{{Brandenburg} {et~al.}(2009){Brandenburg}, {Candelaresi}, \&
  {Chatterjee}}]{BCC09}
{Brandenburg}, A., {Candelaresi}, S., \& {Chatterjee}, P. 2009, \mnras, 398,
  1414

\bibitem[{{Brandenburg} {et~al.}(2005){Brandenburg}, {Chan}, {Nordlund}, \&
  {Stein}}]{BCNS05}
{Brandenburg}, A., {Chan}, K.~L., {Nordlund}, {\AA}., \& {Stein}, R.~F. 2005,
  AN, 326, 681

\bibitem[{{Brandenburg} {et~al.}(1992){Brandenburg}, {Moss}, \&
  {Tuominen}}]{BMT92}
{Brandenburg}, A., {Moss}, D., \& {Tuominen}, I. 1992, \aap, 265, 328

\bibitem[{{Brandenburg} {et~al.}(1998){Brandenburg}, {Saar}, \&
  {Turpin}}]{BST98}
{Brandenburg}, A., {Saar}, S.~H., \& {Turpin}, C.~R. 1998, \apjl, 498, L51

\bibitem[{{Brown} {et~al.}(2008){Brown}, {Browning}, {Brun}, {Miesch}, \&
  {Toomre}}]{BBBMT08}
{Brown}, B.~P., {Browning}, M.~K., {Brun}, A.~S., {Miesch}, M.~S., \& {Toomre},
  J. 2008, \apj, 689, 1354

\bibitem[{{Brown} {et~al.}(2010){Brown}, {Browning}, {Brun}, {Miesch}, \&
  {Toomre}}]{BBBMT10}
{Brown}, B.~P., {Browning}, M.~K., {Brun}, A.~S., {Miesch}, M.~S., \& {Toomre},
  J. 2010, \apj, 711, 424

\bibitem[{{Brown} {et~al.}(2011){Brown}, {Miesch}, {Browning}, {Brun}, \&
  {Toomre}}]{BMBBT11}
{Brown}, B.~P., {Miesch}, M.~S., {Browning}, M.~K., {Brun}, A.~S., \& {Toomre},
  J. 2011, \apj, 731, 69

\bibitem[{{Brun} {et~al.}(2004){Brun}, {Miesch}, \& {Toomre}}]{BMT04}
{Brun}, A.~S., {Miesch}, M.~S., \& {Toomre}, J. 2004, \apj, 614, 1073

\bibitem[{{Brun} {et~al.}(2011){Brun}, {Miesch}, \& {Toomre}}]{BMT11}
{Brun}, A.~S., {Miesch}, M.~S., \& {Toomre}, J. 2011, \apj, 742, 79

\bibitem[{{Busse}(2002)}]{Bu02}
{Busse}, F.~H. 2002, PhFl, 14, 1301

\bibitem[{{Chatterjee} {et~al.}(2004){Chatterjee}, {Nandy}, \&
  {Choudhuri}}]{CNC04}
{Chatterjee}, P., {Nandy}, D., \& {Choudhuri}, A.~R. 2004, \aap, 427, 1019

\bibitem[{{Choudhuri}(1990)}]{Ch90}
{Choudhuri}, A.~R. 1990, \apj, 355, 733

\bibitem[{{Choudhuri} {et~al.}(2007){Choudhuri}, {Chatterjee}, \&
  {Jiang}}]{CCJ07}
{Choudhuri}, A.~R., {Chatterjee}, P., \& {Jiang}, J. 2007, Phys. Rev. Lett.,
  98, 131103

\bibitem[{{Christensen} \& {Aubert}(2006)}]{CA06}
{Christensen}, U.~R., \& {Aubert}, J. 2006, GeoJI, 166, 97

\bibitem[{{Cole} {et~al.}(2013){Cole}, {K{\"a}pyl{\"a}}, {Mantere}, \&
  {Brandenburg}}]{CKMB13}
{Cole}, E., {K{\"a}pyl{\"a}}, P.~J., {Mantere}, M.~J., \& {Brandenburg}, A.
  2013, submitted to ApJL, arXiv:1309.6802

\bibitem[{{Del Sordo} {et~al.}(2013){Del Sordo}, {Guerrero}, \&
  {Brandenburg}}]{DSGB13}
{Del Sordo}, F., {Guerrero}, G., \& {Brandenburg}, A. 2013, \mnras, 429, 1686

\bibitem[{{Dikpati} \& {Charbonneau}(1999)}]{DC99}
{Dikpati}, M., \& {Charbonneau}, P. 1999, \apj, 518, 508

\bibitem[{{Dikpati} {et~al.}(2004){Dikpati}, {de Toma}, {Gilman}, {Arge}, \&
  {White}}]{DdTGAW04}
{Dikpati}, M., {de Toma}, G., {Gilman}, P.~A., {Arge}, C.~N., \& {White}, O.~R.
  2004, \apj, 601, 1136

\bibitem[{{Dikpati} \& {Gilman}(2006)}]{DG06}
{Dikpati}, M., \& {Gilman}, P.~A. 2006, \apj, 649, 498

\bibitem[{{Donati} {et~al.}(2003){Donati}, {Collier Cameron}, {Semel},
  {Hussain}, {Petit}, {Carter}, {Marsden}, {Mengel}, {L{\'o}pez Ariste},
  {Jeffers}, \& {Rees}}]{Donati03}
{Donati}, J.-F., {et~al.} 2003, \mnras, 345, 1145

\bibitem[{{D'Silva} \& {Choudhuri}(1993)}]{DSC93}
{D'Silva}, S., \& {Choudhuri}, A.~R. 1993, \aap, 272, 621

\bibitem[{{Gastine} {et~al.}(2012){Gastine}, {Duarte}, \& {Wicht}}]{GDW12}
{Gastine}, T., {Duarte}, L., \& {Wicht}, J. 2012, \aap, 546, A19

\bibitem[{{Ghizaru} {et~al.}(2010){Ghizaru}, {Charbonneau}, \&
  {Smolarkiewicz}}]{GCS10}
{Ghizaru}, M., {Charbonneau}, P., \& {Smolarkiewicz}, P.~K. 2010, \apjl, 715,
  L133

\bibitem[{{Gilman}(1983)}]{Gi83}
{Gilman}, P.~A. 1983, \apjs, 53, 243

\bibitem[{{Glatzmaier}(1985)}]{Gl85}
{Glatzmaier}, G.~A. 1985, \apj, 291, 300

\bibitem[{{Glatzmaier}(1987)}]{Gl87}
{Glatzmaier}, G.~A. 1987, ASSL, 137, 263

\bibitem[{{Goudard} \& {Dormy}(2008)}]{GD08}
{Goudard}, L., \& {Dormy}, E. 2008, EL, 83, 59001

\bibitem[{{Guerrero} \& {K{\"a}pyl{\"a}}(2011)}]{GK11}
{Guerrero}, G., \& {K{\"a}pyl{\"a}}, P.~J. 2011, \aap, 533, A40

\bibitem[{{Hackman} {et~al.}(2001){Hackman}, {Jetsu}, \& {Tuominen}}]{Thomas01}
{Hackman}, T., {Jetsu}, L., \& {Tuominen}, I. 2001, \aap, 374, 171

\bibitem[{{Hathaway}(2011)}]{Ha11}
{Hathaway}, D.~H. 2011, arXiv:1103.1561

\bibitem[{{Jouve} \& {Brun}(2007)}]{JB07}
{Jouve}, L., \& {Brun}, A.~S. 2007, \aap, 474, 239

\bibitem[{{K{\"a}pyl{\"a}} {et~al.}(2009){K{\"a}pyl{\"a}}, {Korpi}, \&
  {Brandenburg}}]{KKB09a}
{K{\"a}pyl{\"a}}, P.~J., {Korpi}, M.~J., \& {Brandenburg}, A. 2009, \aap, 500,
  633

\bibitem[{{K{\"a}pyl{\"a}} {et~al.}(2010{\natexlab{a}}){K{\"a}pyl{\"a}},
  {Korpi}, \& {Brandenburg}}]{KKB10}
{K{\"a}pyl{\"a}}, P.~J., {Korpi}, M.~J., \& {Brandenburg}, A.
  2010{\natexlab{a}}, \mnras, 402, 1458

\bibitem[{{K{\"a}pyl{\"a}} {et~al.}(2010{\natexlab{b}}){K{\"a}pyl{\"a}},
  {Korpi}, {Brandenburg}, {Mitra}, \& {Tavakol}}]{KKBMT10}
{K{\"a}pyl{\"a}}, P.~J., {Korpi}, M.~J., {Brandenburg}, A., {Mitra}, D., \&
  {Tavakol}, R. 2010{\natexlab{b}}, AN, 331, 73

\bibitem[{{K{\"a}pyl{\"a}} {et~al.}(2006{\natexlab{a}}){K{\"a}pyl{\"a}},
  {Korpi}, {Ossendrijver}, \& {Stix}}]{KKOS06}
{K{\"a}pyl{\"a}}, P.~J., {Korpi}, M.~J., {Ossendrijver}, M., \& {Stix}, M.
  2006{\natexlab{a}}, \aap, 455, 401

\bibitem[{{K{\"a}pyl{\"a}} {et~al.}(2004){K{\"a}pyl{\"a}}, {Korpi}, \&
  {Tuominen}}]{KKT04}
{K{\"a}pyl{\"a}}, P.~J., {Korpi}, M.~J., \& {Tuominen}, I. 2004, \aap, 422, 793

\bibitem[{{K{\"a}pyl{\"a}} {et~al.}(2006{\natexlab{b}}){K{\"a}pyl{\"a}},
  {Korpi}, \& {Tuominen}}]{KKT06}
{K{\"a}pyl{\"a}}, P.~J., {Korpi}, M.~J., \& {Tuominen}, I. 2006{\natexlab{b}},
  AN, 327, 884

\bibitem[{{K{\"a}pyl{\"a}} {et~al.}(2011{\natexlab{a}}){K{\"a}pyl{\"a}},
  {Mantere}, \& {Brandenburg}}]{KMB11}
{K{\"a}pyl{\"a}}, P.~J., {Mantere}, M.~J., \& {Brandenburg}, A.
  2011{\natexlab{a}}, AN, 332, 883

\bibitem[{{K{\"a}pyl{\"a}} {et~al.}(2012){K{\"a}pyl{\"a}}, {Mantere}, \&
  {Brandenburg}}]{KMB12a}
{K{\"a}pyl{\"a}}, P.~J., {Mantere}, M.~J., \& {Brandenburg}, A. 2012, \apjl,
  755, L22

\bibitem[{{K{\"a}pyl{\"a}} {et~al.}(2011{\natexlab{b}}){K{\"a}pyl{\"a}},
  {Mantere}, {Guerrero}, {Brandenburg}, \& {Chatterjee}}]{KMGBC11}
{K{\"a}pyl{\"a}}, P.~J., {Mantere}, M.~J., {Guerrero}, G., {Brandenburg}, A.,
  \& {Chatterjee}, P. 2011{\natexlab{b}}, \aap, 531, A162

\bibitem[{{Kitchatinov} \& {Mazur}(2000)}]{KM00}
{Kitchatinov}, L.~L., \& {Mazur}, M.~V. 2000, \solphys, 191, 325

\bibitem[{{Kitchatinov} \& {Olemskoy}(2012)}]{KO11}
{Kitchatinov}, L.~L., \& {Olemskoy}, S.~V. 2012, \solphys, 276, 3

\bibitem[{{Kitchatinov} {et~al.}(1994){Kitchatinov}, {Pipin}, \&
  {R\"udiger}}]{KPR94}
{Kitchatinov}, L.~L., {Pipin}, V.~V., \& {R\"udiger}, G. 1994, Astron. Nachr.,
  315, 157

\bibitem[{{Kitchatinov} \& {R\"udiger}(1995)}]{KR95}
{Kitchatinov}, L.~L., \& {R\"udiger}, G. 1995, \aap, 299, 446

\bibitem[{{Kochukhov} {et~al.}(2013){Kochukhov}, {Mantere}, {Hackman}, \&
  {Ilyin}}]{Oleg11}
{Kochukhov}, O., {Mantere}, M.~J., {Hackman}, T., \& {Ilyin}, I. 2013, \aap,
  550, A84

\bibitem[{{Korhonen} {et~al.}(2007){Korhonen}, {Berdyugina}, {Hackman},
  {Ilyin}, {Strassmeier}, \& {Tuominen}}]{Heidi07}
{Korhonen}, H., {Berdyugina}, S.~V., {Hackman}, T., {Ilyin}, I.~V.,
  {Strassmeier}, K.~G., \& {Tuominen}, I. 2007, \aap, 476, 881

\bibitem[{Krause \& R{\"a}dler(1980)}]{KR80}
Krause, F., \& R{\"a}dler, K.-H. 1980, {Mean-field Magnetohydrodynamics and
  Dynamo Theory} (Oxford: Pergamon Press)

\bibitem[{{K{\"u}ker} {et~al.}(2001){K{\"u}ker}, {R{\"u}diger}, \&
  {Schultz}}]{KRS01}
{K{\"u}ker}, M., {R{\"u}diger}, G., \& {Schultz}, M. 2001, \aap, 374, 301

\bibitem[{{Leighton}(1969)}]{Lei69}
{Leighton}, R.~B. 1969, \apj, 156, 1

\bibitem[{{Lindborg} {et~al.}(2011){Lindborg}, {Korpi}, {Hackman}, {Tuominen},
  {Ilyin}, \& {Piskunov}}]{Marjaana11}
{Lindborg}, M., {Korpi}, M.~J., {Hackman}, T., {Tuominen}, I., {Ilyin}, I., \&
  {Piskunov}, N. 2011, \aap, 526, A44

\bibitem[{{Malkus} \& {Proctor}(1975)}]{MP75}
{Malkus}, W.~V.~R., \& {Proctor}, M.~R.~E. 1975, JFM, 67, 417

\bibitem[{{Miesch} {et~al.}(2006){Miesch}, {Brun}, \& {Toomre}}]{MBT06}
{Miesch}, M.~S., {Brun}, A.~S., \& {Toomre}, J. 2006, \apj, 641, 618

\bibitem[{{Miesch} {et~al.}(2000){Miesch}, {Elliott}, {Toomre}, {Clune},
  {Glatzmaier}, \& {Gilman}}]{METCGG00}
{Miesch}, M.~S., {Elliott}, J.~R., {Toomre}, J., {Clune}, T.~L., {Glatzmaier},
  G.~A., \& {Gilman}, P.~A. 2000, \apj, 532, 593

\bibitem[{{Miesch} {et~al.}(2012){Miesch}, {Featherstone}, {Rempel}, \&
  {Trampedach}}]{MFRT12}
{Miesch}, M.~S., {Featherstone}, N.~A., {Rempel}, M., \& {Trampedach}, R. 2012,
  \apj, 757, 128

\bibitem[{{Mitra} {et~al.}(2009){Mitra}, {Tavakol}, {Brandenburg}, \&
  {Moss}}]{MTBM09}
{Mitra}, D., {Tavakol}, R., {Brandenburg}, A., \& {Moss}, D. 2009, \apj, 697,
  923

\bibitem[{{Mitra} {et~al.}(2010){Mitra}, {Tavakol}, {K{\"a}pyl{\"a}}, \&
  {Brandenburg}}]{MTKB10}
{Mitra}, D., {Tavakol}, R., {K{\"a}pyl{\"a}}, P.~J., \& {Brandenburg}, A. 2010,
  \apjl, 719, L1

\bibitem[{{Moss} \& {Brandenburg}(1995)}]{MB95}
{Moss}, D., \& {Brandenburg}, A. 1995, GApFD, 80, 229

\bibitem[{{Nelson} {et~al.}(2013){Nelson}, {Brown}, {Brun}, {Miesch}, \&
  {Toomre}}]{NBBMT13}
{Nelson}, N.~J., {Brown}, B.~P., {Brun}, A.~S., {Miesch}, M.~S., \& {Toomre},
  J. 2013, \apj, 762, 73

\bibitem[{{Ol{\'a}h} {et~al.}(2000){Ol{\'a}h}, {Koll{\'a}th}, \&
  {Strassmeier}}]{OKS00}
{Ol{\'a}h}, K., {Koll{\'a}th}, Z., \& {Strassmeier}, K.~G. 2000, \aap, 356, 643

\bibitem[{{Ossendrijver}(2003)}]{O03}
{Ossendrijver}, M. 2003, \aapr, 11, 287

\bibitem[{{Parker}(1955{\natexlab{a}})}]{Pa55b}
{Parker}, E.~N. 1955{\natexlab{a}}, \apj, 122, 293

\bibitem[{{Parker}(1955{\natexlab{b}})}]{Pa55a}
{Parker}, E.~N. 1955{\natexlab{b}}, \apj, 121, 491

\bibitem[{{Parker}(1987{\natexlab{a}})}]{Pa87b}
{Parker}, E.~N. 1987{\natexlab{a}}, \apj, 321, 984

\bibitem[{{Parker}(1987{\natexlab{b}})}]{Pa87a}
{Parker}, E.~N. 1987{\natexlab{b}}, \solphys, 110, 11

\bibitem[{{Pipin}(2013)}]{Pi13}
{Pipin}, V.~V. 2013, in IAU Symposium, Vol. 294, IAU Symposium, ed. A.~G.
  {Kosovichev}, E.~{de Gouveia Dal Pino}, \& Y.~{Yan}, 375--386

\bibitem[{{Pipin} \& {Kosovichev}(2011)}]{PK11}
{Pipin}, V.~V., \& {Kosovichev}, A.~G. 2011, \apjl, 727, L45

\bibitem[{{Pipin} \& {Seehafer}(2009)}]{PS09}
{Pipin}, V.~V., \& {Seehafer}, N. 2009, \aap, 493, 819

\bibitem[{{Pouquet} {et~al.}(1976){Pouquet}, {Frisch}, \& {L\'eorat}}]{PFL76}
{Pouquet}, A., {Frisch}, U., \& {L\'eorat}, J. 1976, JFM, 77, 321

\bibitem[{{Racine} {et~al.}(2011){Racine}, {Charbonneau}, {Ghizaru}, {Bouchat},
  \& {Smolarkiewicz}}]{RCGBS11}
{Racine}, {\'E}., {Charbonneau}, P., {Ghizaru}, M., {Bouchat}, A., \&
  {Smolarkiewicz}, P.~K. 2011, \apj, 735, 46

\bibitem[{{R\"adler} \& {Br\"auer}(1987)}]{Ra87}
{R\"adler}, K.-H., \& {Br\"auer}, H.-J. 1987, AN, 308, 101

\bibitem[{{Rempel}(2005)}]{Re05}
{Rempel}, M. 2005, \apj, 622, 1320

\bibitem[{{Rempel}(2006)}]{Re06}
{Rempel}, M. 2006, \apj, 647, 662

\bibitem[{{Robinson} \& {Chan}(2001)}]{RC01}
{Robinson}, F.~J., \& {Chan}, K.~L. 2001, \mnras, 321, 723

\bibitem[{{R\"udiger}(1989)}]{R89}
{R\"udiger}, G. 1989, {Differential Rotation and Stellar Convection. Sun and
  Solar-type Stars} (Berlin: Akademie Verlag)

\bibitem[{{R\"udiger} \& {Brandenburg}(1995)}]{RB95}
{R\"udiger}, G., \& {Brandenburg}, A. 1995, \aap, 296, 557

\bibitem[{{R{\"u}diger} \& {Hollerbach}(2004)}]{RH04}
{R{\"u}diger}, G., \& {Hollerbach}, R. 2004, {The Magnetic Universe:
  Geophysical and Astrophysical Dynamo Theory} (Weinheim: Wiley-VCH)

\bibitem[{{Saar} \& {Brandenburg}(1999)}]{SB99}
{Saar}, S.~H., \& {Brandenburg}, A. 1999, \apj, 524, 295

\bibitem[{{Schou} {et~al.}(1998){Schou}, {Antia}, {Basu}, {Bogart}, {Bush},
  {Chitre}, {Christensen-Dalsgaard}, {di Mauro}, {Dziembowski}, {Eff-Darwich},
  {Gough}, {Haber}, {Hoeksema}, {Howe}, {Korzennik}, {Kosovichev}, {Larsen},
  {Pijpers}, {Scherrer}, {Sekii}, {Tarbell}, {Title}, {Thompson}, \&
  {Toomre}}]{Schouea98}
{Schou}, J., {et~al.} 1998, \apj, 505, 390

\bibitem[{{Schrinner} {et~al.}(2012){Schrinner}, {Petitdemange}, \&
  {Dormy}}]{SPD12}
{Schrinner}, M., {Petitdemange}, L., \& {Dormy}, E. 2012, \apj, 752, 121

\bibitem[{{Schrinner} {et~al.}(2005){Schrinner}, {R{\"a}dler}, {Schmitt},
  {Rheinhardt}, \& {Christensen}}]{SRSRC05}
{Schrinner}, M., {R{\"a}dler}, K.-H., {Schmitt}, D., {Rheinhardt}, M., \&
  {Christensen}, U. 2005, AN, 326, 245

\bibitem[{{Schrinner} {et~al.}(2007){Schrinner}, {R{\"a}dler}, {Schmitt},
  {Rheinhardt}, \& {Christensen}}]{SRSRC07}
{Schrinner}, M., {R{\"a}dler}, K.-H., {Schmitt}, D., {Rheinhardt}, M., \&
  {Christensen}, U.~R. 2007, GApFD, 101, 81

\bibitem[{{Simard} {et~al.}(2013){Simard}, {Charbonneau}, \& {Bouchat}}]{SCB13}
{Simard}, C., {Charbonneau}, P., \& {Bouchat}, A. 2013, \apj, 768, 16

\bibitem[{{Spruit}(2000)}]{Spr00}
{Spruit}, H.~C. 2000, \ssr, 94, 113

\bibitem[{{Spruit}(2003)}]{Spr03}
{Spruit}, H.~C. 2003, \solphys, 213, 1

\bibitem[{{Stefani} \& {Gerbeth}(2003)}]{SG03}
{Stefani}, F., \& {Gerbeth}, G. 2003, \pre, 67, 027302

\bibitem[{{Stix}(1976)}]{St76}
{Stix}, M. 1976, \aap, 47, 243

\bibitem[{{Tobias}(1998)}]{To98}
{Tobias}, S.~M. 1998, \mnras, 296, 653

\bibitem[{{Warnecke} {et~al.}(2011){Warnecke}, {Brandenburg}, \&
  {Mitra}}]{WBM11}
{Warnecke}, J., {Brandenburg}, A., \& {Mitra}, D. 2011, \aap, 534, A11

\bibitem[{{Warnecke} {et~al.}(2012){Warnecke}, {K{\"a}pyl{\"a}}, {Mantere}, \&
  {Brandenburg}}]{WKMB12}
{Warnecke}, J., {K{\"a}pyl{\"a}}, P.~J., {Mantere}, M.~J., \& {Brandenburg}, A.
  2012, \solphys, 280, 299

\bibitem[{{Warnecke} {et~al.}(2013){Warnecke}, {K{\"a}pyl{\"a}}, {Mantere}, \&
  {Brandenburg}}]{WKMB13}
{Warnecke}, J., {K{\"a}pyl{\"a}}, P.~J., {Mantere}, M.~J., \& {Brandenburg}, A.
  2013, \apj, 778, 141

\bibitem[{{Yoshimura}(1976)}]{Yo76}
{Yoshimura}, H. 1976, \solphys, 50, 3

\bibitem[{{Zhao} {et~al.}(2013){Zhao}, {Bogart}, {Kosovichev}, {Duvall}, \&
  {Hartlep}}]{ZBKDH13}
{Zhao}, J., {Bogart}, R.~S., {Kosovichev}, A.~G., {Duvall}, Jr., T.~L., \&
  {Hartlep}, T. 2013, \apj, 774, L29

\end{thebibliography}

\end{document}